\documentclass[onecolumn, superscriptaddress, prb,11pt]{revtex4-2}

\usepackage{hyperref}
\hypersetup{backref,colorlinks=true,allcolors=prussianblue}

\usepackage{amsmath,amssymb}

\usepackage{graphicx}
\graphicspath{{./Figures/}}
\usepackage[caption=false]{subfig}
\usepackage{setspace}
\usepackage{float}

\usepackage{rotating}

\DeclareUnicodeCharacter{2009}{\,} 
\DeclareUnicodeCharacter{2212}{-}
\DeclareUnicodeCharacter{202F}{\,}

\usepackage{xcolor}
\definecolor{sangria}{rgb}{0.57, 0.0, 0.04}
\definecolor{arsenic}{rgb}{0.23, 0.27, 0.29}
\definecolor{prussianblue}{rgb}{0.0, 0.19, 0.33}
\definecolor{phthalogreen}{rgb}{0.07, 0.21, 0.14}
\definecolor{dgreen}{rgb}{0.0, 0.4, 0.2}

\usepackage[most]{tcolorbox}
\usepackage{tikz}
\usetikzlibrary{calc,arrows,shapes,positioning}
\usetikzlibrary{patterns,snakes}
\usetikzlibrary{positioning,decorations.pathreplacing}

\tikzstyle{block0} = [rectangle, draw=arsenic!50,
    text width=10em, text centered,  minimum height=1.5cm,line width=1mm]
\tikzstyle{block1} = [rectangle, draw=arsenic!50, fill=arsenic!20, 
    text width=10em, text centered,  minimum height=1.5cm,line width=1mm]
\tikzstyle{block2} = [rectangle, draw=sangria!50, fill=sangria!20, 
    text width=10em, text centered,  minimum height=1.5cm,line width=1mm]    
\tikzstyle{block3} = [rectangle, draw=prussianblue!50, fill=prussianblue!20, 
    text width=10em, text centered, minimum height=1.5cm,line width=1mm]    
\tikzstyle{block4} = [rectangle, draw=phthalogreen!50, fill=phthalogreen!20, 
    text width=10em, text centered, minimum height=1.5cm,line width=1mm]    
    
\tikzstyle{line} = [draw=arsenic, -latex', line width=0.1cm]

\usepackage{atbegshi}
\def\ep{\mathcal{E}}
\def\der#1{d\mathbf{#1}\;}
\def\im{\mathit{i}}
\def\gw{G$_0$W$_0$}

\def\fig#1{Fig.~\ref{#1}}
\def\eq#1{Eq.~\eqref{eq:#1}}
\def\tab#1{Table~\ref{#1}}

\begin{document}

\author{Okan K. Orhan}
\affiliation{Department of Mechanical Engineering, University of British Columbia, 2054 - 6250 Applied Science Lane, Vancouver, BC, V6T 1Z4, Canada}

\author{Mauricio Ponga}
\affiliation{Department of Mechanical Engineering, University of British Columbia, 2054 - 6250 Applied Science Lane, Vancouver, BC, V6T 1Z4, Canada}

\title{Surface-plasmon properties of noble metals with exotic phases}

\begin{abstract} 
Noble-metal nanoparticles have been the industry standard for plasmonic applications due to their highly populated plasmon generations.
Despite their remarkable plasmonic performance, their widespread use in plasmonic applications is commonly hindered due to limitations on the available laser sources and relatively low operating temperatures needed to retain mechanical strength in these materials. 
Motivated by recent experimental works, in which exotic hexagonal-closed-packed (HCP) phases have been identified in gold (Au), silver (Ag) and copper (Cu), we present the plasmonic performance of two HCP polytypes in these materials using high-accuracy first-principles simulations.
The isolated HCP phases commonly reach thermal and mechanical stability at high temperatures due to monotonically decreasing Gibbs free energy differences compared to the face-centered cubic (FCC) phases.  
We find that several of these polytypes are harder and produce bulk plasmons at lower energies with comparable lifetimes than their conventional FCC counterparts. 
It also leads to the localized surface-plasmon resonance (LSPR) in perfectly spherical HCP-phased  nanoparticles, embedded onto dielectric matrices,  at substantially lower energies  with comparable lifetimes to their FCC counterparts.
LSPR peak locations and lifetimes can be tuned by controlling the operational temperature, the dielectric permittivity of hosting matrix and the grain size.
Our work suggests that noble-metal nanoparticles can be tailored to develop exotic HCP phases to obtain novel plasmonic properties.

\end{abstract}

\maketitle

\section{Introduction}\label{sec:s1}
Nobel metal nanoparticles (NPs) and thin films are the most commonly used plasmonic media in diverse opto-electronic applications such as data recording~\cite{black2000magnetic,Mansuripur:09,sato2009heat,zou2014recording}, bio-sensing and imaging~\cite{Alagiri2017,bengali2018gold,RODRIGUES201874,doi:10.1021/ar7002804,doi:10.1021/jp062536y}, photo-induced and heat-assisted drug delivery~\cite{Medici2015329,JAIN200718,Zhang2015,Tran2017}.
Most of these applications are commonly operated above room temperature (around $\sim 300-400$~K) where these metals tend to easily deform due to their relatively low stacking fault energy and high ductility\cite{Abbott2019}.
There have been considerable efforts to improve the mechanical properties of noble-metal-based NPs while retaining or improving plasmonic properties with different strategies such as alloying~\cite{Blaber_2010,ADOM:ADOM201500446,PhysRevB.6.1209,Nishijima2014,doi:10.1021/acsphotonics.5b00586,hashimoto2016ag}. 
However, substitutional alloying of noble metals commonly reduces plasmonic performance at the most common operational wavelengths~\cite{Orhan_2019,Bello:19}. 
Therefore, the quest for finding novel materials capable of generating strong plasmons with significant lifetimes (in the order of $\geq 1$~fs) and at lower energies remains an open challenge.  

\begin{figure}[t]
\centering

\includegraphics[width=0.5\textwidth]{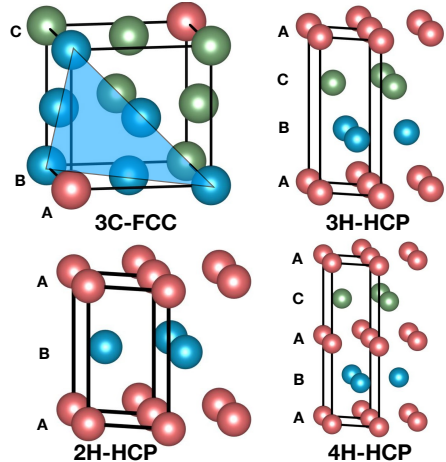}

\caption{Representative face-centered cubic structure (top-left) with the \{111\} plane highlighted in blue. The 3H-hexagonal-closed-packed equivalent structure with the ABC stacking sequence (top-right). 2H-HCP and 4H-HCP polytypes are shown at the bottom left and right panels, respectively. The relative stacking of \{111\} planes is also shown.}
\label{fig:Rep-Str}
\end{figure}

Noble metals such as Au, Ag and Cu most commonly appear in the face-centered cubic (FCC) phase which has a closed-packed atomic-plane arrangement of the ABC-stacking sequence along the \{111\} crystallographic plane, as illustrated in \fig{fig:Rep-Str}.
However, there has been a considerable effort to develop more exotic phases such 2H-HCP (AB-stacking sequence) and 4H-HCP (ABAC-stacking sequence) phases, illustrated in \fig{fig:Rep-Str}.
Huang \textit{et al.} has shown that Au-2H-HCP thin films can be grown on graphene oxide sheets~\cite{Huang2011}, while Bian \textit{et al.} has successfully produced Au-2H-HCP NPs in controlled-diffusion super-crystals~\cite{Bian2018}.
Fan \textit{et al.} has also synthesized FCC-2H-HCP-FCC hetero-phase nanoribbons~\cite{Fan2020}. 
Similarly, it has been shown that the Au-4H-HCP polytype can be stabilized into nanoribbons~\cite{Fan2015} and metastable nanorods~\cite{Han2020}. 
Benaissa and Ferhat \cite{BENAISSA2017170} have demonstrated that 2H-, 4H-, and 6H-HCP Au phases can be thermally and dynamically can be stabilized using first-principles methods.

Chakraborty \textit{et al.} have managed to grow the meta-stable thick films of Ag-4H-HCP and two-dimensional Ag-2H-HCP using the controlled electrochemical deposition on the silicon and glass substrates~\cite{Chakraborty_2011}. 
They have also shown that the 4H-HCP phase is relatively stable using the first-principles simulations.
Recent experimental work carried out by Thevamaran \textit{et al.} have shown that the 2H-HCP and 4H-HCP phases of Ag can be achieved by a simultaneous martensitic phase transformation when Ag micro-particles are impacted onto an obstacle at a high velocity using laser-induced projectile impact test (LIPIT)~\cite{Thevamaran:2016, Thevamaran:2019}. 
This martensitic phase transformation has been further corroborated by Funes \textit{et al.}~\cite{ROJAS2021116892} 
In these works, Ag NPs formed gradient nano-grained (GNG) structures~\cite{Fang:2011} due to a decaying shock pressure providing a vast domain of rich microstructures to engineer material properties. 
Other works have documented Ag with HCP structure, in particular when extracted from Ag/Au ores of north Russia, when synthesized using a high-pressure magnetron, in nanorods (diameter $\sim10-100$ nm), nanoribbons, and bulk films~\cite{doi:10.1080/00206818109455083,WETLI1997876,doi:10.1021/acs.jpcc.6b02169}. 
Moreover, using solution-phase epitaxial growth under ambient conditions, 4H hexagonal Cu was grown on 4H Au nanoribbons~\cite{Fan2015}. 
This epitaxial growth was also extended to other materials such as Ir, Rh, Os, Ru as well~\cite{C6SC02953A}.

However, plasmonic properties of such architected materials with HCP-based polytypes remain unavailable since it is relatively difficult to synthesize these polytypes. 
This issue hinders the experimental characterization of such materials for plasmonic properties. %
Fan \textit{et al.} has performed the monochromated electron-energy loss spectroscopy (EELS) measurements on the thin 2H-HCP and 4H-HCP Au nanoribbons and calculated the dielectric functions within a semi-empirical approach~\cite{Fan2015,Fan2020}.
In that work, they have used the experimentally available data for the FCC Au to describe the low-energy spectral range.
Bian \textit{et al.} \cite{} has predicted a localized surface plasmon resonance (LSPR) of the Au-2H-HCP super-crystals in the ordered arrays and found a strong plasmonic peak at $\sim630$~nm ($\sim1.97$~eV)\cite{Bian2018}.

We herein present a systematic and comprehensive study of the stability assessment and temperature-dependent material properties of the 2H-HCP and 4H-HCP phases of Au, Ag and Cu within the fully first-principles approaches. 
We first investigate the thermal and mechanical stabilities of the isolated HCP crystals using density-functional theory (DFT)~\cite{PhysRev.136.B864,PhysRev.140.A1133} and its perturbative extension called the density-functional perturbation theory (DFPT)~\cite{PhysRevB.55.10337,PhysRevB.55.10355}.
It is shown that the underlying KS band-structure is not sufficiently well-defined when the random-phase approximation (RPA) ~\cite{PhysRev.82.625,PhysRev.85.338,PhysRev.92.609,RevModPhys.36.844} is used to describe the inter-band transitions in noble metals.
The \gw~\cite{Aryasetiawan_1998,Orhan_2019,RANGEL2020107242} method is applied to obtain the approximate quasi-particle (QP) band-structures~\cite{landau1957theory,landau1957oscillations,landau1959theory} to improve accuracy of the optical simulations.
The intra-band part, also called the Drude plasmon, is included in the Drude-Lorentz model~\cite{ANDP:ANDP19003060312,ANDP:ANDP19003081102,lorentz1909theory,fox2002optical} using the first-principles temperature-dependent plasmonic parameters.

Inspired by these recent experimental works on stabilizing NPs of the exotic noble-metal phases, we analyze the LSPR in perfectly spherical NPs of 2H-HCP and 4H-HCP phases embedded onto a dielectric substrate such as silica and compare them to the FCC phase. 
The effects of hosting matrix and NP size are analyzed. 
An operational wavelength and temperature of $830$~nm and $400$~K are chosen as they are the most common operational condition in the emerging heat-assisted magnetic recording (HAMR)~\cite{black2000magnetic, zou2014recording,Abadia:18} where the high-power near-field-transducer (NFT)~\cite{Datta2017,YANG2017159,Abadia:18} is used as the plasmonic antenna to achieve temporary and instant heating of a local region in the recording medium above its Curie temperature.  
The current NFT designs are mainly based on the FCC Au, which is soft and may protrude or deform for temperature rises as little as a few tens of Kelvin~\cite{Abbott2019}.
We investigate the thermal and mechanical properties of the HCP-based polytypes to assess their relative performance compared to the FCC Au.

The discoveries mentioned above and technologically relevant problems are the motivation behind this work. 
However, beyond that, this work provides a well-defined workflow and practical computational experiments for plasmonic materials. 
A comprehensive  Supplementary Information (SI) is available for the readers to provide a thorough background on the electronic structure, theoretical spectroscopy, thermal and mechanical properties of metallic solids, and further data analysis on the $100-500$~K temperature range. 

\section{Theoretical and Computational Methods}
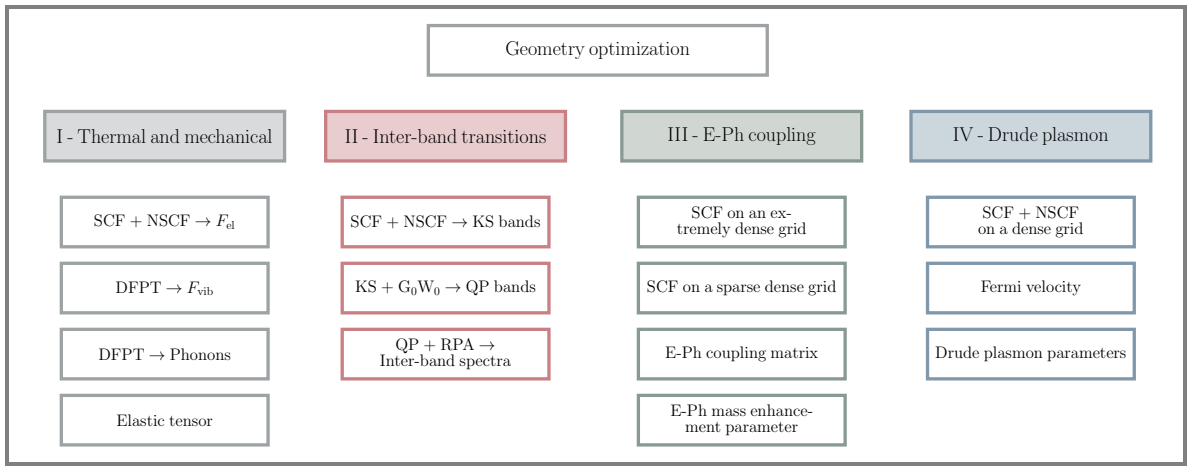
\begin{figure}[b]
\centering
\tcbox[sharp corners, boxsep=0.0mm, boxrule=0.5mm,  colframe=gray, colback=white]{
\resizebox{0.9\textwidth}{!}{
\begin{tikzpicture}
    \node [block0,text width=10cm,] (vcr) {\LARGE Geometry optimization};
    
    \node [block1,text width=7cm,below left=1 and 4.25cm of vcr] (thermo){\LARGE I - Thermal and mechanical};   
    
    \node [block1,fill=white, text width=6cm,below=1 of thermo] (t1){\Large SCF + NSCF $\rightarrow$ $F_\mathrm{el}$};   
    
    \node [block1,fill=white, text width=6cm,below=3 of thermo] (t2){\Large DFPT $\rightarrow$ $F_\mathrm{vib}$};
    
    \node [block1,fill=white, text width=6cm,below=5 of thermo] (t3){\Large DFPT $\rightarrow$ Phonons};
    
     \node [block1,fill=white, text width=6cm,below=7 of thermo] (t4){\Large  Elastic tensor};

	\node [block2,text width=7cm,below left=1 and -4.25cm of vcr] (gwrpa){\LARGE II - Inter-band transitions};    
	
	 \node [block2,fill=white, text width=6cm,below=1 of gwrpa] (g1){\Large SCF + NSCF $\rightarrow$ KS bands};  
	 
	 \node [block2,fill=white, text width=6cm,below=3 of gwrpa] (g3){\Large KS + G$_0$W$_0$ $\rightarrow$ QP bands};  
	 
	  \node [block2,fill=white, text width=6cm,below=5 of gwrpa] (g3){\Large QP + RPA $\rightarrow$ Inter-band spectra};

    \node [block4,text width=7cm,below right=1 and -4.5cm of vcr] (elph){\LARGE III - E-Ph coupling};  
    
   	\node [block4,fill=white, text width=6cm,below=1 of elph] (d1){\Large SCF on an extremely dense grid};     
    
    \node [block4,fill=white, text width=6cm,below=3 of elph] (d1){\Large SCF on a sparse dense grid};

     \node [block4,fill=white, text width=6cm,below=5 of elph] (d1){\Large E-Ph coupling matrix};  
     
     \node [block4,fill=white, text width=6cm,below=7 of elph] (d1){\Large E-Ph mass enhancement parameter};

    \node [block3,text width=7cm,below right=1 and 4.25cm of vcr] (dp){\LARGE IV - Drude plasmon};
    
   	\node [block3,fill=white, text width=6cm,below=1 of dp] (d1){\Large SCF + NSCF on a dense grid}; 
   	  
    \node [block3,fill=white, text width=6cm,below=3 of dp] (d2){\Large Fermi velocity}; 
    
    \node [block3,fill=white, text width=6cm,below=5 of dp] (d3){\Large Drude plasmon parameters};

\end{tikzpicture}}}
\caption{The schematic illustration of workflow used for the first-principle simulations. SCF:Self-consistent field, NSCF: Non-SCF, $F_\mathrm{el}$: Electronic Helmholtz free energy, $F_\mathrm{vib}$: Vibrational Helmholtz free energy, DFPT: Density-functional perturbation theory, E-Ph: Electron-phonon, KS: Kohn-Sham, QP: Quasi-particle
}
\label{fig:wf}
\end{figure}

In this section, we introduce the theoretical concepts, their computational implementations and limitations.
A schematic illustration of the workflow used for the first-principle simulations is presented in \fig{fig:wf}.

Kohn-Sham DFT (KS-DFT)~\cite{PhysRev.136.B864,PhysRev.140.A1133} using semi-local electronic exchange-correlation (xc-) functionals~\cite{PhysRev.140.A1133,PhysRevB.21.5469,PhysRevB.33.8822,PhysRevB.46.6671} is an almost-ubiquitous approach for simulating ground-state properties of solids.
It also offers a computationally feasible framework to obtain the second-order elastic tensor within Hooke's law and the lattice dynamics within the density-functional perturbation theory (DFPT)~\cite{PhysRevB.55.10337,PhysRevB.55.10355}.
It also provides a computationally feasible starting point for theoretical spectroscopy simulations such the random-phase approximation (RPA)~\cite{PhysRev.82.625,PhysRev.85.338,PhysRev.92.609,RevModPhys.36.844}.

\subsection{Figures of merits for stability assessment}
Three stability assessments are necessary to ensure stability of the HCP-based structures. 
It is important to note that the stability assessment of the HCP phases are performed for the isolated phases. 
Commonly, they have been grown within complex hetero-phased structures for which the thermal and mechanical stability is achieved through complex mechanisms beyond the scope of this work.  

For the sake of a direct comparability, we  alternatively use a HCP unit cell with the ABC-stacking sequence to represent FCC phase, as shown in \fig{fig:Rep-Str}.
This FCC-equivalent phase is conveniently called 3H-HCP to achieve a more systematic and simplified notation throughout thus work. 
To ensure that the 3H-HCP unit cell are structurally equivalent to the FCC unit cell, the corresponding structural parameters of their relaxed geometries are compared in  \tab{tab:FCC-3H-Crys-Comp}. 
The percentage relative differences are $\leq \%1$ for the lattice parameters and the angle between A-B-C atoms in \fig{fig:Rep-Str}. 
The first line of the stability assessment is the thermal stability. 
The Gibbs free energy (GFE) is commonly used as the figure of merit (FOM) to assess formation of the competing phases in a solid. 
Using the GFE of 3H-HCP phases as the reference systems, the relative GFEs of the HCP-based phases are evaluated by $\Delta G=G^\mathrm{nH-HCP}-G^\mathrm{3H-HCP}$ where $n=2,4$ (see SI for a brief summary of GFE of a non-magnetic, elemental pristine solid). 
The second line is the elastic-stability assessment which is evaluated according to the Born-Huang-stability criteria~\cite{Born:224197}(see SI for further details).
These criteria are indeed a special case of the thermal stability condition ensuring that the GFE of an unstrained crystal is a minimum compared to any other states achieved by infinitesimal strains~\cite{DEMAREST1977281}.
Otherwise, there is a high possibility of phase transformation to a lower symmetry. 
Finally, the dynamics-stability  assessment is evaluated by analyzing phonon dispersions. 
An unstressed crystal is expected to have only real (positive) phonon modes.
Harder phonon modes (higher frequencies) indicate superior dynamical stability as softer phonon modes signal possible phase transitions to lower symmetry  structures at a finite temperature~\cite{PhysRevB.97.134114}.
The necessary quantities are calculated through the first column in \fig{fig:wf}.

\subsection{Optical and plasmonic properties}
At the low-energy spectral range, the optical response of metallic solids is almost entirely electronic and due to two main physical phenomena, namely the inter-band and intra-band transitions, which can be symbolically expressed via the macroscopic complex dielectric function given by $\ep(\omega)=\ep^\mathrm{inter}(\omega)+\ep^\mathrm{intra}(\omega)=\ep_1(\omega)+ \im \ep_2(\omega)$. 
The macroscopic complex dielectric function often serves as the central optical function due to its well-established analytical connections with optical measurements.

The inter-band transitions are due to the vertical transitions of electrons between the occupied and unoccupied electronic states. 
This part can be perturbatively calculated using RPA build on the highly well-defined single-particle electronic states (see SI for further details).
It has been shown that RPA fails to accurately describe the inter-band transitions in Au, Ag, and Cu~\cite{Orhan_2019}.
This is due to the missing electron-hole quasiparticle (QP) screening within the approximate KS-DFT~\cite{landau1957theory,landau1957oscillations,landau1959theory}, leading to an inadequate description of the fully filled $d$-bands of bulk Au, Ag, and Cu~\cite{PhysRevB.86.125125, Orhan_2019,RANGEL2020107242}. 
The most common corrective approach to improve the accuracy of underlying electronic structures is to use the QP formalism~\cite{pines1966elementary,aulbur2000quasiparticle} due to its high compatibility with the approximate KS-DFT (see SI for further details). 
Within this formalism, the inter-band transitions are obtained through the second column in \fig{fig:wf}.

The intra-band transition, also called \emph{the Drude plasmon}, is due to the collective oscillations of the nearly free electrons in phase with the longitudinal part of the driving electromagnetic radiation.
Unlike the inter-band transition simulations, it requires a highly dense Brillouin zone sampling at and around the Fermi surface, i.e., at least around $\sim16000$ grid points~\cite{marini2001optical}, which is not feasible for high-throughput simulations. 
Thus, the Drude plasmon is classically treated within the Drude-Lorentz model given by~\cite{ANDP:ANDP19003060312,ANDP:ANDP19003081102,lorentz1909theory,fox2002optical}
\begin{align}\label{eq:eq1}
\ep^{\mathrm{intra}}(\omega)=
1-\frac{\omega_\mathrm{p}^2}{\omega^2+\im \eta_\mathrm{p} \omega},
\end{align}
where $\omega_\mathrm{p}$ and $\eta_\mathrm{p}$ are the Drude plasmon energy (plasma frequency) and the inverse lifetime, respectively. 
The Drude plasmon energy is approximately given by
\begin{align}\label{eq:eq2}
\omega_\mathrm{p}^2=\frac{4\pi}{3} N(E_\mathrm{F})\langle v^2_\mathrm{F} \rangle,
\end{align}
where $N(E_\mathrm{F})$ and $ v_\mathrm{F}$ are the density-of-state (DOS) at the Fermi level and the Fermi velocity, respectively. 
Within Matthiessen's rule, the inverse lifetime for a \textit{pristine} solid  can be expressed as the sum of the electron-electron (e-e) and the electron-phonon (e-ph) scatterings terms given by 
\begin{align}\label{eq:eq3}
\eta_\mathrm{p}=\eta_\mathrm{e-e}+\eta_\mathrm{e-ph}.
\end{align}
In general, the e-ph scattering is the predominant phenomenon; however, the e-e scattering is still relatively significant for noble metals compared to other metals~\cite{PhysRevB.13.673}.
The e-e scattering contribution to the inverse lifetime can be approximated within the Fermi liquid theory at the absolute-zero temperature by the expression (in the Hz unit) ~\cite{doi:10.1080/00018739300101514,PhysRevB.55.10869,DalForno2018}
\begin{align}\label{eq:eq4}
\eta_\mathrm{e-e}=\frac{m e^4 (E-E_\mathrm{F})^2}{64 \pi^3 \hbar^3 \epsilon_0 E_s^{3/2}E_\mathrm{F}^{1/2}}
\left(\frac{2\sqrt{E_sE_\mathrm{F}}}{4E_\mathrm{F}+E_s}+\arctan\sqrt{\frac{4E_\mathrm{F}}{E_s}}\right),
\end{align}
where $m$, $e$, $\hbar$, and $\epsilon_0$ are the electronic mass, the unit charge, the reduced Plank constant and the vacuum permittivity, respectively. 
$E_s$ is the kinetic energy associated with  the Thomas-Fermi screening length $q_s=e \sqrt{N(E_\mathrm{F})/\epsilon_0}$ given by $E_s=\hbar^2 q_s^2/(2m)$.
This term accounts for the scattering of electrons around the Fermi level. 
Hence, $(E-E_\mathrm{F})$ is consistently set to represent the smearing parameter used for the double-delta integral during the e-ph coupling simulations. 
Within the Debye model, the e-ph scattering term is approximately given by~\cite{PhysRevB.71.174302,Hofmann_2009}
\begin{align}\label{eq:eq5}
\eta_\mathrm{e-ph}=\frac{2\pi k_\mathrm{B} \lambda T }{3},
\end{align}
where $T$ is the absolute temperature, and $\lambda$ is the e-ph mass enhancement parameter.
We refer the reader to Refs.~\citenum{Hofmann_2009}~and~\citenum{PONCE2016116} for details on how to obtain the  e-ph mass enhancement parameter using DFPT, summarized in the third column in \fig{fig:wf}.

Assuming that the metallic bands have a parabolic dispersion normal to the Fermi surface, the averaged Fermi-velocity square in \eq{eq2} is approximated by~\cite{doi:10.1080/14786435808237011}
\begin{align}\label{eq:eq6}
\langle v^2(E_\mathrm{F})\rangle=
\frac{\Big(\sum_i\int_{S_\mathrm{F_i}}\der{k} \left|\frac{\partial E_{i,\mathbf{k}}}{\partial \mathbf{k}}\right|^2
 \Big)}
{\Big( \sum_j \int_{S_\mathrm{F_j}}\der{k'} \Big)},  
\end{align}
where $S_\mathrm{F_i}$ and $E_{i,\mathbf{k}}$ are the Fermi surface and the KS energy of the $i-\textrm{th}$ metallic band, respectively.
The average of the Fermi-velocity square is practically evaluated on a slab with a thickness of $d_\mathrm{Fermi}$. 
In principle, at the absolute zero, $d_\mathrm{Fermi}\rightarrow 0$; however, at a finite temperature, the Fermi surface is smeared due to excitation of electrons close to the Fermi surface.
The finite-temperature effect can be carried to the average of the Fermi-velocity square by setting $d_\mathrm{Fermi}=k_\mathrm{B}T$ representing the electronic temperature.
By doing so, the average of the Fermi-velocity becomes consistent with the inverse plasmon lifetime given by \eq{eq4} and \eq{eq5}.
Furthermore, the Fermi velocity is re-normalized by $\lambda$ and becomes $v_\mathrm{F} \rightarrow v_\mathrm{F}/(1+\lambda)$~\cite{Grimvall_1976}. 
The final Drude parameters are obtained by the fourth column in \fig{fig:wf}.

\subsection{Localized surface-plasmon generation in the perfectly spherical nanoparticles}
The condition of the LSPR generation in a metallic NP smaller than the incident wavelength and embedded onto a dielectric  hosting matrix is given by~\cite{Kreibig1995}.
\begin{align}\label{eq:eq7}
\ep_1=-2\ep_m,
\end{align}
where $\ep_1$ is the real part of the bulk macroscopic dielectric function of  the metallic NP and $\ep_m$ is the dielectric permittivity of the hosting matrix.
By this condition, the LSPR of a metallic NP becomes~\cite{YESHCHENKO2013275}
\begin{align}\label{eq:eq8}
\omega_\mathrm{sp}=\sqrt{\frac{\omega_\mathrm{p}^2}{1 +2\ep_\mathrm{m}+\ep_1^\mathrm{inter}}-\eta^2},
\end{align}
where $\eta$ is the inverse lifetime of the LSP generated on the NP.
This constant can be given for an arbitrarily shaped NP  given by~\cite{doi:10.1063/1.1587686,B604856K}
\begin{align}\label{eq:eq9}
\eta=\eta_\mathrm{p}+\alpha\frac{v_\mathrm{F}}{L_\mathrm{eff}}. 
\end{align}
$L_\mathrm{eff}=4V/A$ is the effective path length of hot carriers~\cite{doi:10.1063/1.1587686} where $V$ and $A$ are the volume and the surface of the NP, respectively.
$\alpha$ is the theory-dependent constant which is in the order of $1$.

\subsection{Figures of merits for mechanical properties}
In addition to high plasmonic performance, it is crucial for a candidate plasmonic to have good workability for easy integration in any design, mechanical strength, and hardness for durability.
Mechanical properties of complex materials such as the GNG structure are determined by a large number of parameters such as grain-boundary energies, grain sizes and gradient~\cite{argon2008strengthening,Wu7197,ZHOU20186}.
Despite that, the elastic constant of a perfect crystal can provide an initial assessment of its mechanical performance.  
Three simple FOMs can be used to assess workability, strength, and hardness, i.e., Poisson's ratio, Pugh ratio (ratio between bulk and shear moduli), and hardness. 
Commonly, a critical minimum-value (CMV) of $0.25$ for Poisson's ratio~\cite{doi:10.1080/14786440808520496} and a CMV of $1.75$ for the Pugh ratio~\cite{BOUCETTA201459} are expected for materials with large plastic deformation and ductility, respectively. 
The final FOM is the Vickers hardness ($H_\mathrm{V}$), which can be used as a measure for the yield strength ($\sigma_\mathrm{y}$) since they are approximately related by $H_\mathrm{V} \approx 3 \; \sigma_\mathrm{y}$~\cite{BROOKS2008412}.
$H_\mathrm{V}$ is commonly approximated by its semi-empirical relation to the shear modulus ($S$) by $H_\mathrm{V}^\mathrm{Teter}=0.151 S$~\cite{teter_1998}.

\subsection{Approximate thermal conductivity}
During the high-power NFT operations, the feedback heat has to be quickly dissipated to avoid excessive thermal softening and  plastic deformation~\cite{Abadia:18}. 
High thermal conductivity is desirable to endure high operational temperatures. 
Free electrons in metals are the main mediator for heat conduction. 
The electronic thermal conductivity can be approximated by substituting the electronic conductivity ($\sigma$) within the Drude model into the Wiedemann-Franz law (including the quantum effects), given by
\begin{align}\label{eq:eq10}
\kappa_\mathrm{e}&=\frac{\pi^2 k_\mathrm{B}^2 T \sigma}{3 e^2}, \quad \mathrm{where} \quad \sigma= \frac{2 e^2}{3} \frac{N(E_\mathrm{F})}{V_0}\frac{\langle v^2_\mathrm{F} \rangle}{(1+\lambda)^2}\eta_\mathrm{p}^{-1} \Rightarrow \nonumber \\
\kappa_\mathrm{e}&= \frac{2 \pi^2 k_\mathrm{B}^2 T}{9}\frac{N(E_\mathrm{F})}{V_0} \frac{\langle v^2_\mathrm{F} \rangle}{(1+\lambda)^2} \eta_\mathrm{p}^{-1},
\end{align}
where $V_0$ is the equilibrium volume and $2$ accounts for the spin-channels.

Second significant mediator of the heat conduction above the Debye temperature is the lattice thermal conductivity  mediated via phonons.
It is given by the semi-empirical formula~\cite{Morelli2006}
\begin{align}\label{eq:eq11}
\kappa_\mathrm{lat}=f(\gamma) \frac{m_\mathrm{avr}V_0^{1/3}\Theta_\mathrm{D}^3}{\gamma N^{2/3} T},
\end{align}
where $m_\mathrm{avr}$, $N$, $\Theta_\mathrm{D}$, and $\gamma$ are the average mass, the number of atoms, the Debye  temperature, and the Gr\"uneisen parameter, respectively (see SI for further details).
Finally, the scaling function is given by
\begin{align}\label{eq:eq12}
f(\gamma)=\frac{2.42814 \times 10^7}{1 + 0.228/\gamma^2 - 0.514/\gamma},
\end{align}
when $m_\mathrm{avr}$ in (kg) unit and $V_0$ in (m$^3$) unit for $\kappa_\mathrm{lat}$ in (W$\cdot$m$^{-1}$ $\cdot$ K$^{-1}$) unit.

\section{Result and Discussion}
\subsection{Stability assessment}
\begin{figure}[H]
\centering

\includegraphics[width=0.45\textwidth]{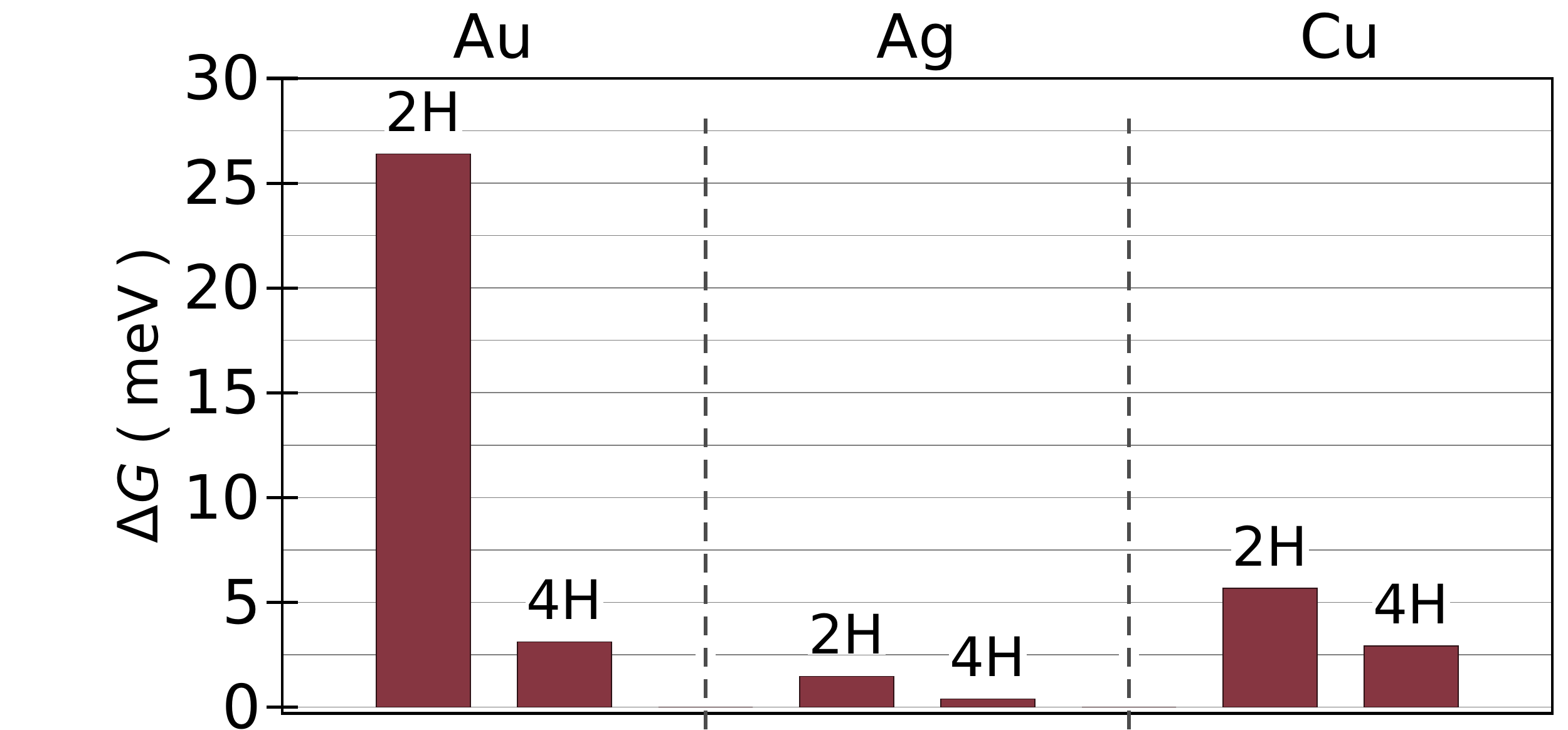}

\caption{Gibbs free-energy differences per atom ($\Delta G=\Delta E_0+\Delta F_\mathrm{el}+\Delta F_\mathrm{vib}$) of the isolated 2H-HCP and 4H-HCP phases of Au, Ag and Cu with respect to their  FCC-equivalent 3H-HCP phases at $400$~K.}
\label{fig:ThermFOM300K}
\end{figure}

We start with the stability assessment of the isolated crystals. 
In \fig{fig:ThermFOM300K}, the GFE differences ($\Delta G$) of the 2H-HCP and 4H-HCP phases with respect to the 3H-HCP counterparts are shown for $T=400$~K. 
At this operational temperature, the 2H-HCP and 4H-HCP phases are thermally less favorable. However, their $\Delta G$ is in the order of $1$~meV except for Au-2H-HCP. 
This small difference ($1$~meV) is nearly the numerical accuracy of the KS-DFT. 
Hence, those with a $\Delta G$ in this order are pretty likely to be thermally stabilized through an appropriate manufacturing process via residual stresses. 
One example can be the case of HCP-based Ag phases~\cite{Thevamaran:2016, Thevamaran:2019, ROJAS2021116892} stabilized via the severe plastic deformation or controlled electrochemical deposition~\cite{Chakraborty_2011}. 
General trends in $\Delta G$ are predominantly determined by the vibrational Helmholtz free-energy differences ($\Delta F_\mathrm{vib}$), while the electronic contribution ($\Delta F_\mathrm{el}$) are negligible (see \fig{fig:ThermFOM} for temperature-dependent free-energy differences).
Except for Au-2H-HCP, $\Delta F_\mathrm{vib}$ monotonically decreases with increasing temperature due to the entropic contributions.
This leads to thermal stabilization of Ag-2H-HCP, Ag-4H-HCP, Cu-2H-HCP, and Cu-4H-HCP at $853$~K, $745$~K, $1350$~K and $1125$~K, respectively.
The isolated Au-4H-HCP is expected to melt before reaching thermal stability as its $\Delta G$ becomes negative at an extreme temperature when extrapolated. 

Finally, the phonon dispersions of the studied noble metal polytypes show positive (real) phonon modes, indicating dynamical stability at the absolute zero temperature (see \fig{fig:PhDisp} for phonon dispersions). 
There is an incomplete phonon softening between $\Gamma\rightarrow$A in the Au-3H-HCP phase, which can be a precursor for a martensitic phase transformation~\cite{Jin2015}. 
Similarly, the Ag-2H-HCP and the Ag-3H-HCP phases also exhibit incomplete phonon softening between $\Gamma\rightarrow$A, unlike the Ag-4H-HCP phase.
This softening is possibly indicating superior dynamical stability in the Ag-4H-HCP phase.
The Cu polytypes do not have any softening phonon modes, and the general trend is softer phonon modes (lower phonon frequencies) following an increasing trend from 2H-HCP$\rightarrow$3H-HCP$\rightarrow$4H-HCP.

\subsection{Bulk optical and plasmonic response}

\begin{figure}[H]
\centering

\includegraphics[width=0.8\textwidth]{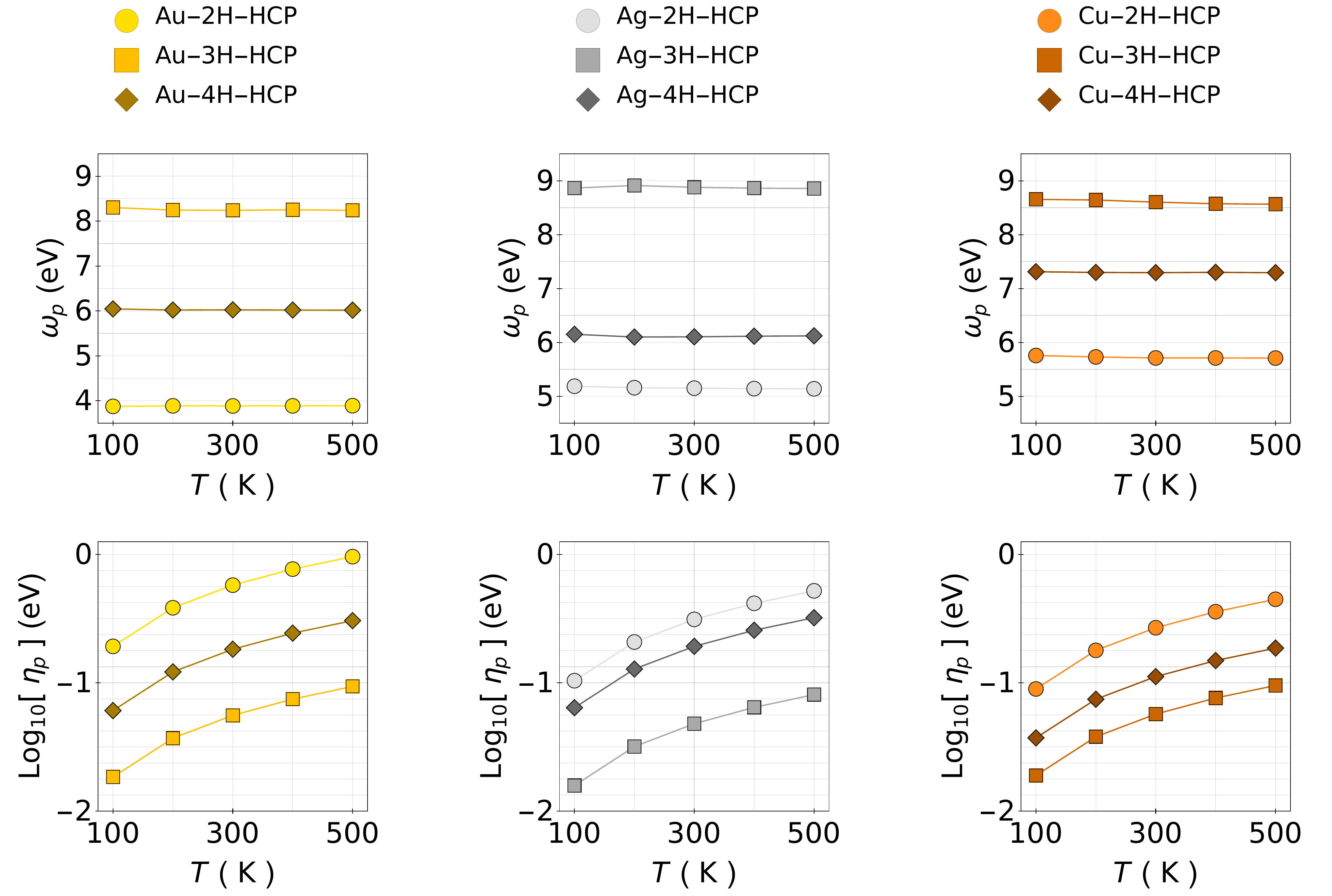}

\caption{Temperature-dependence of the Drude plasmon energy and the inverse lifetime of the isolated  Au, Ag and Cu polytypes.}
\label{fig:DrudePT}
\end{figure}

In \fig{fig:DrudePT}, the temperature-dependence of the Drude plasmon energies ($\omega_p$) and inverse lifetimes ($\eta_p$) are shown. These values are also listed in \tab{tab:DPList}.
For any given polytype, $\omega_p$ has a subtle temperature-dependence due to the similar temperature-dependence of its Fermi velocity, $v_\mathrm{F}$ (see  \fig{fig:PreDP} for the temperature-dependence of $v_\mathrm{F}$).
On the other hand, the inverse lifetime ($\eta_\mathrm{p}$) is linearly increasing with the increasing temperature due to the almost-linear temperature-dependence of $\eta_\mathrm{e-ph}$ in \eq{eq5}. 
The e-e term ($\eta_{e-e}$) generally has an exponentially increasing trend with the increasing temperature; however, this term is in the order of $10^{-4}-10^{-3}$ eV.
Drude plasmon energies follow an ordering of $\omega_\mathrm{p}^\mathrm{3H}>\omega_\mathrm{p}^\mathrm{4H}>\omega_\mathrm{p}^\mathrm{2H}$, while the inverse lifetimes have an opposite trend.
The ordering of $\omega_\mathrm{p}$ among the polytypes of a given metal follows the ordering of $v_\mathrm{F}$.
Since $\omega_\mathrm{p}\propto \sqrt{N(E_\mathbf{F})}/\lambda$, the identical orderings in DOS and $\lambda$ almost balance themselves (see \fig{fig:PreDP}).
The ordering in $v_\mathrm{F}$ (so in $\omega_\mathrm{p}$) is due to the steeper band dispersion in the 3H-HCP polytypes.

The calculated Drude plasmon energies (listed in \tab{tab:DPList}) of the 3H-HCP phases are between $8.0-9.0$~eV, which are in close agreement with the previous experimental and theoretical works ranging between $\sim 7.5 - 9.5 $~eV. 
We refer the reader to Ref.~\citenum{Orhan_2019} for further details. 
The Au-2H-HCP phase produces the Drude plasmon at a very low energy, i.e., $\sim3.8$~eV. 
However, it is a relatively short-living with a $7.17$~fs lifetime at the room temperature (listed in \tab{tab:DPList}).
The other HCP-based polytypes also produce Drude plasmons at lower energies compared to their FCC counterparts. 
This result is quite promising as these polytypes may give much lower energies (higher frequencies) to generate bulk plasmons and enable more diverse applications. 
For instance, heat-assisted magnetic recording devices commonly use $830$~nm wavelength at the higher temperatures~\cite{Bello:19}, while most medical and bio-sensing applications use the typical red laser ($\sim 650$~nm) at  lower temperatures~\cite{Orhan_2019}.

\begin{figure}[t]
\centering

\includegraphics[width=1.0\textwidth]{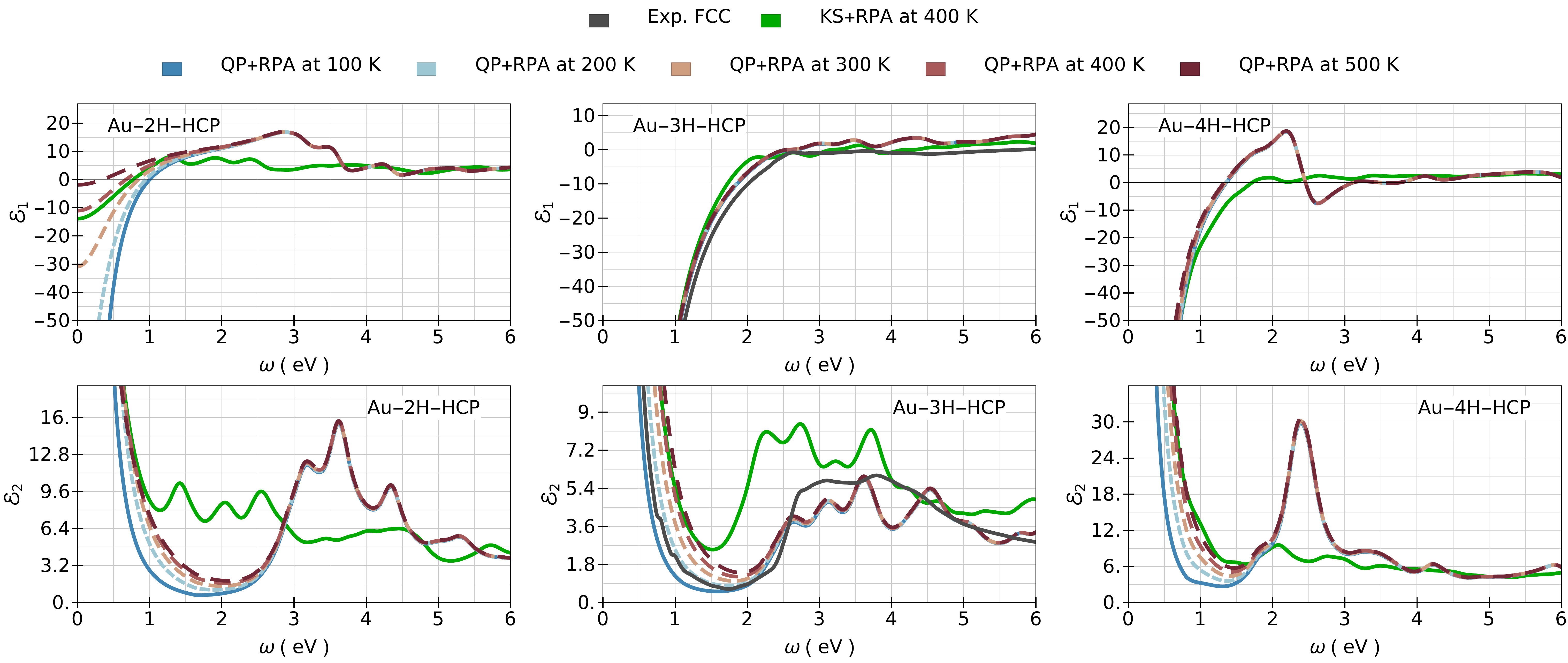}

\caption{Temperature dependence of the macroscopic dielectric functions of Au polytypes, calculated using the random-phase approximation  build on the quasi-particle band-structures (QP+RPA).
The macroscopic dielectric functions build on the Kohn-Sham band-structures (KS+RPA) at $400$~K are shown for comparison. 
The experimentally available data for the FCC phase were extracted from Ref.~\citenum{Babar:15}.
}

\label{fig:Au-EPS}
\end{figure}

\begin{figure}[t]
\centering

\includegraphics[width=1.0\textwidth]{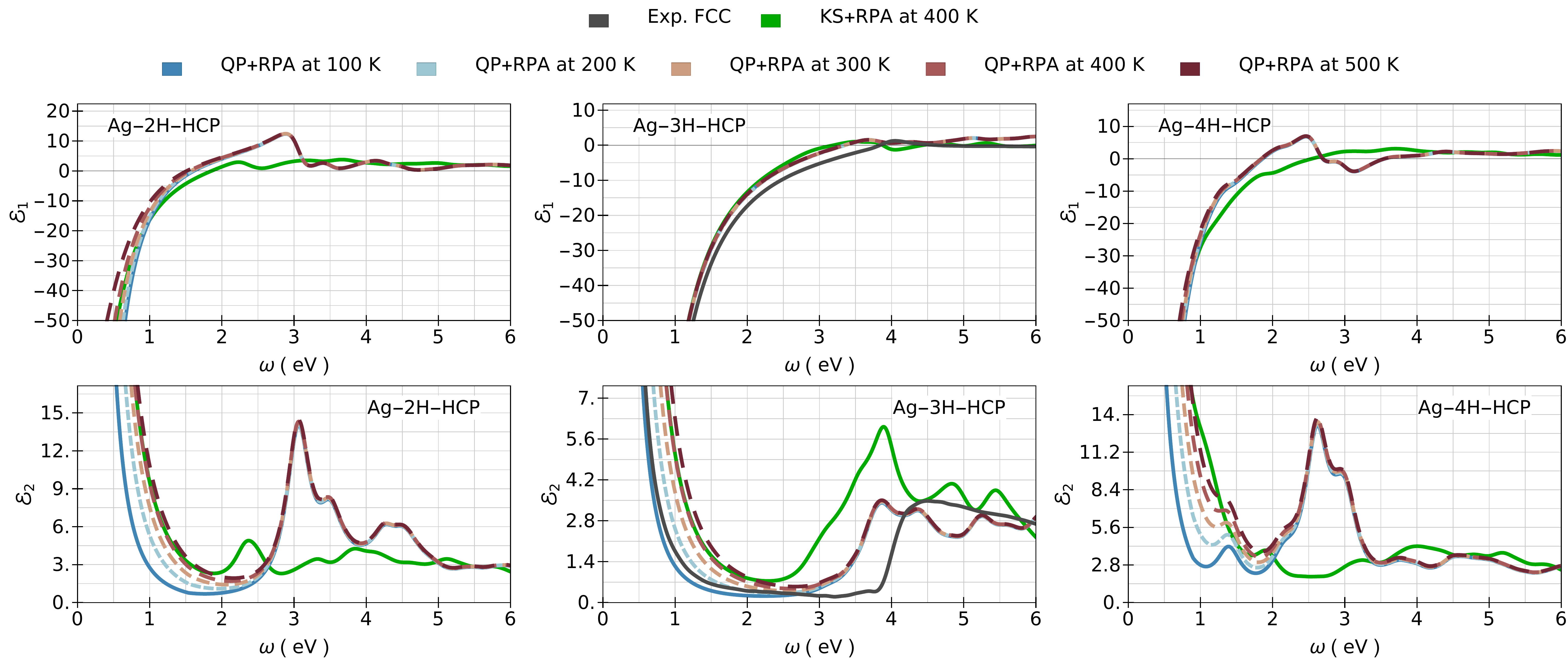}

\caption{Temperature dependence of the macroscopic dielectric functions of Ag polytypes, calculated using the random-phase approximation  build on the quasi-particle band-structures (QP+RPA).
The macroscopic dielectric functions build on the Kohn-Sham band-structures (KS+RPA) at $400$~K are shown for comparison. 
The experimentally available data for the FCC phase were extracted from Ref.~\citenum{Babar:15}.
}

\label{fig:Ag-EPS}
\end{figure}

\begin{figure}[b]
\centering

\includegraphics[width=1.0\textwidth]{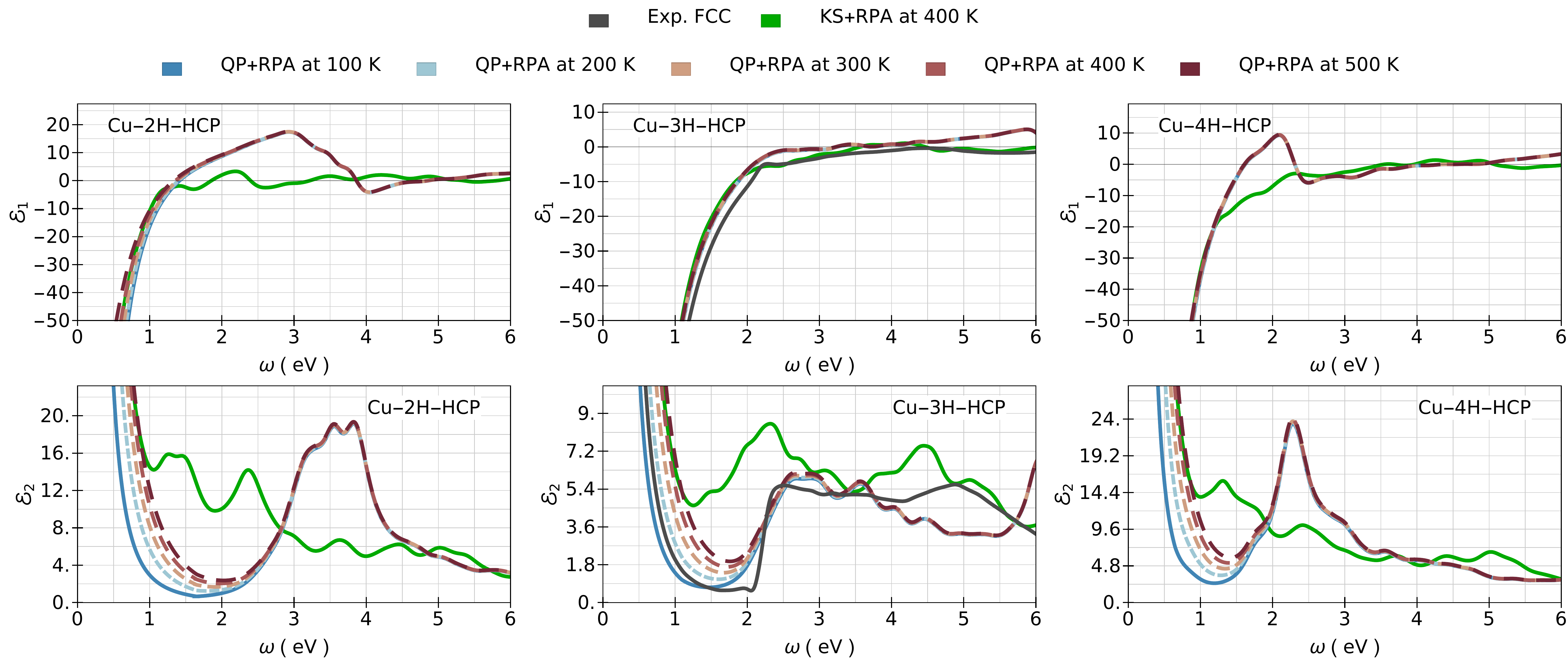}

\caption{Temperature dependence of the macroscopic dielectric functions of Cu polytypes, calculated using the random-phase approximation  build on the quasi-particle band-structures (QP+RPA).
The macroscopic dielectric functions build on the Kohn-Sham band-structures (KS+RPA) at $400$~K are shown for comparison. 
The experimentally available data for the FCC phase were extracted from Ref.~\citenum{Babar:15}.
}

\label{fig:Cu-EPS}
\end{figure}

In Figs.~\ref{fig:Au-EPS}~-~\ref{fig:Cu-EPS}, the macroscopic dielectric functions within RPA starting from the KS band-structures (KS+RPA) for $400$~K and the QP band-structures for $100-500$~K are presented alongside the available experimental spectra for the 3H-HCP phase, extracted from Ref.~\citenum{Babar:15}. 
The temperature dependence of the Drude plasmons predominantly affects the near-infrared spectral range of the absorption spectrum where $\ep_2 \approx \omega_\mathrm{p}^2\eta_\mathrm{p}/(\omega^3 + \omega \eta_\mathrm{p}^2)$. 
Hence, it is scaled by $\eta_\mathrm{p}$ which has a strong temperature-dependence. 
The real part, $\ep_1$ has a more subtle temperature dependence due to $\ep_1\approx 1-\omega_\mathrm{p}/(\omega^2 +\eta_\mathrm{p}^2)$ where $\eta_\mathrm{p}$-dependence rapidly vanishes for the increasing $\omega$.  
This is due to the fact that $\ep_2 = \omega_\mathrm{p}^2\eta_\mathrm{p}^/(\omega^3 + \omega \eta_\mathrm{p}^2)$, which is scaled by $\eta_\mathrm{p}$.
The only exception is the Au-2H-HCP phase for which the asymptotic tail at $0$~K becomes finite for $\omega \rightarrow 0$ with the increasing temperature. 
This is due to the fact that its $\eta_\mathrm{p}$ is converging to the order of $\omega_\mathrm{p}$ and $\ep_1=1-\omega_\mathrm{p}^2/\eta_\mathrm{p}^2$ for $\omega \rightarrow 0$.

Figs.~\ref{fig:Au-EPS}~-~\ref{fig:Cu-EPS} show the dielectric functions for the three polytypes as computed by the QP+RPA, and the KS-RPA. For the cases of the 3H-HCP (shown in the middle top and bottom panels of Figs.~\ref{fig:Au-EPS}~-~\ref{fig:Cu-EPS}), experimental values of the  dielectric function are also provided. 
It can be observed that the dielectric functions of the 3H-HCP phases within the QP+RPA are in good agreement with the experimental absorption spectra. 
In particular, the Drude tails match the experimental curves nicely at around $\sim 100-200$~K and the predicted QP+RPA captures most of the trends observed in experiments. 
Some discrepancies are evident, in particular, where the Drude tails merge to the inter-band parts. 
The experimental curves in Ref.~\citenum{Babar:15} are indeed obtained by combining the different experimental data from Ref.~\citenum{palik1998handbook} as it is not feasible to obtain the entire spectral range within a single experiment.  
For instance, the low-energy spectral range and the inter-band transitions were separately obtained and later combined~\cite{Babar:15}.
Hence, there are discontinuities where the intra-band and inter-band parts merge in the experimental curves. 
These also explain the sharper inter-band transition edges in the experimental curves compared to the more smeared edges in our calculated spectra.  
Despite that, the QP+RPA successfully predicted the low-energy spectral range while capturing the trends of the different experimental approaches at high-energy ranges.
For instance, for Au at $\omega = 2$ eV, QP+RPA predicts $\ep_1 = -6.9$ and $\ep_2 = 0.82$ while the experimental values are $\ep_1 = -10.3$ and $\ep_2 = 0.87$ at $100$~K. 
The discrepancy is much higher within KS+RPA as it predicts $\ep_1 = -3.6$ and $\ep_2 = 5.0$. 
Similar behavior can be seen for Ag and Cu, as shown in Figs. \ref{fig:Ag-EPS}~-~\ref{fig:Cu-EPS} middle panels.
It is also important to note that the experimental Drude plasmon parameters substantially vary in the literature.
We refer to the reader Refs.~\citenum{Babar:15}~and~\citenum{Orhan_2019} for a more comprehensive discussion on the experimental measurements of the Drude plasmons for the FCC phases.

There are also some discrepancies at the higher energies where only the inter-band transitions occur.
It is important to note that the experimental spectra tend to be more smeared due to the experimental conditions and the defects in samples. 
In the calculations, we use a fixed value of $0.2$~eV for the inverse lifetime of the inter-band transitions rather than larger smearing to have a better resolution.
Despite these discrepancies, the QP+RPA trends are in line with the experiments, while the KS+RPA spectra are substantially more inaccurate. 
Having gain confidence in the absorption spectra predictions using the QP+RPA, we now focus on comparing the absorption spectra within KS+RPA and QP+RPA, as shown in the bottom panels of Figs.~\ref{fig:Au-EPS}~-~\ref{fig:Cu-EPS}.
Different trends can be observed for the macroscopic dielectric function due to the \gw ~correction. 
For the 2H-HCP and 4H-HCP phases, the absorption peaks are located at higher energies with considerably high intensities within QP+RPA compared to the KS+RPA. 
This is due to two main effects of the \gw ~correction on the underlying electronic bands (see \fig{fig:ElBands} for the KS and QP band-structures). 
The first effect is that the \gw ~correction steepens the band dispersion.  
As the exchange part of the self-energy is only effective on the valence bands, the average stretching is larger on the valence bands compared to the conduction bands.
This shifts the absorption peaks to higher energies. 
The second effect is that a total shift of the $d$-band centers to higher energies.
This migrates the higher energy peaks to lower energies, contributing to the intense absorption peaks in the 2H-HCP and 4H-HCP phases. 
In the 3H-HCP phases, the first effect is also present; however, the $d$-band centers are not substantially shifted. 
It leads to the total shifts in their absorption spectra to higher energies with less intensity. 
The shift is not fully sufficient in the Ag-3H-HCP case for which QP+RPA slightly underestimates the absorption-spectrum edge compared to the experimental spectrum, as can be observed in Fig.~\ref{fig:Ag-EPS} in the middle bottom panel. 
This partial discrepancy can be due to the missing spin-orbit coupling in our work as \gw ~and the phonon dispersion simulations lack spin-orbit coupling. 
By all means, it is far more complex as the quantum many-body effects due to both electrons and excitations play significant roles.

Despite being purely a bulk property, the dielectric function determines the LSPR population. 
It sets crucial preliminary conditions for LSPR for which $\ep_1 \approx -1$ and $\ep_2 \approx 0$ need to be satisfied~\cite{ROCCA19951}. 
The first condition is to ensure sufficient free electron density, while the second condition is to avoid a high electron-hole-pair population. 
With these conditions in mind, the 2H-HCP phases, followed by the 4H-HCP, satisfy the first condition at lower energies than the 3H-HCP phases.
In particular, the $\ep_1$ of Au-2H-HCP polytype approaches to first-instance where $\ep_1 \approx -1$ at a quite-low energy of $\sim 0.6$~eV.
This predicts a high LSP generation at lower energies; however, $\ep_2 >> 0$ indicating a rapid decay to electron-hole-pair.
Furthermore, $\ep_2$ of the 2H-HCP and 4H-HCP phases becomes quite large at higher energies than the 3H-HCP phases indicating shorter-living LSPR despite the enhanced intensity. 
The only exception is the Cu-4H-HCP, where $\ep_2$ approaches its local minimum around when $\ep_1 \approx -1$ where it may produce similar LSP generated by the Cu-3H-HCP. 
In summary, the overall shifted trend in the 2H-HCP and 4H-HCP phases indicates more intense LSPR where higher absorption intensities at these lower energies imply shorter-living ones.

\subsection{Localized surface-plasmon generation in the perfectly spherical nanoparticles}
We now turn our attention to the case where perfectly spherical NPs are embedded onto a dielectric hosting matrix such as silica. We point out that this problem definition is a very simplistic scenario, as the hosting matrix does not strongly interacts with the NPs. 
\begin{figure}[t]
\centering
\includegraphics[width=1.0\textwidth]{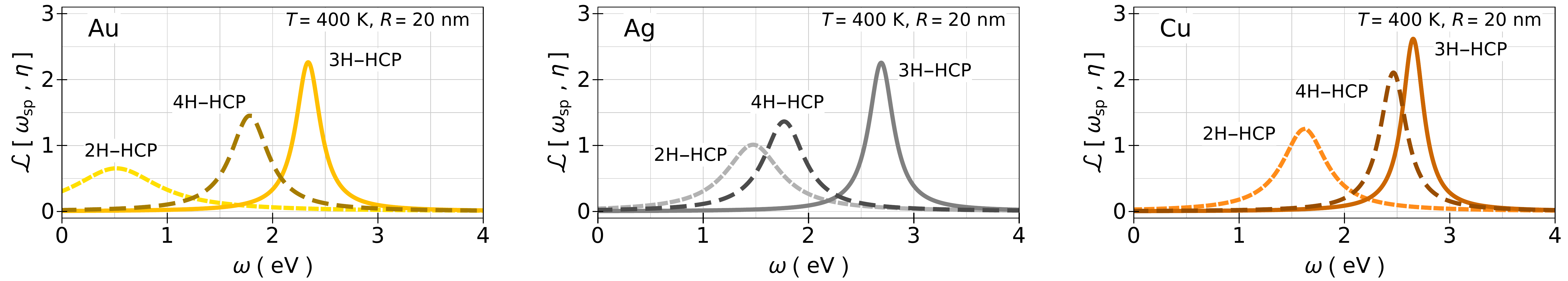}
\caption{Localized surface-plasmon resonance in the spherical nanoparticles of the noble-metal polytypes with radii of $R=20$~nm  at $T=400$~K, shown using the Lorentzian line shape. 
The NPs are embedded onto a hosting matrix with a dielectric permittivity of $\ep_\mathrm{m}=4$.    
}
\label{fig:NM-SurfPlas400K20nm}
\end{figure}

In \fig{fig:NM-SurfPlas400K20nm}, LSPR  in the spherical NPs with $R=20$~nm embedded onto a representative dielectric hosting matrix are shown at an operating temperature of $T=400$~K, using the Lorentzian line shape denoted by $\mathcal{L}[\omega_\mathrm{sp}, \eta]$ (also listed in \tab{tab:SurfPlas400}). 
The dielectric permittivity of the representative dielectric hosting matrix is chosen as $\ep_\mathrm{m}=4.0$ since it is approximately equal to the dielectric constant of silica~\cite{robertson_2004}.  
The 2H-HCP-phased NPs have LSPR peaks at the lowest energies compared to their 3H-HCP and 4H-HCP counterparts.  
The LSPR peak of the Au-2H-HCP-phased NP is at a very low energy of $\sim 0.90$~eV with the shortest lifetime of $4.25$~fs. 
The considerable short lifetime is mostly due to the strong e-ph coupling in Au-2H-HCP, indicating mainly rapid-decaying of LSPR to lattice vibrations. 
Moreover, the absorption-spectrum edge of Au-2H-HCP in \fig{fig:Au-EPS} also hints a rapid-decaying of plasma into the electron-hole pairs.
The Ag-2H-HCP NP generates LSPR at the closest energy of $1.63$~eV with a lifetime of $6.58$~fs with respect to the operational wavelength of  $830$~nm ($\sim 1.494$~eV).
The low-energy LSRP peaks in the 2H-HCP-phased NPs are primarily due to the strong e-ph mass enhancement parameters of the 2H-HCP phase which re-normalize the Fermi velocities and lead to the lower Drude plasmon energies as shown in \fig{fig:DrudePT}. 
The 4H-HCP-phased NPs have their LSPR peaks at higher energies compared to their 2H-HCP-phased counterparts while at lower energies compared to their 3H-HCP counterparts. 
The general trend in the LSPR energies is an increasing order of $\omega_\mathrm{sp}^\mathrm{3H-HCP}> \omega_\mathrm{sp}^\mathrm{4H-HCP}> \omega_\mathrm{sp}^\mathrm{2H-HCP}$. 
The LSPR lifetimes also follow a similar trend. 
In the case of Cu NPs, the 3H-HCP-phased and 4H-HCP-phased NPs generate LSPR peaks at the considerable close energies with comparable lifetimes of $13.68$ \emph{vs.} $17.00$~fs, respectively. 

The LSPR peak location and lifetime can be further tuned by controlling the operational temperature, the hosting matrix's dielectric permittivity and NP's size.
In order to show these effects, we start by analyzing the temperature-dependence by setting $\ep_\mathrm{m}=1.0$ in \eq{eq8} and $\alpha=0$ in \eq{eq9}.
This setting eliminates the effects of the hosting matrix and the NP size,  while the operational temperature is varied between $100-500$~K as shown in \fig{fig:SurfPlasT}. 
The increasing temperature shorten the plasmon lifetime due to $\tau_\mathrm{p}\propto 1/T$ through \eq{eq5}.  
Moreover, it noticeably shifts the LSPR peak to lower energies for the Au-2H-HCP, Au-3H-HCP and 4H-HCP phases, while for the remaining polytypes the effect is only minor.

Next, the operating temperature was set at $400$~K to assess the effects of the hosting matrix by varying $\ep_\mathrm{m}=1.0-2.0$ and the grain size by varying $R=10-20$ nm as shown in \fig{fig:SurfPlasT400K}. 
Increasing $\ep_\mathrm{m}$ shifts the LSPR peaks to lower energies without having any effect on the lifetimes. 
Introducing the size effects by setting $\alpha=1.0$ primarily lead to substantial shortening of the LSPR lifetimes. 
In the case of Au-2H-HCP-phased NP, it also shifts the LSPR peak to lower energies where such shift is higher for smaller NPs.
Increasing NP size lengthens the LSPR lifetimes in all cases, since \eq{eq9} is inversely proportional to the effective path length or simply to $R$ for the perfectly spherical NPs. 
In summary, the increasing grains size has an opposite effect on the LSPR peaks compared to the increasing temperature and the increasing hosting-matrix dielectric permittivity.

\subsection{Bulk mechanical properties }
\begin{figure}[H]
\centering
\includegraphics[width=0.8\textwidth]{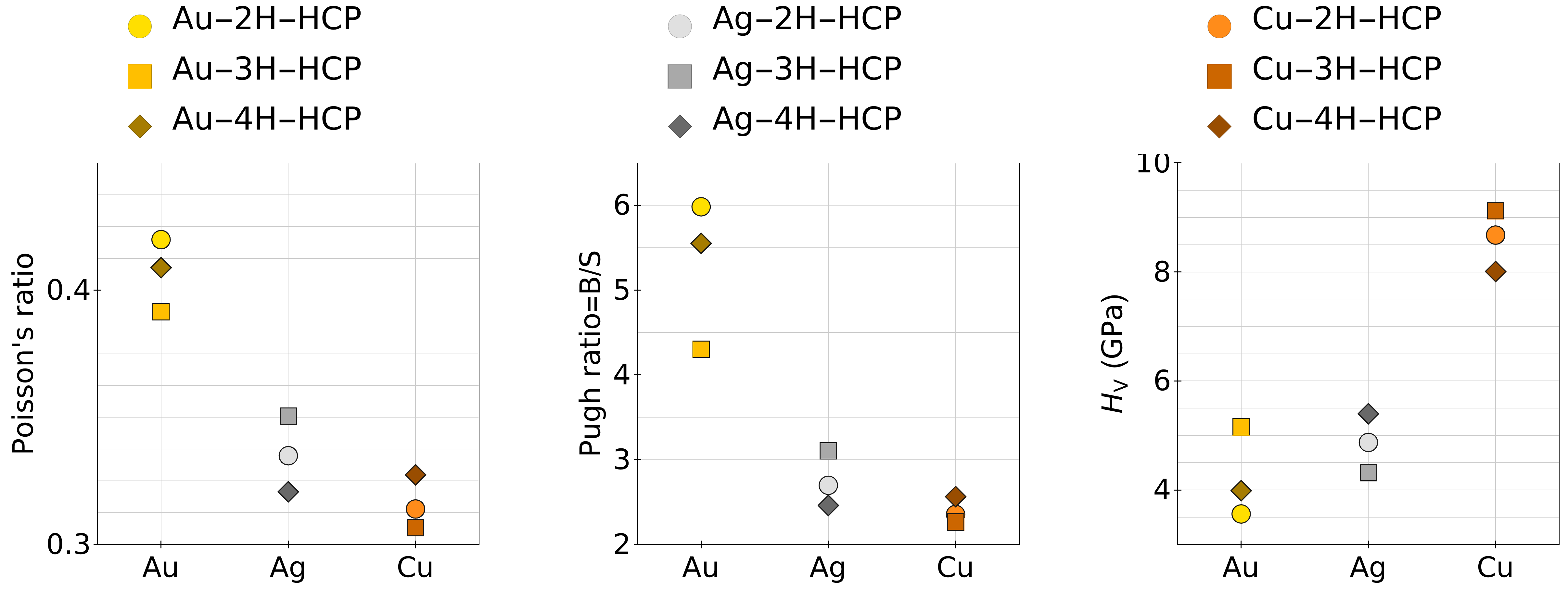}
\caption{Poisson's ratios,  Pugh ratios and Vickers hardness for Au, Ag and Cu polytypes.}
\label{fig:MechProp}
\end{figure}
The mechanical FOMs are presented in \fig{fig:MechProp}.
As expected, Au polytypes have the largest Poisson's and Pugh ratios, indicating a superior plastic deformation and ductility among all noble metal polytypes. However, these properties come at the expense of lowered hardness ($H_V$) and strength.  
It can be also seen when comparing their Vickers hardness to that of the Au-3H-HCP.
Unlike Au, the 2H-HCP and 4H-HCP phases of Ag are predicted to have improved mechanical strength due to their lower Pugh ratios compared to the Ag-3H-HCP. 
Their Poisson's ratios above the CMV of $0.25$ indicate that they are also posses good workability. 
In terms of mechanical strength, the Cu polytypes have shown superior properties among the studied metals, i.e., low Pugh ratio and high hardness.
Although Cu-4H-HCP has the lowest Vickers hardness among the other Cu polytypes ($\sim 0.7$~GPa, and $\sim 1.1$~GPa lower compared to the Cu-2H-HCP and Cu-3H-HCP, respectively), it has has a much higher Vickers hardness compared to Ag and Au polytypes. 

\subsection{Approximate thermal conductivity}
\begin{figure}[H]
\centering
\includegraphics[width=0.8\textwidth]{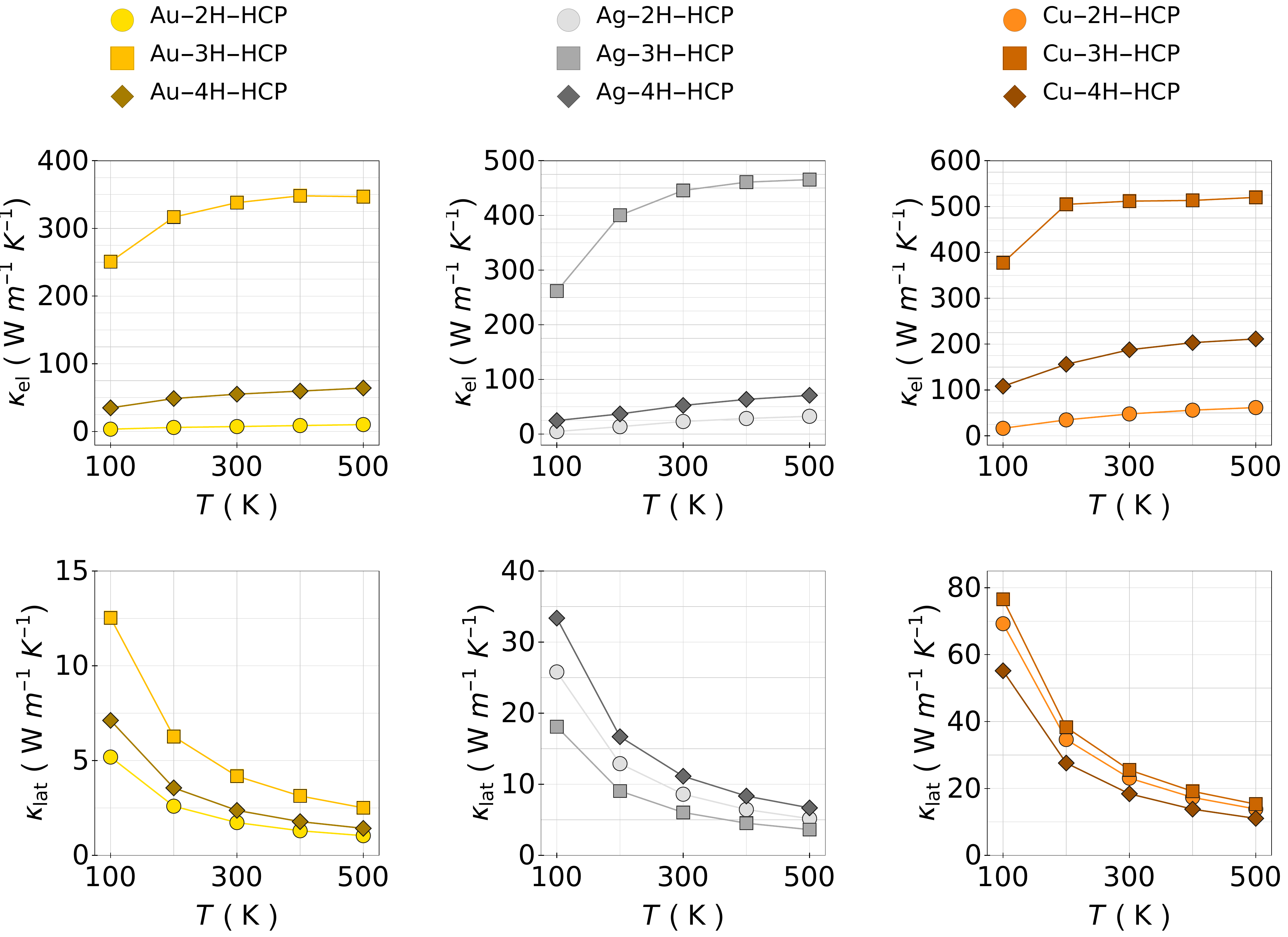}
\caption{Temperature-dependence of lattice (left panel) and electronic (right panel) thermal conductivity of the isolated bulk Au, Ag and Cu polytypes.}
\label{fig:ThermCondT}
\end{figure}
The temperature-dependent  electronic and lattice thermal conductivities are shown in \fig{fig:ThermCondT}. 
The total thermal conductivities are also listed in \tab{tab:TotThermCond}. 
It is evident that the electronic conductivity is the most important contribution to thermal conductivity for all polytypes. 
The electronic conductivity shows an exponentially increasing trends with the increasing temperature due to its $T/(1+\lambda)$ dependence in \eq{eq10} where $N(E_\mathrm{F}$ and $v_\mathrm{F}$ are weakly temperature-dependent (see \fig{fig:PreDP}).
On the other hand, lattice thermal conductivity exponentially decreases with temperature due to its $(1/T)$ dependence. 

The total thermal conductivities of the Au-3H-HCP and Ag-3H-HCP are $351$ and $465$~(W$\cdot$m$^{-1}$ $\cdot$K$^{-1}$) at $400$~K, respectively.
They are in reasonably good agreement with the experimental values $312$ and $420$~(W$\cdot$m$^{-1}\cdot$K$^{-1}$)~\cite{Touloukian1970} despite slight overestimation by our simulations. 
In the case of the Cu-3H-HCP, the total thermal conductivity is $532$~(W$\cdot$m$^{-1}\cdot$K$^{-1}$) which is quite overestimated compared to the experimental value of $392$~(W$\cdot$m$^{-1}\cdot$K$^{-1}$) at $400$~K~\cite{Touloukian1970}.
Such overestimation is expected as we assume that our systems are perfect crystals.
Hence, in \eq{eq3}, we do not include the electron-impurity scattering term, which can become significant above the room temperatures and reduce the plasmon lifetime, $\tau_\mathrm{p}$, depending on the sample~\cite{PhysRevB.16.5277}. Among all sources of error, $\tau_\mathrm{p}$ is the most significant contributor to an overestimation of the thermal conductivity for Cu. 
It is worth mention Cu tends to quickly oxidize compared to the other noble metals, which can reduce the thermal conductivity significantly compared to the pristine metal. 

In the case of  2H-HCP and 4H-HCP polytypes, the electronic thermal conductivity become much lower due to their large e-ph couplings which make difficult for the free electrons to mediate the heat dissipation. 

\section{Conclusion}
In this work, we have presented a comprehensive stability assessment and temperature-dependent material properties of 2H-HCP, 3H-HCP (FCC-equivalent) and 4H-HCP polytypes of Au, Ag and Cu. 
It has been found that the isolated 2H-HCP and 4H-HCP crystals become thermally more-favorable at high temperatures with respect to their 3H-HCP counterparts, with the exception of the Au-2H-HCP case.
The isolated Au-2H-HCP crystal has a linearly increasing vibrational Helmholtz energy difference ($\Delta F_\mathrm{vib}$) with respect to the Au-3H-HCP phase. 
However, it is important to note that this phase has been stabilized in the hetero-phased nanoribbons.
The studied polytypes successfully satisfy the elastic and dynamical stability criteria.
The former one was demonstrated by applying the Born-Huang-stability criteria for the hexagonal symmetry, whilst the later one was assessed by analyzing the phonon dispersions which have positive phonon modes.

The increasing temperature has a very subtle effect on the Drude plasmon energy ($\omega_\mathrm{p}$) due to the weak temperature-dependence of the Fermi velocity ($v_\mathrm{F}$) and the ratio $N(E_\mathrm{F})/(1+\lambda)^2$ where  $N(E_\mathrm{F})$ is the DOS at the Fermi level and $\lambda$ is the e-ph mass enhancement parameter. 
On the contrary, the Drude plasmon inverse lifetime ($\eta_\mathrm{p}$) is strongly temperature-dependent as $\eta_\mathrm{p} \propto  T$ via \eq{eq5}.

It has been shown that the semi-local functionals fail to construct accurate band-structures in the case of noble metals and lead to erroneous inter-band transition spectra.
The over-flattening of the electronic bands (particularly the fully filled $d$-bands) due to missing non-local quantum many-body effects within the semi-local functionals were corrected by the real-time \gw. 
Through this corrective measure, the valence bands are more steepen than the conduction bands as the self-energy exchange term  mostly acts on the occupied bands. 
This leads to an overall shift on the inter-band transitions to the higher energies. 
Comparison between the absorption spectra within the KS+RPA, QP+RPA and the experimental absorption spectra of the 3H-HCP polytypes  showed that the KS+RPA erroneously predicts the absorption peaks at the lower energies, while the QP+RPA shifted them closer to the experimental values due to steepening of the electronic-band dispersions by the \gw ~correction. 
As a result, the first-principles Drude tails nicely match  the experimental curves of the 3H-HCP phases at around $100-200$~K as shown in Figs.~\ref{fig:Au-EPS}~-~\ref{fig:Cu-EPS}.

We have investigated LSRPs in the perfectly spherical NPs embedded onto the silica-like hosting matrices. 
The increasing temperature shortens the LSPR lifetimes and generally shifts the LSPR peaks to the lower energies. 
This is due to the significant temperature-dependence of $\eta_\mathrm{p}$ which mostly determine the LSPR inverse lifetime ($\eta$) in \eq{eq9} and lower the peak position through \eq{eq8}.
The finite-size effect ($\alpha \neq 0$ in \eq{eq9}) further shortens the LSPR lifetime. 
As the finite-size contribution in \eq{eq9} is inversely proportional to the effective path length, the increasing NP radius ($R$) increases the LSPR lifetimes. 
The increasing dielectric permittivity of the hosting matrix ($\ep_\mathrm{m}$) shifts the LSPR peaks to the lower energies.
These mechanisms alongside the available polytype give a wide-range control over the tuning the surface-plasmons for a designated application.
At the technologically relevant  $830$~nm wavelength and for an operational temperature of $400$~K, the 2H-HCP  and 4H-HCP spherical NPs with $20$~nm radius generate substantially stronger LSPR compared to their 3H-HCP counterparts (see \tab{tab:SurfPlas400})
The Cu-4H-HCP NP is an exception with quite similar LSPR with the Cu-3H-HCP NP.
Despite the more intense LSPR at $830$~nm, the LSPR lifetimes in the 2H-HCP and 4H-HCP NPs are shorter compared to their 3H-HCP counterparts.
However, their LSPR are still reasonably long-living, i.e., in the order of $1$~fs. 

The 2H-HCP and 4H-HCP phases of Ag is mechanically stronger compared to the Ag-3H-HCP, whilst the 3H-HCP phases of Au and Ag are superior compared to their other phases. 
Without any exception, the 2H-HCP and 4H-HCP phases exhibited lower thermal conductivity compared to their 3H-HCP counterpart. 

The exotic HCP-based phases of noble metals can be promising alternatives for a wider-range plasmonic applications in terms of operational wavelengths, temperatures, ambient stress and thermal conditions. 
This work provides a reference work for these materials and a systematic workflow to study similar structures using the expedient first-principle methods. 

\section{Acknowledgements}

We acknowledge the support from the rom the Natural Sciences and Engineering Research Council of Canada (NSERC) through the Discovery Grant under Award Application Number RGPIN-2016-06114, and the New Frontiers in Research Fund (NFRFE-2019-01095).
This research was supported in part through computational resources and services provided by Advanced Research Computing at the University of British Columbia.

\clearpage
\onecolumngrid
\raggedbottom
\setcounter{section}{0}
\setcounter{equation}{0}
\setcounter{figure}{0}
\setcounter{table}{0}
\setcounter{page}{1}
\makeatletter
\renewcommand{\theequation}{S\arabic{equation}}
\renewcommand{\thefigure}{S\arabic{figure}}
\renewcommand{\thetable}{S\arabic{table}}

\renewcommand{\bibnumfmt}[1]{[S#1]}
\renewcommand{\citenumfont}[1]{S#1}
\begin{center}
\textbf{\large Supporting Information \\[0.25cm]"Surface-plasmon properties of noble metals with exotic phases"}
\end{center}

\section{Approximate Gibbs free energy}\label{sec:SI-GFE}
For a non-magnetic elemental and pristine solid, the Gibbs free energy (GFE) is given by 
\begin{align}\label{eq:si1}
G(V,T)= H(V)+F_\mathrm{el}(V,T)+F_\mathrm{vib}(V,T),
\end{align}
where  $H(V)$ is the formation enthalpy,  $F_\mathrm{el}(V,T)$ and $F_\mathrm{vib}(V,T)$ are the electronic and vibrational  Helmholtz free energies, respectively.
At the equilibrium volumes, the formation enthalpy can be simply approximated by the ground-state energy within the approximate KS-DFT such as $H(V)=E_0$.

The electronic Helmholtz free energy for metallic systems is approximated by~\cite{WANG20042665,landaystat,10.3389/fmats.2017.00036}

\begin{align}\label{eq:si2}
F_\mathrm{el} = &
\left(\int dE\; N(E,V)f E - \int^{E_\mathrm{F}} dE\; N(E,V)E\right)\\ \nonumber
&-T \left(-k_\mathrm{B}\int dE\; N(E,V)\left[f ln(f)+(1-f ) ln(1-f) \right]\right),
\end{align}
where $k_\mathrm{B}$ is the Boltzmann constant, $N(E,V)$ is the electron density of states (DOS), $f=f(E,E_\mathrm{F},T)$ is the Fermi-Dirac distribution function around the Fermi level with the Fermi energy, $E_\mathrm{F}$. 
The first term is the energy of free electrons, whilst the second term is due to electronic excitation above the Debye temperature by ignoring magnetic and electron-phonon (e-ph) couplings. 
The vibrational Helmholtz free energy for lattice ions within the quasi-harmonic approximation using the phonon dispersion is given by~\cite{RevModPhys.74.11,SHANG20101040}
\begin{align}\label{eq:si3}
F_\mathrm{vib}(T,V)=k_\mathrm{B}T \int_0^\infty d\omega \; \ln\left[2\; \sinh \left(\frac{\hbar \omega }{2 k_\mathrm{B}T}\right)\right]g(\omega),
\end{align}
where  $g(\omega)$ is the phonon DOS. $k_\mathrm{B}T$ is the usual thermodynamic factor, where $k_\mathrm{B}$ is the Boltzmann's constant and $T$ the absolute temperature, and $\omega$ is the vibration frequency of the phonon modes.

\section{Elastic properties}\label{sec:SI-Elastic}

The necessary and sufficient conditions for hexagonal symmetry  is given by~\cite{PhysRevB.90.224104} 
\begin{align}\label{eq:si4}
&c_{11} > |c_{12}|,\quad  c_{33}\left(c_{11}+c_{12}\right) > 2c_{13}^2 ,\; \mathrm{and} \quad  c_{44} >0 , 
\end{align}
where the components $c_{ij},~i=j=1,2,3$ of the second-order elastic tensor
The second-order elastic tensor is commonly calculated by performing a series of ground-state calculations on the infinitesimal-strained structures. 
The elastic constants are then calculated from the changes of the energy density with respect to the strain components.  
The bulk and shear modulus can be derived using the second-order elastic tensor within the the Voigt-Reuss-Hill (VRH) averaging~\cite{doi:10.1002/andp.18892741206,doi:10.1002/zamm.19290090104,Hill_1952}.
Young's modulus, the approximate sound velocities, the Poisson ratio and the approximate Debye temperature through their relation to the bulk and shear modulus. 
The Voigt-Reuss-Hill averaged bulk and shear modulus, symbolically $B$ and $S$, respectively, are given by
\begin{align}\label{eq:si5}
B_\mathrm{VHR}=\frac{B_\mathrm{V}+B_\mathrm{R}}{2},\quad \mathrm{and}\quad S_\mathrm{VHR}=\frac{S_\mathrm{V}+S_\mathrm{R}}{2},
\end{align}
for which the Voigt and Reuss averaging provide an upper-bound and a lower-bound, respectively.
The Voigt- and Ruess-averaged bulk and shear modulus for hexagonal systems are related to the elements of the second-order elastic tensor, symbolically $\mathbb{C}$, by~\cite{tromans2011elastic}
\begin{align}\label{eq:si6}
B_\mathrm{V}&=\frac{2(c_{11}+c_{12}+0.5 c_{33}+2c_{13})}{9},\quad \mathrm{and} \quad B_\mathrm{R}=\frac{1}{3\alpha+6\beta}; \nonumber \\
S_\mathrm{V}&=\frac{7c_{11}-5c_{12}+12c_{44}+2c_{33}-4c_{13}}{30},\quad \mathrm{and} \quad
S_\mathrm{R}=\frac{5}{4\alpha-4\beta+3\lambda},
\end{align}
where $3\alpha=s_{11}+s_{11}+s_{33}$, $3\beta=s_{23}+s_{31}+s_{12}$, and $3\lambda=s_{44}+s_{55}+s_{66}$.
$s_{ij}$ are elements of the elastic compliance tensor given by $\mathbb{S}=\mathbb{C}^{-1}$.
Using these two modulus, Young's modulus ($Y$) and the Poisson ratio ($\nu_\mathrm{P}$)  are
\begin{align}\label{eq:si7}
Y =\frac{9 B S}{3B + S}, \quad \mathrm{and}\quad \nu_\mathrm{P}=\frac{3B-2S}{6B+2S}. 
\end{align}
Furthermore, the  transverse (shear) and the longitudinal (compressional) wave velocities ($v_\mathrm{t}$ and $v_\mathrm{l}$, respectively) are given by~\cite{ANDERSON1963909}
\begin{align}\label{eq:si8}
v_\mathrm{l}=\left[\left(KB\frac{4}{3}S\right) \frac{1}{\rho}\right]^{1/2},\quad \mathrm{and}\quad v_{t}=\left[\frac{S}{\rho} \right]^{1/2},
\end{align} 
where $\rho$ is the density of solid. 
Finally, the Debye temperature can be approximated for equilibrium volume  by~\cite{ANDERSON1963909}
\begin{align}\label{eq:si9}
\Theta_\mathrm{D}=\frac{h}{k_\mathrm{B}}\left[\frac{3}{4\pi}\frac{N}{V_0}\right]^{1/3}v_\mathrm{s},
\end{align}
where $N$ is the number of atoms in the unit cell and $v_\mathrm{av}$ is the average constant sound velocity.
The average sound velocity, $v_s$, can be determined either by solving the Christoffel equation in each direction~\cite{JAEKEN2016445}, or approximately by the relation~\cite{ANDERSON1963909}
\begin{align}\label{eq:si10}
v_s=\left(\frac{1}{3}\right)^{-1/3}\left[\frac{2}{v_\mathrm{t}^3}+\frac{1}{v_\mathrm{l}^3}\right]^{-1/3}. 
\end{align}
Finally, the Gr\"uneisen parameter is expediently given in terms of the Poisson's ratio ($\nu$) by~\cite{JIAO201647}
\begin{align}\label{eq:eq17}
\gamma=\frac{3(1+\nu)}{2(2-3\nu)}.
\end{align}

\section{Random-phase approximation}\label{sec:SI-RPA}
In the linear-response regime, the microscopic dielectric function is given by
\begin{align}\label{eq:si11}
\varepsilon^{-1}_{\mathbf{G},\mathbf{G}'}(\mathbf{q},\omega)=
\delta_{\mathbf{G},\mathbf{G}'} + v_{\mathbf{G},\mathbf{G}'}(\mathbf{q})\chi_{\mathbf{G},\mathbf{G}'}(\mathbf{q},\omega),
\end{align}
where $v_{\mathbf{G},\mathbf{G}'}(\mathbf{q})=4\pi/\left(|\mathbf{q}+\mathbf{G}||\mathbf{q}+\mathbf{G}'|\right)$ is the bare Coulomb interaction and  $\chi_{\mathbf{G},\mathbf{G}'}(\mathbf{q},\omega)$ is the interacting response function which is given within RPA by~\cite{PhysRev.82.625,PhysRev.85.338,PhysRev.92.609,RevModPhys.36.844} 
\begin{align}\label{eq:si12}
\chi_{\mathbf{G}\mathbf{G}'}(\mathbf{q},\omega)=
\chi^0_{\mathbf{G}\mathbf{G}'}(\mathbf{q},\omega)\left(1+v_{\mathbf{G},\mathbf{G}'}(\mathbf{q})\chi_{\mathbf{G}\mathbf{G}'}(\mathbf{q},\omega)\right),
\end{align}
which is s Dyson equation.
$\chi^0_{\mathbf{G}\mathbf{G}'}$ is the non-interacting response function, given within Fermi's golden rule~\cite{dirac1927quantum} by 
\begin{align}\label{eq:si13}
 \chi^0_{\mathbf{G}\mathbf{G}'}(\mathbf{q},\omega) =  2\sum_{c,v,\mathbf{k}}& 
(f_{v,\mathbf{k}}-f_{c,\mathbf{k}-\mathbf{q}})
\frac{| \langle\psi_{c,\mathbf{k}-\mathbf{q}}|e^{\im (\mathbf{q}+\mathbf{G})\cdot\mathbf{r}}|\psi_{v,\mathbf{k}}\rangle |^2}{\omega-\omega_{cv,\mathbf{k}\mathbf{q}}+\im\Gamma},
\end{align}
where $\psi_{i,\mathbf{k}}$ is the well-defined single-particle electronic states, $f_{i,\mathbf{k}}$ is its occupancy; the indices $c$ and $v$ are running over the conduction and valence bands, respectively. 
The inter-band part of the macroscopic dielectric function  is obtained by summing over the first Brillouin zone at the vanishing momentum limit ($\mathbf{q}\rightarrow 0$)~\cite{sottile2003response}.

\section{Perturbative quasi-particle formalism}\label{sec:SI-QP}

The quasi-particle (QP) single-particle electronic wave-functions can be calculated by self-consistently solving the QP equation, which resembles to the KS equation with an additional self-energy terms and the QP re-normalization to the electronic mass.
Assuming that the the KS and the QP eigenstates differ very little such as  $\langle \psi_i|\psi_i^\mathrm{QP}\rangle \approx 1$, a more practical approach is to perturbatively calculate the QP energies  by 
\begin{align}\label{eq:si14}
\epsilon_i^\mathrm{QP} =\epsilon_i^\mathrm{KS}+
Z_i \langle\psi_i^\mathrm{KS}| \hat{\Sigma}\left(\epsilon_i^\mathrm{KS}\right) - \hat{v}_\mathrm{xc}^\mathrm{KS}|\psi_i^\mathrm{KS} \rangle,  \quad \mbox{and}
Z_i =\left[1- \left.\frac{\partial\Sigma'(\omega)}{\partial \omega}\right\vert_{\omega=\epsilon_i^\mathrm{KS}}\right]^{-1},
\end{align}
where $Z_i$ is called the QP re-normalization factor and $\hat{v}_\mathrm{xc}^\mathrm{KS}$ is the KS exchange-correlation operator. 
Within the one-shot, non-self-consistent GW, or simply G$_0$W$_0$, the self-energy operator can be calculated by the single iteration of Hedin's equations~\cite{PhysRev.139.A796,hedin1969solid} when the vertex function is set to the Kronecker-Delta-function product. 
By doing so, it is  simply $\Sigma=\im G_0 W_0$, where $G_0$ and $W_0$ are the non-interacting single-particle Green function and the screened Coulomb interactions, respectively.

\section{Computational details}
The initial crystallographic information for the FCC Au, Ag, Cu were adopted from  Ref.~\citenum{Suh1988}. 
They were used to construct 2H-HCP, 3H-HCP and 4H-HCP structures by following AB, ABC, ABAC stacking sequences, respectively.
The in-house norm-conserving PBE pseudo-potentials were generated by the pseudo-potential generator OPIUM~\cite{opium}.
The initial structures were optimized by using the variable-cell relaxation routine within the  the  Quantum ESPRESSO (QE) software~\cite{0953-8984-21-39-395502,0953-8984-29-46-465901} with a common high kinetic energy cutoff of $150$~Ry and a choice of smearing parameter for the Marzari-Vanderbilt cold smearing~\cite{PhysRevLett.82.3296} of $0.1$~eV.
Structural optimization were performed on the $12\times 12\times 6$, $12\times 12 \times 4$, and  $12\times 12 \times 3$ Monkhorst-Pack-equivalent~\cite{PhysRevB.13.5188} Brillouin zone ($k$-point) samplings  without any shift for the 2H-HCP, 3H-HCP and 4H-HCP structures, respectively.
Convergence thresholds for the total energies, the total forces and the self-consistency were set to  $10^{-7}$~Ry,  $10^{-6}$~(Ry/a$_0^3$) and $10^{-10}$, respectively.
For further simulations, a  kinetic energy cutoff of $100$~Ry was used as the common converged value by less than $1$~meV per atom different in the total energies, whilst the Marzari-Vanderbilt cold smearing parameter was kept same.
Other parameters were adjusted for each set of simulations.

In the first set of simulations the thermodynamic quantities, the elastic properties and the phonon dispersion were calculated using the  \textit{"thermo\_pw"} package~\cite{ThermoPW} within the QE software. 
During these simulations, the Brillouin zone samplings, which were used in the geometry optimizations, were shifted by  $1\times 1 \times 1$.
For the phonon dispersion simulations, the $2 \times 2\times 6$, $2 \times 2\times 4$ and  
$2 \times 2\times 3$  Monkhorst-Pack-equivalent $q$-point samplings were used for the 2H-HCP, 3H-HCP and 4H-HCP structures, respectively.

The e-ph coupling calculations requires multiple simulations with carefully chosen Brillouin zone sampling for feasible computational cost.
The first set of self-consistent-field (SCF) simulations were performed on a dense $24 \times 24 \times 24$ grid.
These simulations were later used for fitting during the e-ph coupling matrix simulations.
The second set of SCF simulations were performed in a less dense $12 \times 12 \times 12$ $k$-point grid.
The e-ph coupling matrices were calculated on the previously used $q$-point samplings during the phonon dispersion simulations.
The choice of the $k$-point and $q$-point samplings during these subsequent simulations, indeed, were dictated by the available methods within the QP software. 
The e-ph coupling matrices  were evaluated for the five Gaussian broadening used in the double-delta integrals.
These broadening parameters were chosen equivalently to a monotonically increasing temperature from $100$~K to $500$~K.

The third set of simulations were performed using the in-house code~\cite{Orhan_2019} to calculate the Fermi velocities and the Drude plasmon energies. 
As these simulations requires the well-converged KS band-structures on a dense grid, the non-self-consistent-field (NSCF) simulations were performed on the $12 \times 12 \times 12 $ $k$-point grids on top of the SCF simulations on the $24 \times 24 \times 24 $ $k$-point grids.
The Fermi levels and corresponding electronic DOS were chosen from the previous e-ph coupling simulations for each Gaussian broadening to match the electronic temperatures.

The last set of simulations were performed to obtain the approximate quasi-particle band-structures and the  inter-band transitions using the Yambo software~\cite{MARINI20091392}. 
Since the G$_0$W$_0$ simulations requires a large number of additional unoccupied KS states during calculating the self-energy correlations, the NSCF simulations were performed for a total of 100 bands for each spin-channel.
For further details, we refer the reader to Ref.~\citenum{RANGEL2020107242}.
The self-energies were calculated for the entire Brillouin zones, occupied bands and equal number of unoccupied bands. 
RPA simulations were performed on the approximate quasi-particle band-structures with a Lorentzian smearing parameter of full-width $0.2$ eV.

\subsection{Comparison of FCC and 3H-HCP crystal structures}
\begin{table}[H]
\renewcommand{\arraystretch}{1.2} \setlength{\tabcolsep}{12pt}
\begin{center}
\begin{tabular}{lcccc}
\hline \hline
   &   \% $\Delta a$  & \% $\Delta c$   & \% $\Delta d_\mathrm{AB}$ & \% $\Delta \alpha_\mathrm{ABC}$   \\ \cline{2-5}

Au & $-0.31$ & $0.62$ & $0.31$ & $0.34$ \\[0.25cm] 

Ag & $-0.11$ & $0.46$ & $0.27$ & $0.21$ \\[0.25cm]

Cu & $-0.19$ & $0.25$ & $0.10$ & $0.16$ \\

 \hline \hline
\end{tabular}
\end{center}
\caption{Percentage relative differences of lattice parameters, bond lengths and angles between the fully relaxed 3H-HCP and FCC structures.
All values are $< \% 1$ indicating that the ideal FCC and 3H-HCP crystals are equivalent within the accuracy of the approximate KS-DFT.}
\label{tab:FCC-3H-Crys-Comp}
\end{table}

\section{Thermal and mechanical assessments}

\begin{figure}[H]
\centering
\includegraphics[width=0.35\textwidth]{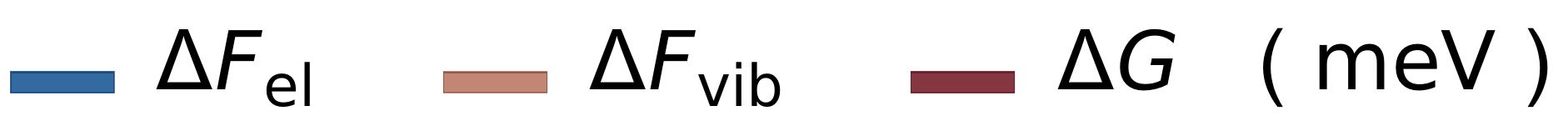}

\subfloat{ \includegraphics[width=0.35\textwidth]{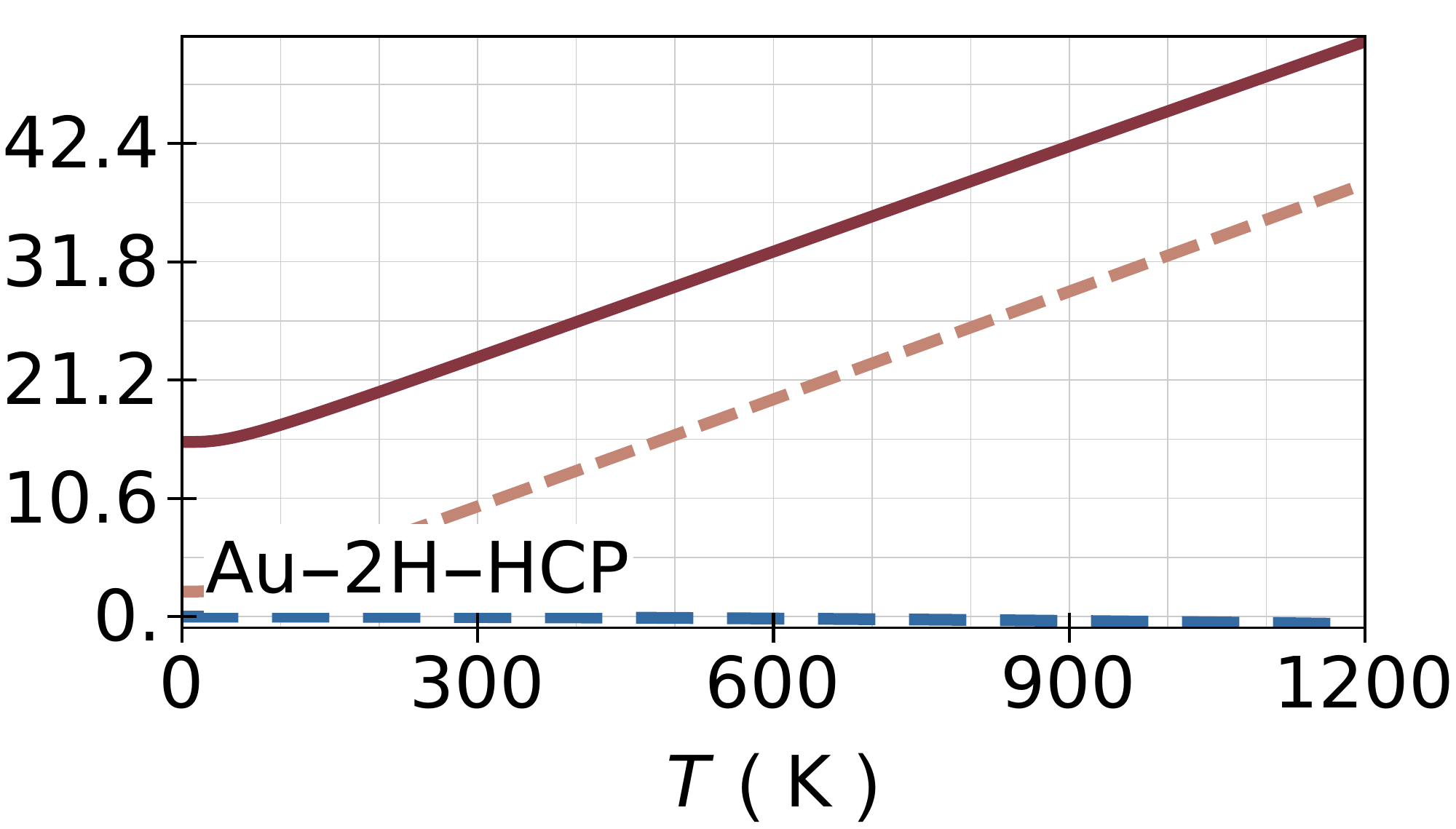}}\hspace{1cm}
\subfloat{ \includegraphics[width=0.35\textwidth]{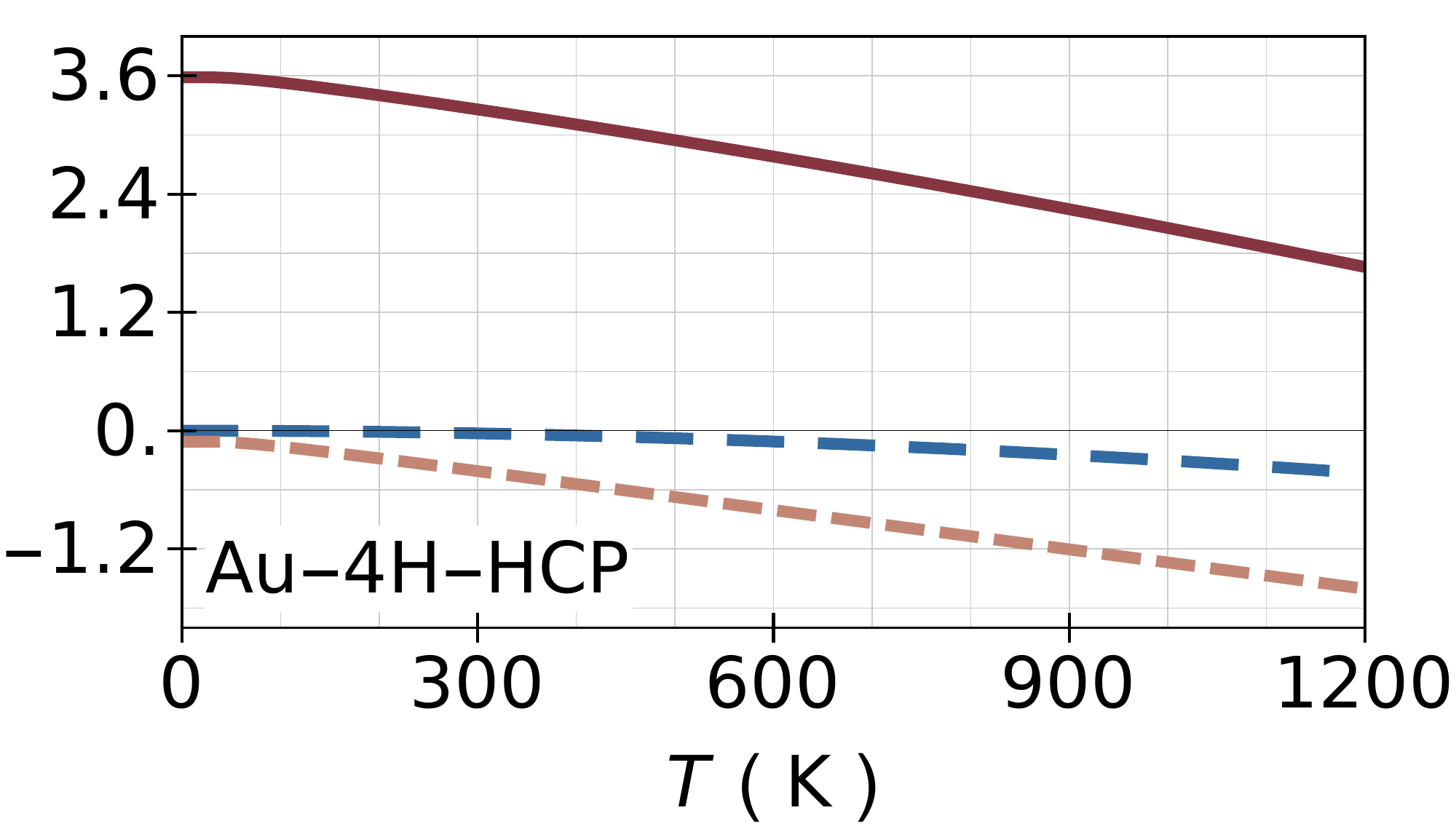}}

\subfloat{ \includegraphics[width=0.35\textwidth]{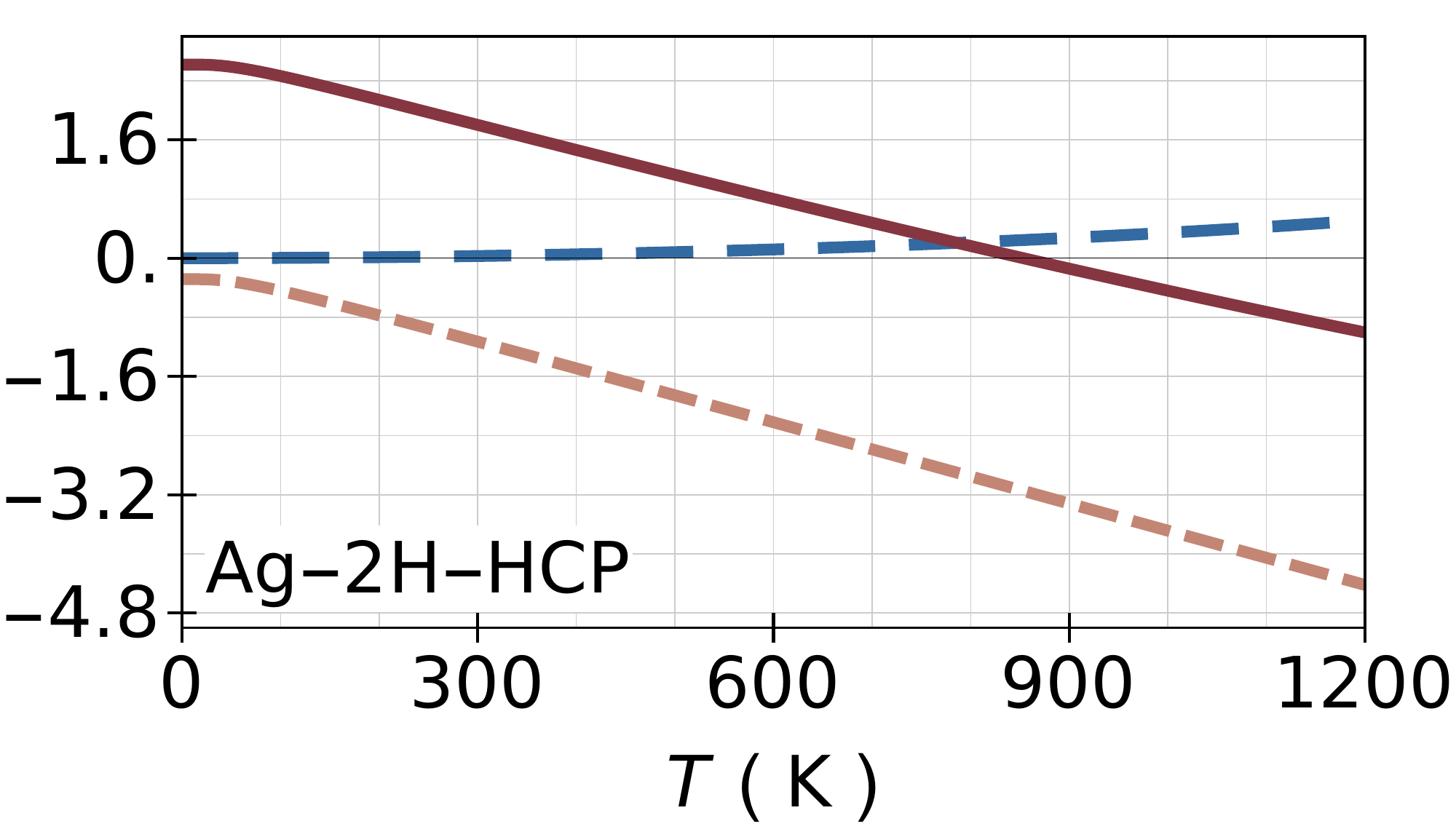}}\hspace{1cm}
\subfloat{ \includegraphics[width=0.35\textwidth]{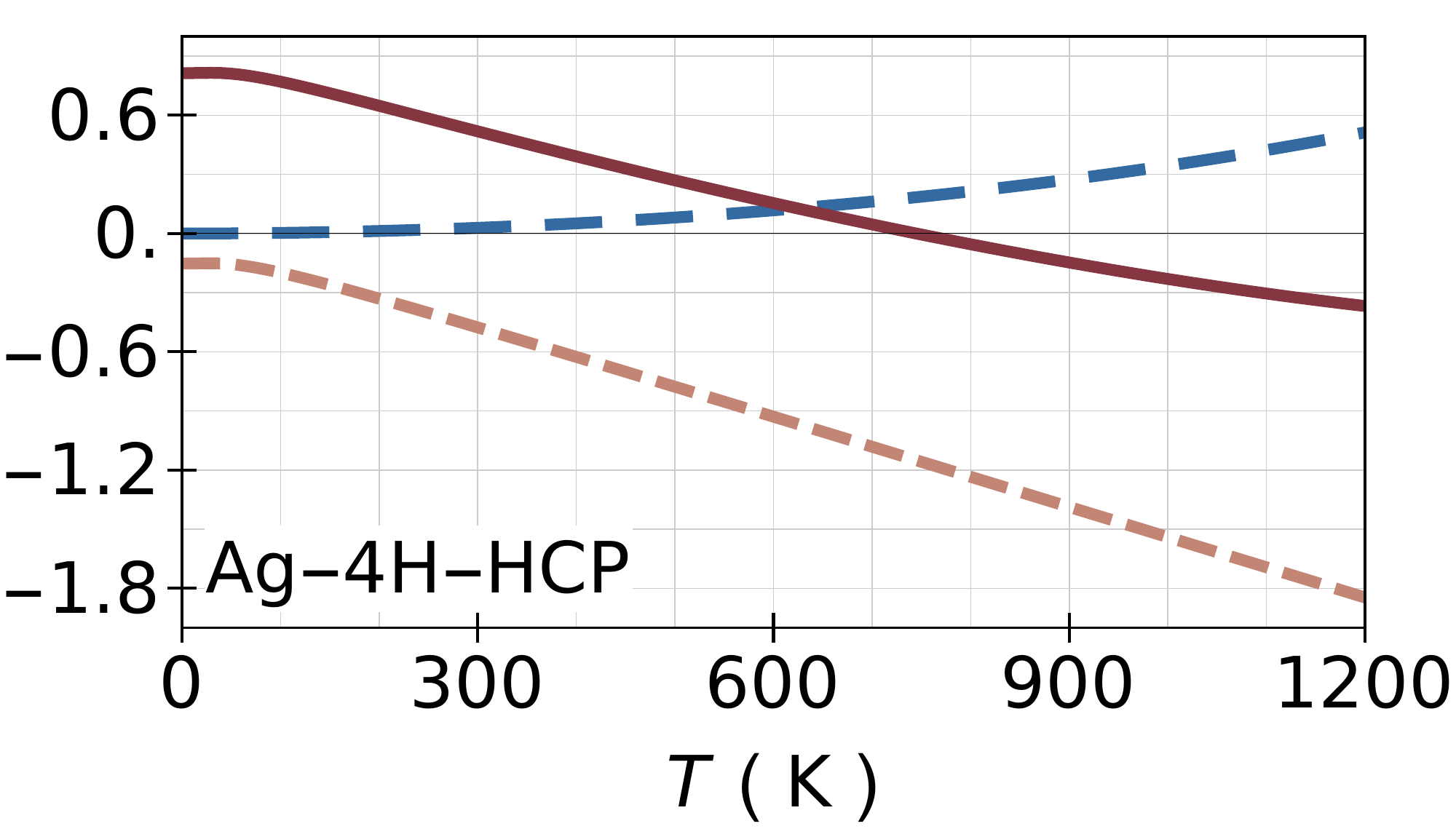}}

\subfloat{ \includegraphics[width=0.35\textwidth]{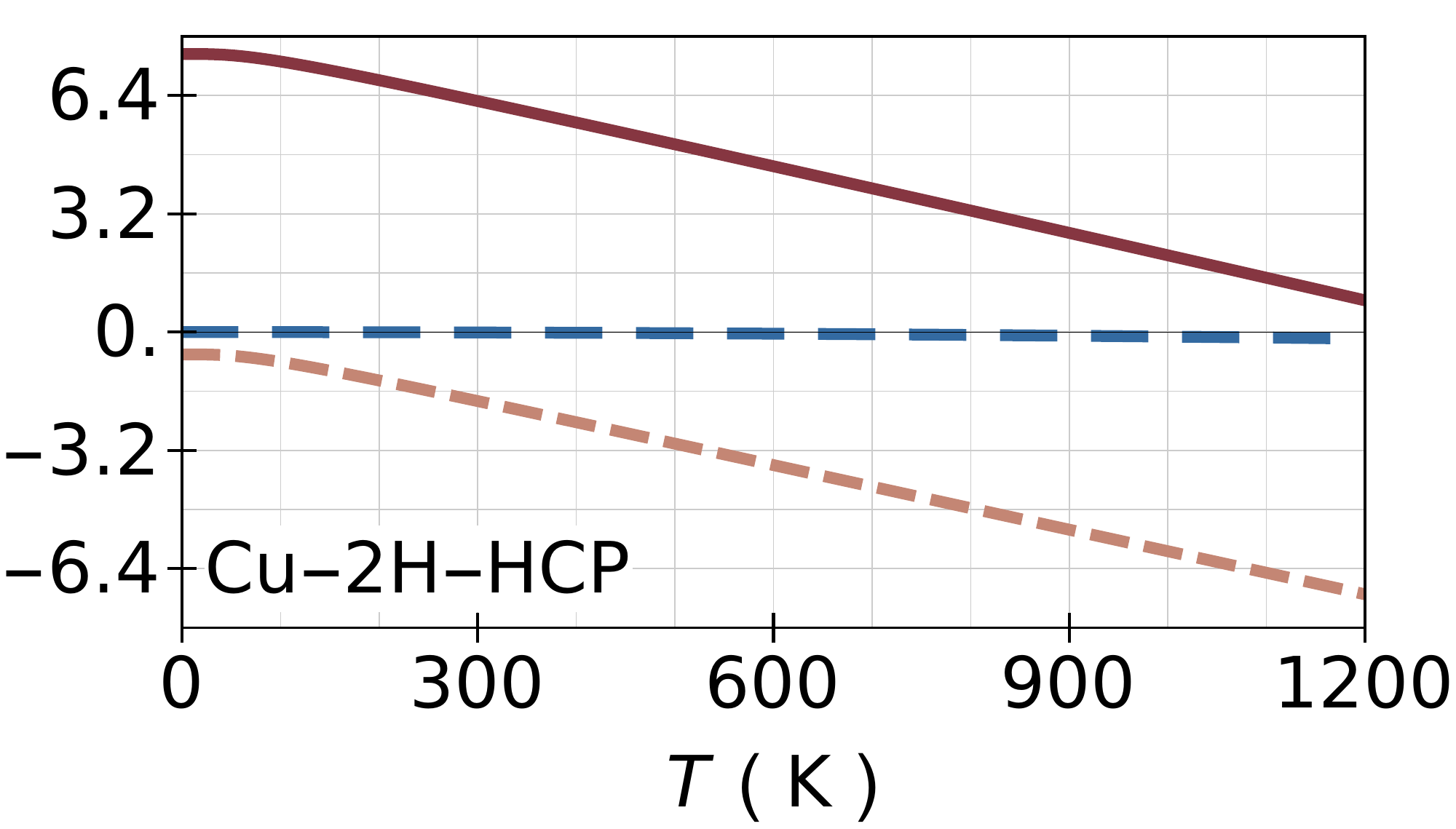}}\hspace{1cm}
\subfloat{ \includegraphics[width=0.35\textwidth]{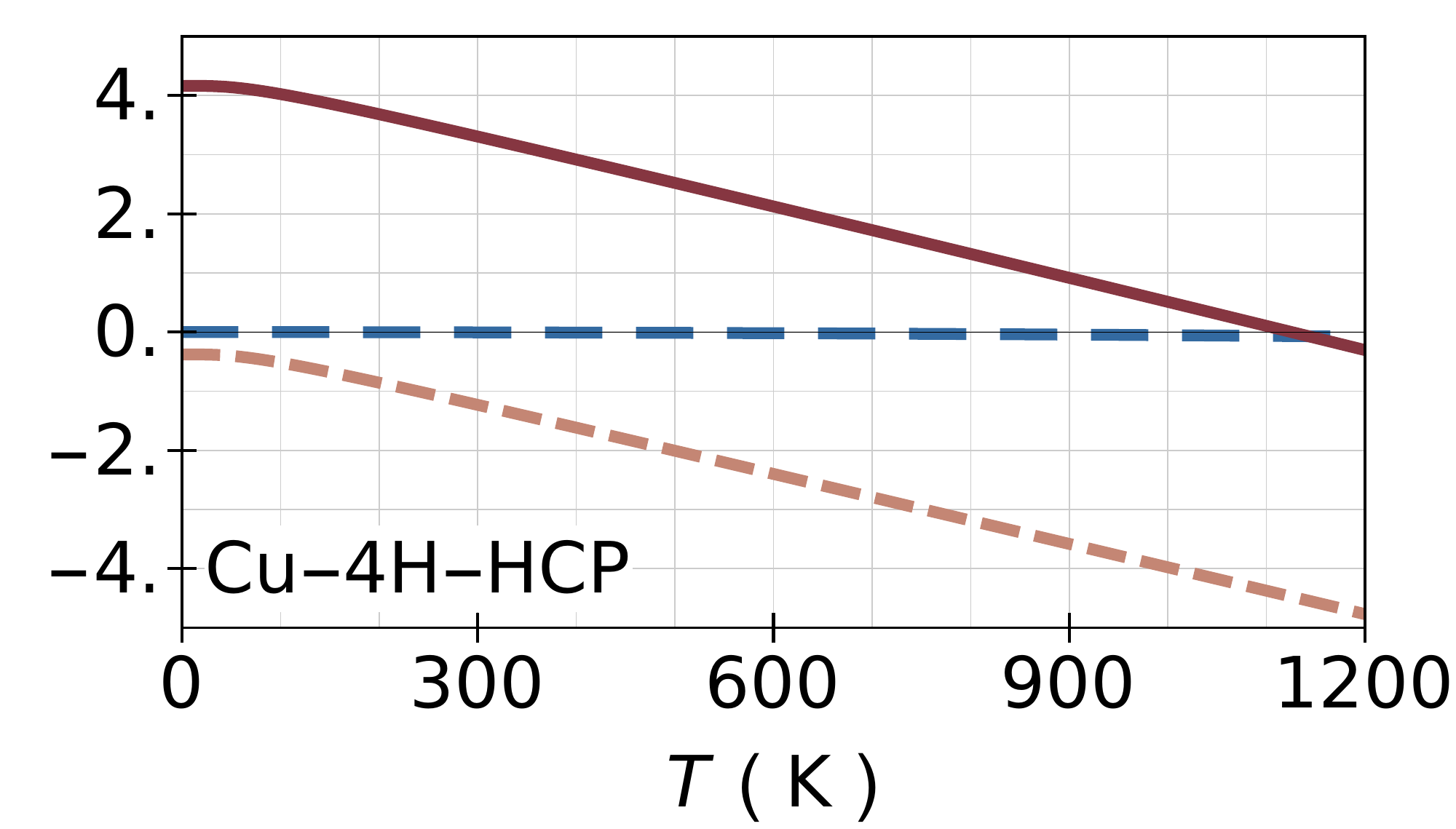}}

\caption{Temperature dependence of the electronic ($\Delta F_\mathrm{el}$) and vibrational ($\Delta F_\mathrm{vib}$) Helmholtz free energy and the Gibbs free energy differences ($\Delta G$) of 2H-HCP and 4H-HCP phases with respect to their 3H-HCP phases.
}
\label{fig:ThermFOM}
\end{figure}

\begin{table}[H]
\renewcommand{\arraystretch}{1.3} \setlength{\tabcolsep}{18pt}
\begin{center}
{\footnotesize
\begin{tabular}{lccccc} \hline \hline 

  &  $c_{11}$  &  $c_{12}$  &  $c_{13}$  &  $c_{33}$  &  $c_{44}$   \\ \cline{2-6} 
Au-2H-HCP  &  201  &  127  &  109  &  184  &  16    \\
Au-3H-HCP  &  205  &  118  &  117  &  209  &  28    \\
Au-4H-HCP  &  202  &  119  &  116  &  213  &  23    \\[0.25cm]
Ag-2H-HCP  &  142  &  65  &  56  &  145  &  24    \\
Ag-3H-HCP  &  144  &  67  &  56  &  152  &  22    \\
Ag-4H-HCP  &  147  &  63  &  52  &  162  &  26    \\[0.25cm]
Cu-2H-HCP  &  215  &  91  &  92  &  238  &  51    \\
Cu-3H-HCP  &  234  &  85  &  85  &  252  &  51   \\
Cu-4H-HCP  &  213  &  102  &  91  &  229  &  49    \\

\hline \hline
\end{tabular}}
\end{center}
\caption{Five independent elastic constants (in GPa units) of the isolated  Au, Ag and Cu polytypes.}
\label{tab:ElasticCon}
\end{table}

\begin{sidewaysfigure}[p]
\centering

\subfloat{ \includegraphics[width=0.33\textwidth]{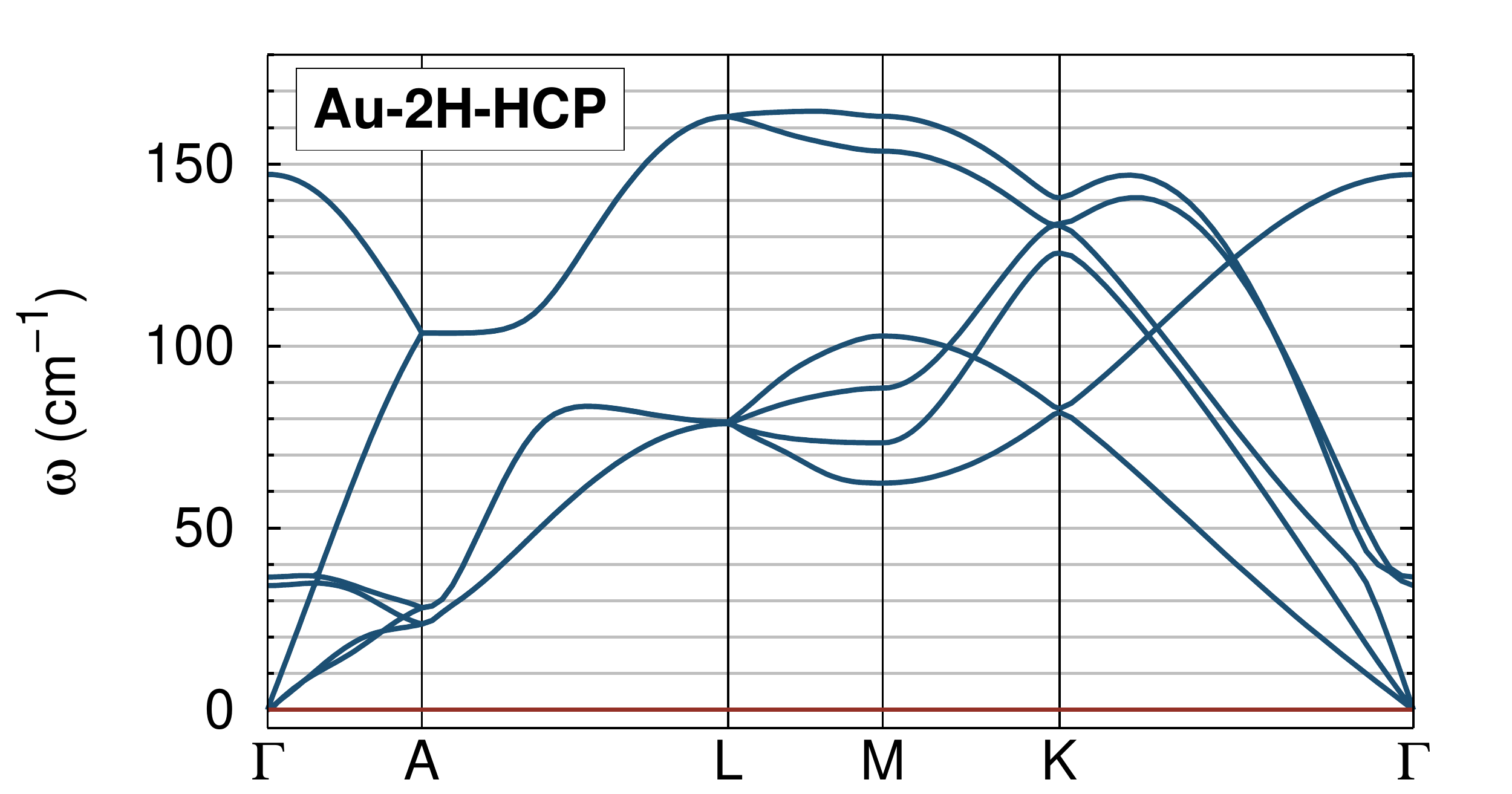}}
\subfloat{ \includegraphics[width=0.33\textwidth]{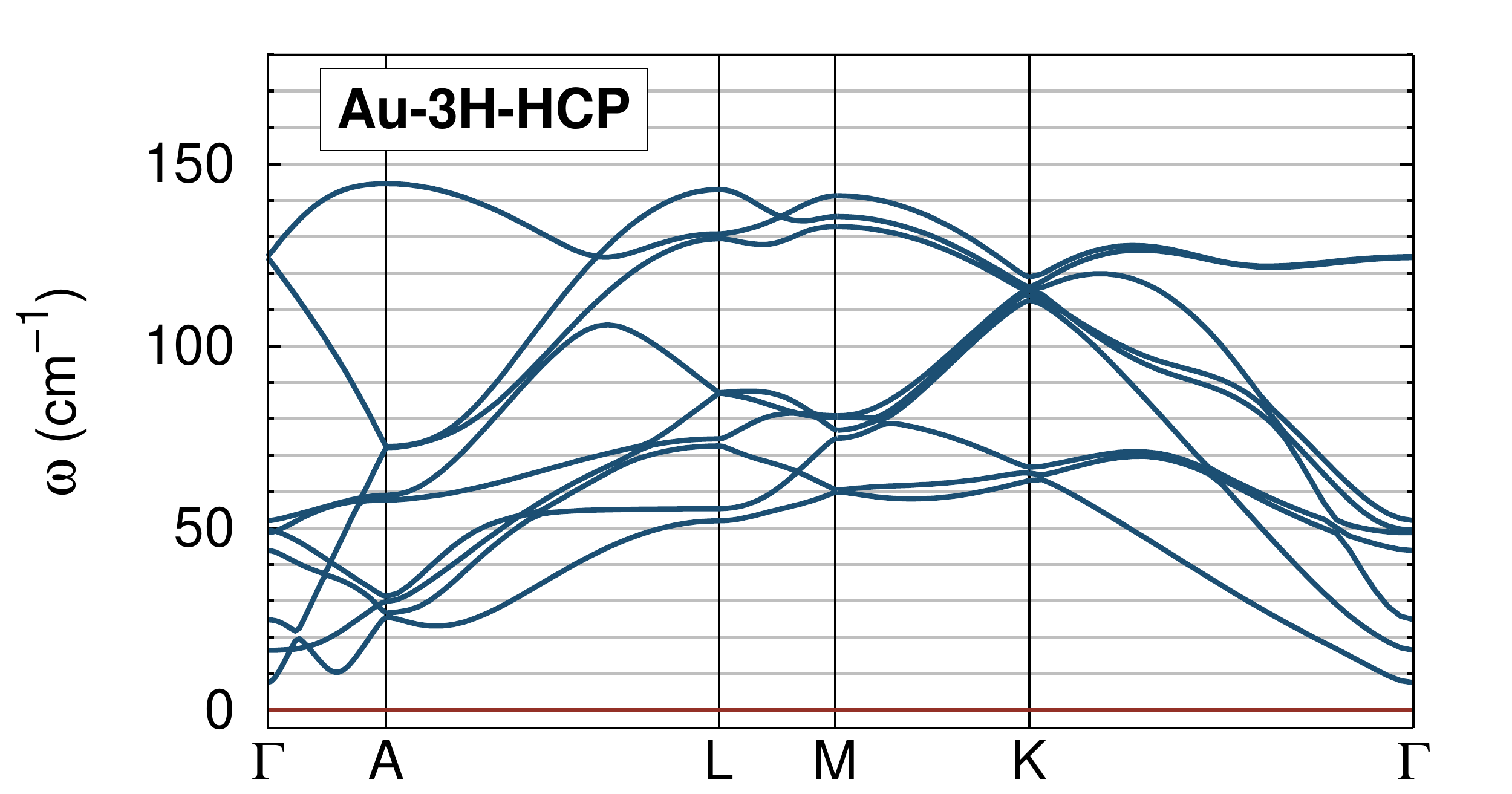}}
\subfloat{ \includegraphics[width=0.33\textwidth]{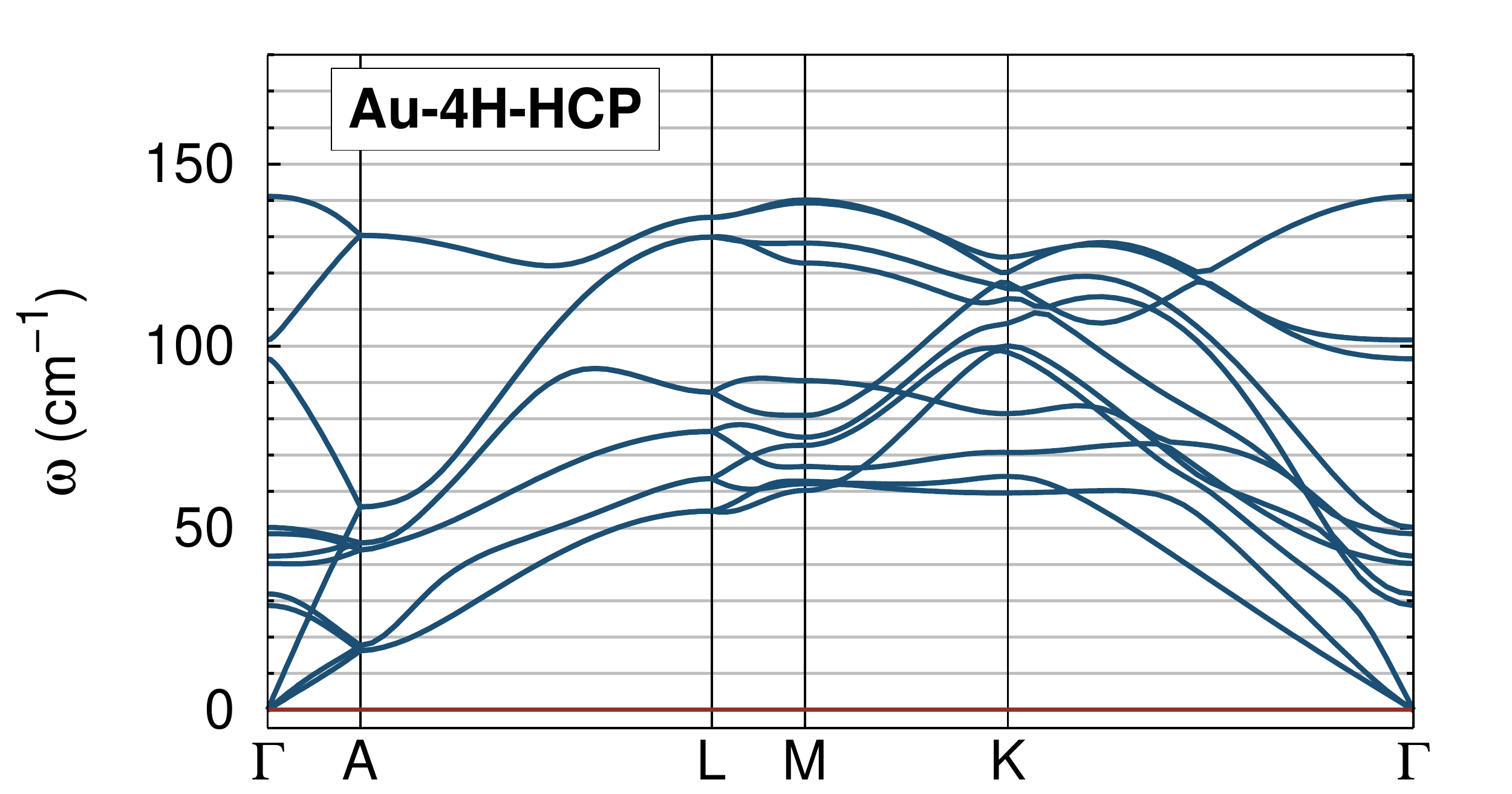}}

\subfloat{ \includegraphics[width=0.33\textwidth]{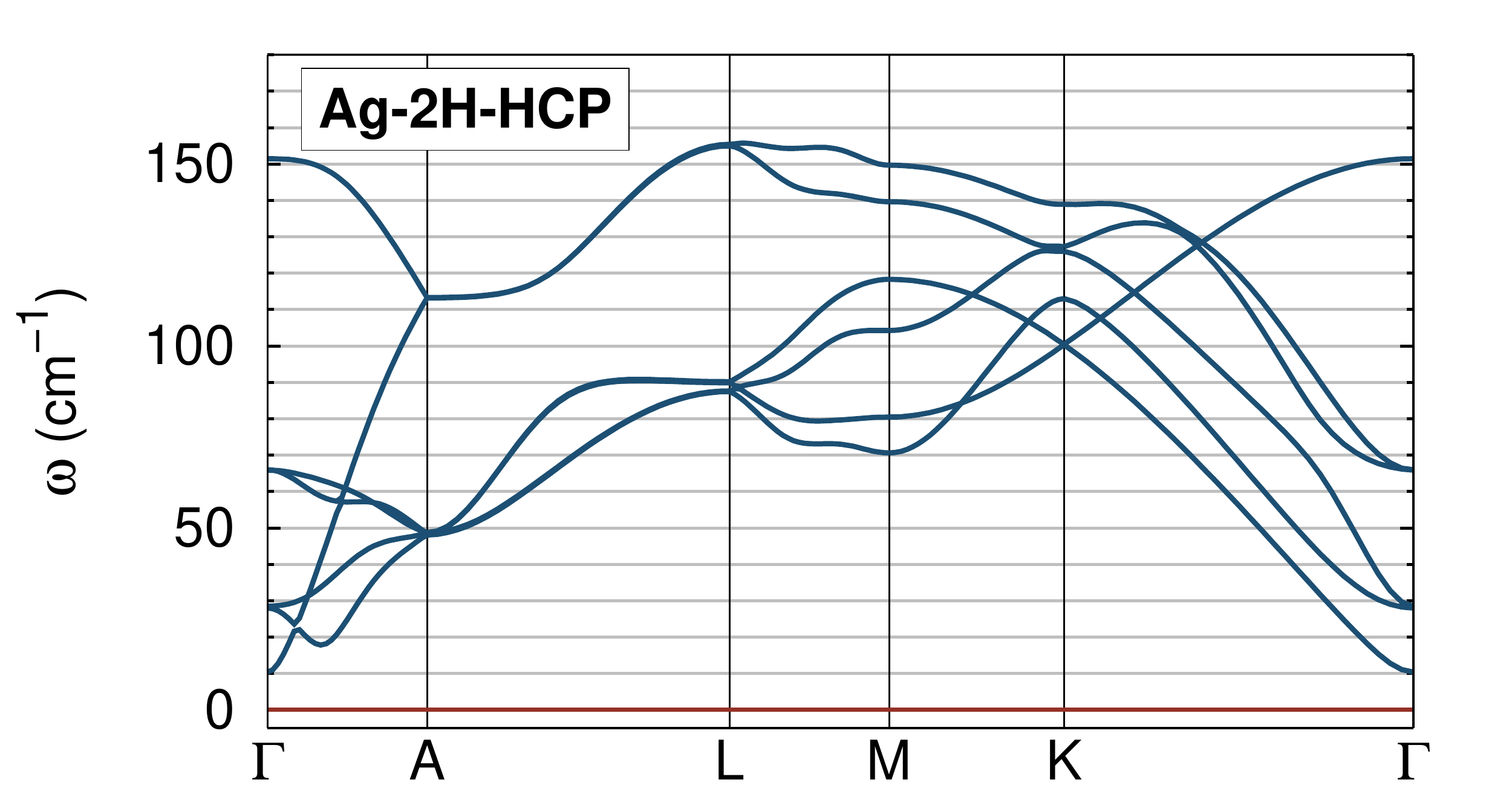}}
\subfloat{ \includegraphics[width=0.33\textwidth]{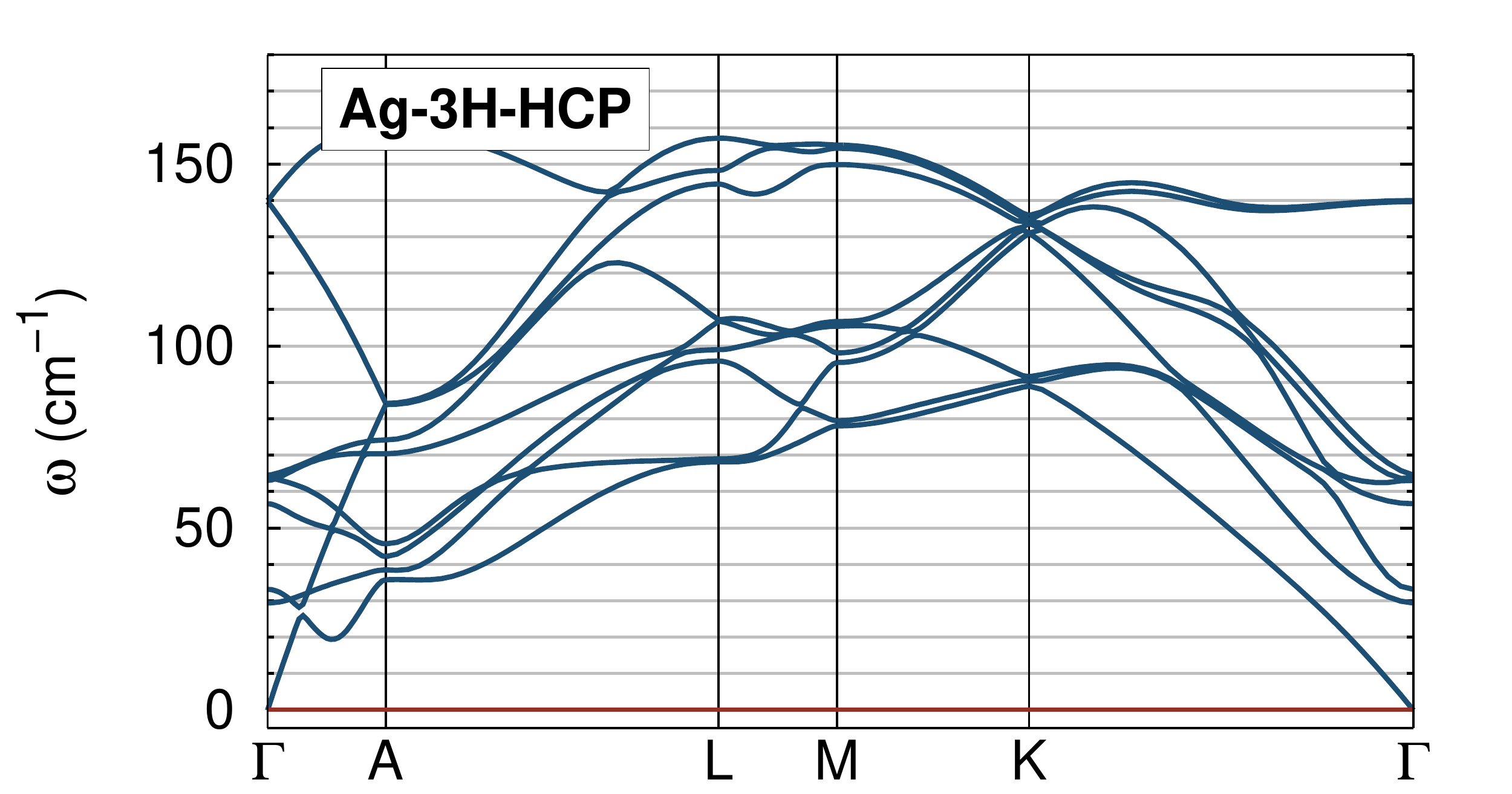}}
\subfloat{ \includegraphics[width=0.33\textwidth]{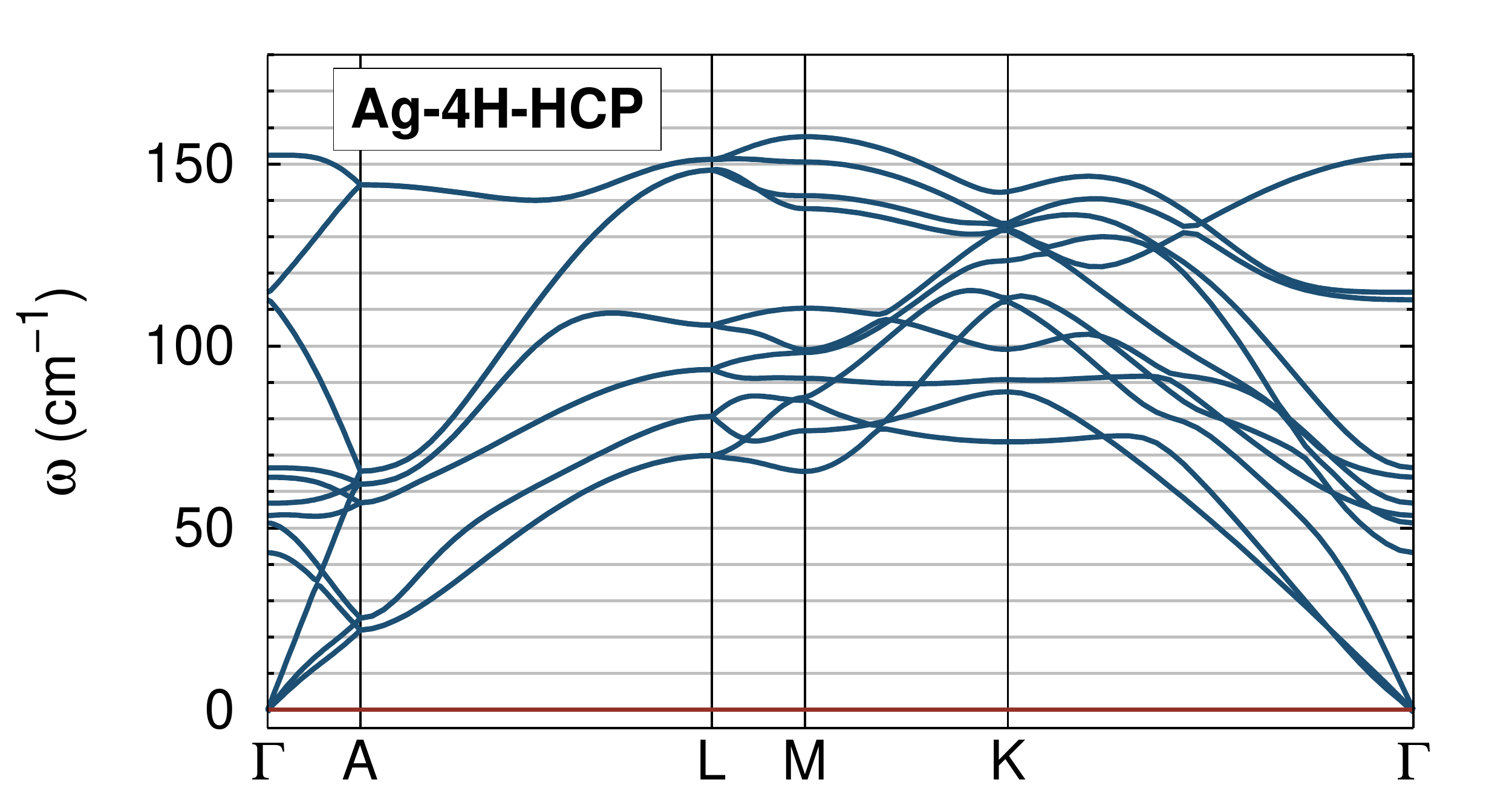}}

\subfloat{ \includegraphics[width=0.33\textwidth]{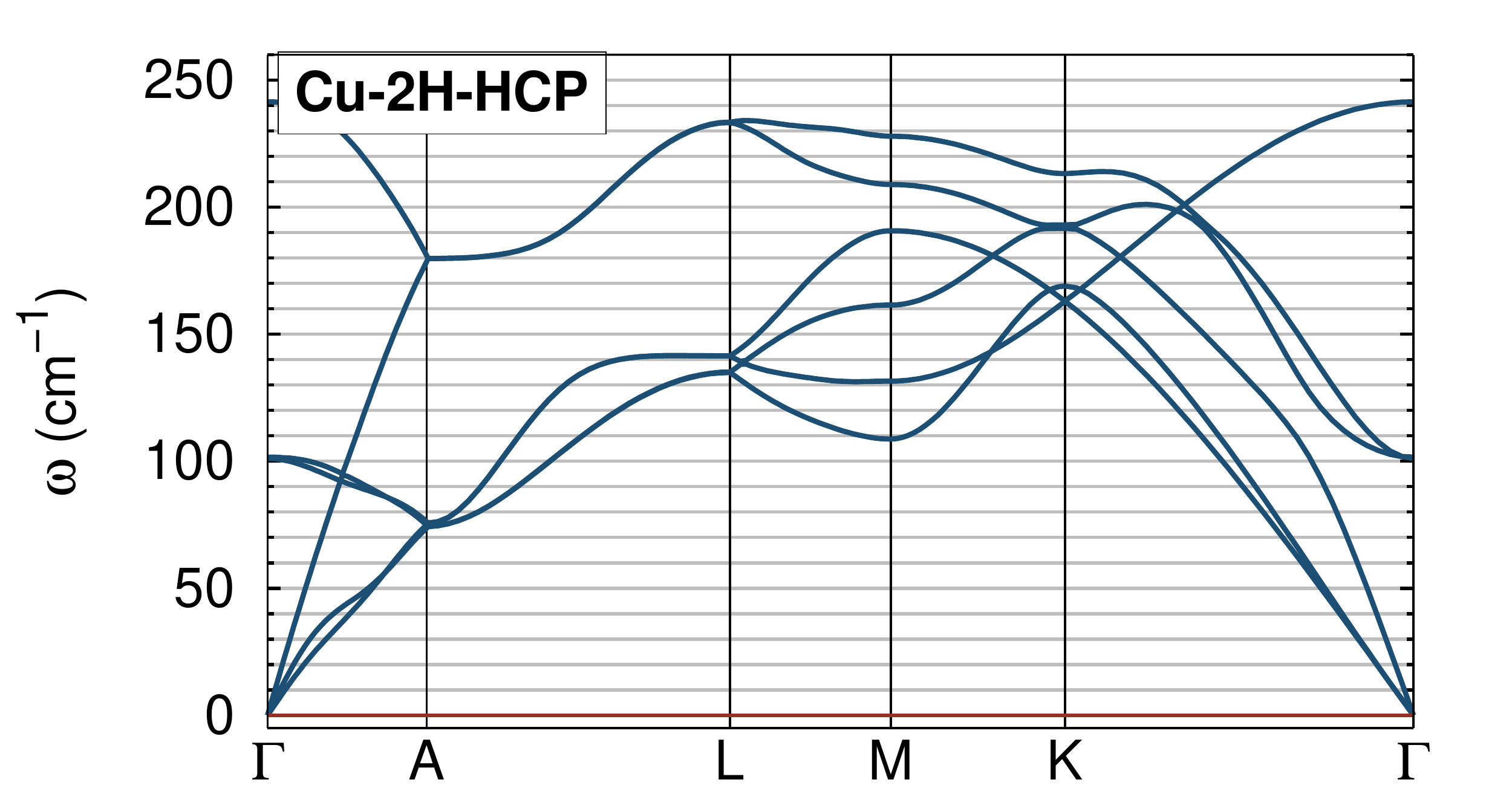}}
\subfloat{ \includegraphics[width=0.33\textwidth]{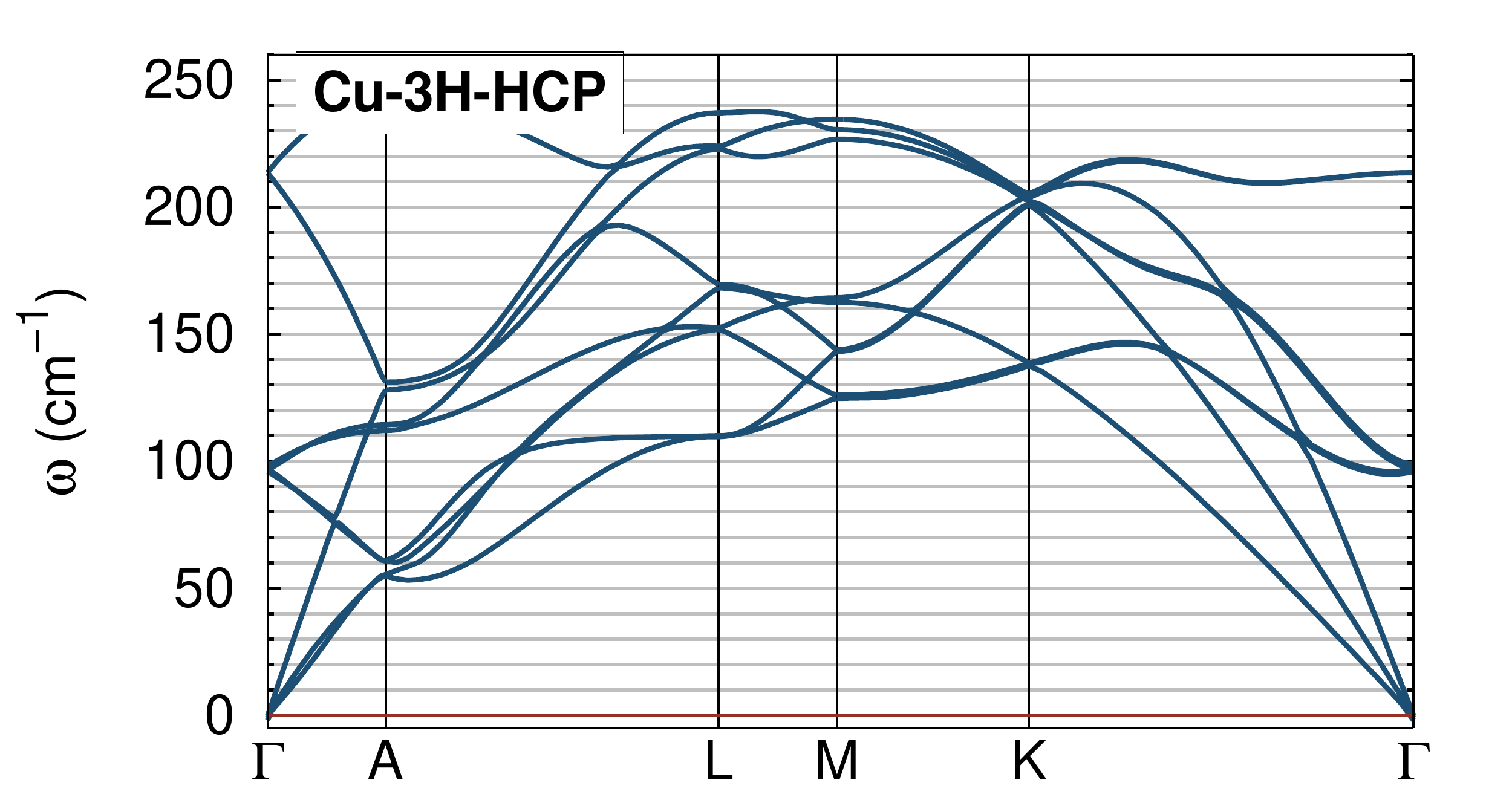}}
\subfloat{ \includegraphics[width=0.33\textwidth]{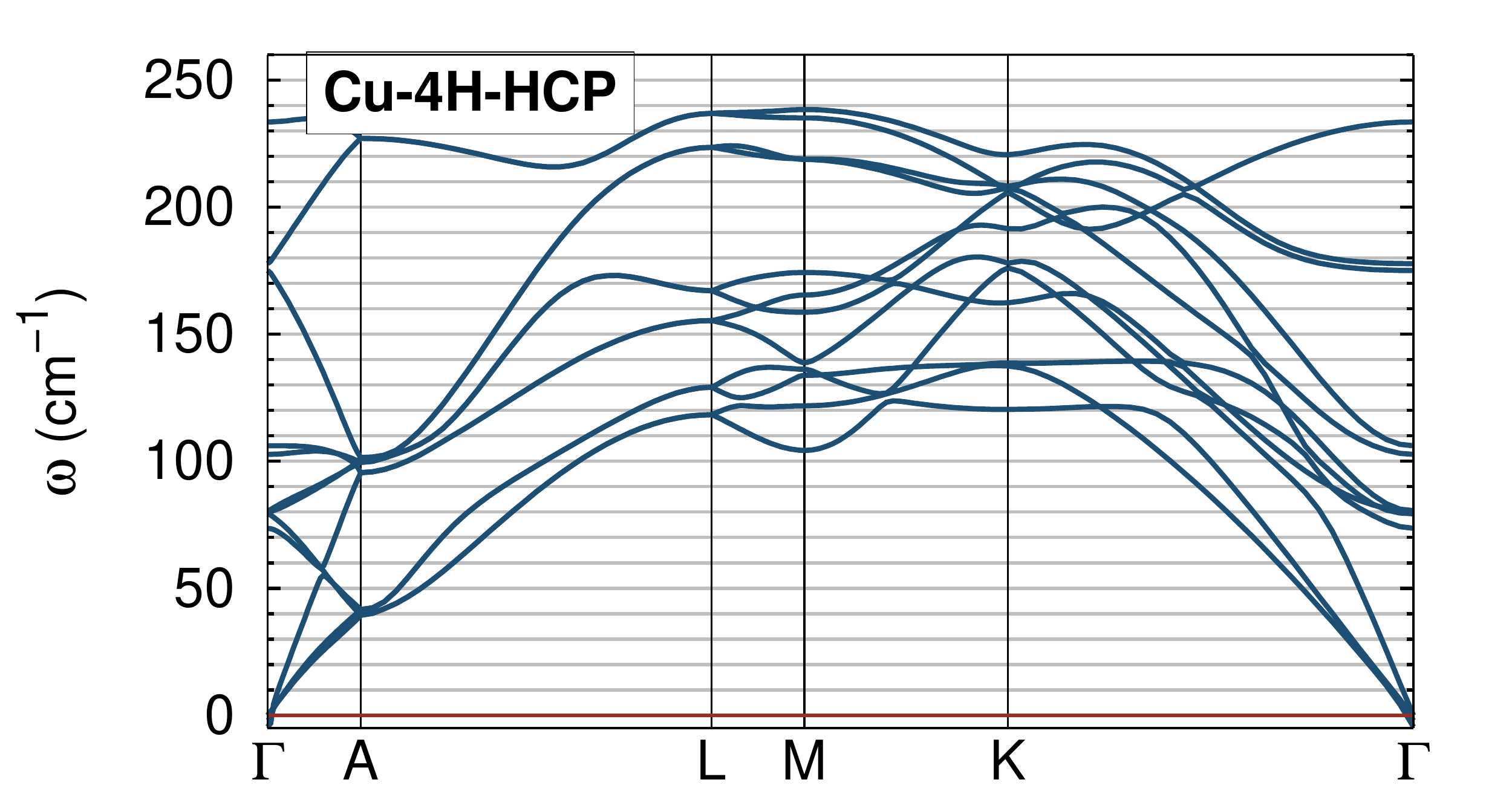}}

\caption{Phonon dispersions of the isolated  Au, Ag and Cu polytypes at their equilibrium volumes.}
\label{fig:PhDisp}
\end{sidewaysfigure}

\section{Temperature-dependence of Fermi velocities, electron-phonon mass enhancement parameters and  Drude plasmon parameters}

\begin{figure}[H]
\centering

\includegraphics[width=0.8\textwidth]{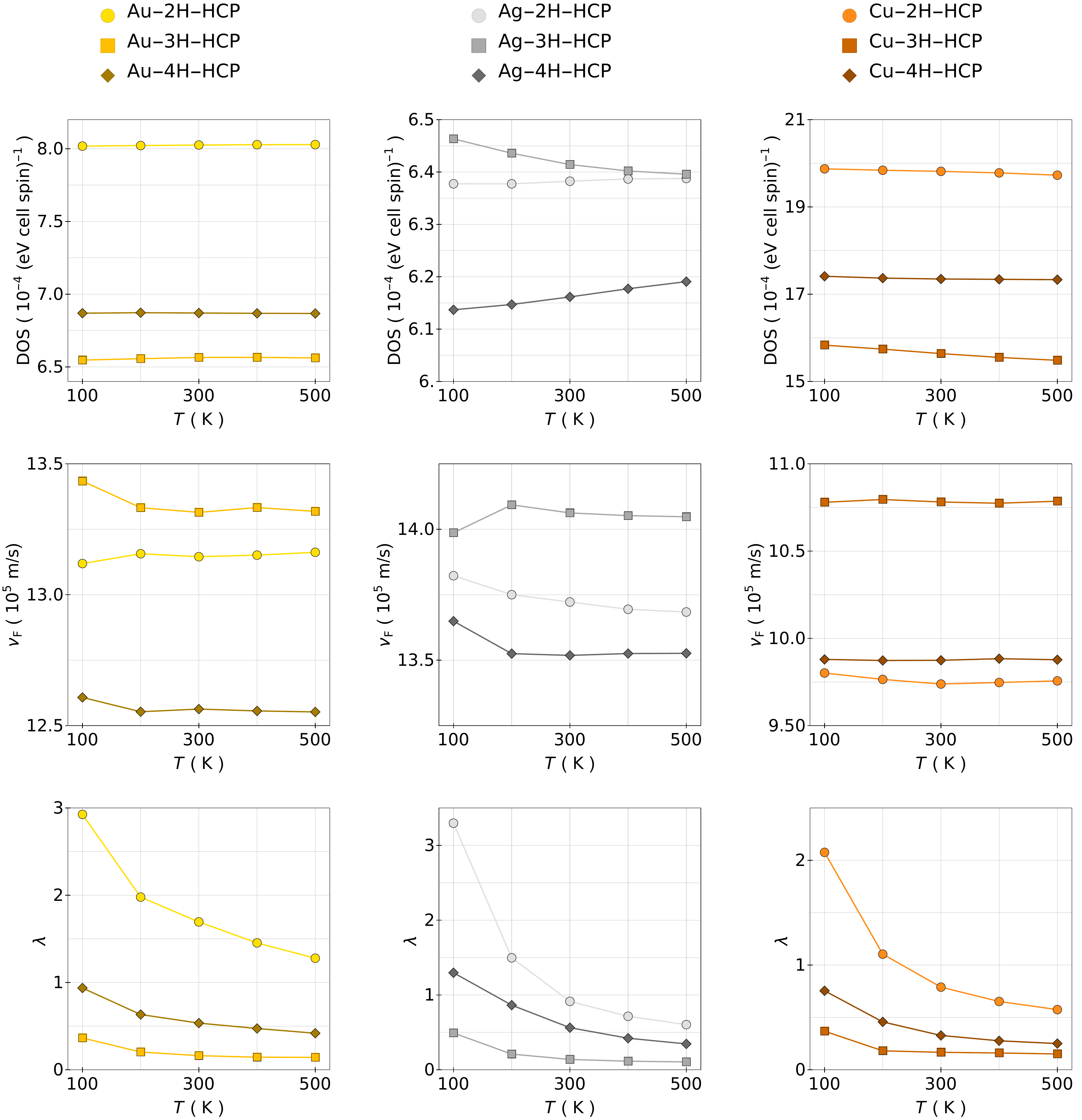}

\caption{Temperature dependence of the electronic density of states (DOS), the Fermi velocities ($v_\mathrm{F}$) and the electron-phonon mass enhancement parameters ($\lambda$) of the isolated Au, Ag and Cu polytypes.}
\label{fig:PreDP}
\end{figure}

\begin{sidewaystable}[p]
\renewcommand{\arraystretch}{1.5} \setlength{\tabcolsep}{5pt}
\begin{center}
{\footnotesize
\begin{tabular}{l ccc|ccc|ccc|ccc|ccc} \hline \hline 

 &  \multicolumn{3}{c}{100}  &  \multicolumn{3}{c}{200}  &  \multicolumn{3}{c}{300}  &  \multicolumn{3}{c}{400}  &  \multicolumn{3}{c}{500} \\ \cline{2-16}
 
 & $\omega_\mathrm{p}$ & $\eta_\mathrm{p}$ & $\tau_\mathrm{p}$ & $\omega_\mathrm{p}$ & $\eta_\mathrm{p}$ & $\tau_\mathrm{p}$ & $\omega_\mathrm{p}$ & $\eta_\mathrm{p}$ & $\tau_\mathrm{p}$ & $\omega_\mathrm{p}$ & $\eta_\mathrm{p}$ & $\tau_\mathrm{p}$ & $\omega_\mathrm{p}$ & $\eta_\mathrm{p}$ & $\tau_\mathrm{p}$ \\ \cline{2-16}

Au-2H-HCP  &  3.875  &  0.192  &  21.519  &  3.887  &  0.384  &  10.757  &  3.884  &  0.577  &  7.169  &  3.887  &  0.769  &  5.375  &  3.890  &  0.962  &  4.299 \\
Au-3H-HCP  &  8.304  &  0.018  &  224.470  &  8.247  &  0.037  &  111.768  &  8.241  &  0.056  &  74.205  &  8.253  &  0.075  &  55.424  &  8.242  &  0.094  &  44.156 \\
Au-4H-HCP  &  6.046  &  0.061  &  68.104  &  6.021  &  0.122  &  34.012  &  6.025  &  0.183  &  22.648  &  6.021  &  0.244  &  16.966  &  6.019  &  0.305  &  13.556 \\[0.25cm]

Ag-2H-HCP  &  5.186  &  0.104 &  39.830  &  5.159  &  0.208 &  19.898  &  5.151  &  0.312  &  13.254  &  5.142  &  0.416 &  9.932  &  5.138  &  0.521  &  7.939 \\
Ag-3H-HCP  &  8.864  &  0.016  &  261.521  &  8.913  &  0.032  &  130.040  &  8.878  &  0.048  &  86.214  &  8.863  &  0.064  &  64.304  &  8.856  &  0.081  &  51.162 \\
Ag-4H-HCP  &  6.151  &  0.064  &  64.699  &  6.100 &  0.128  &  32.302  &  6.104  &  0.192  &  21.504  &  6.115  &  0.257  &  16.105  &  6.122  &  0.321  &  12.865 \\[0.25cm]

Cu-2H-HCP  &  5.757  &  0.090  &  46.204  &  5.730  &  0.179  &  23.098  &  5.711  &  0.269  &  15.396  &  5.711  &  0.358  &  11.545  &  5.709  &  0.448  &  9.235 \\
Cu-3H-HCP  &  8.655  &  0.019  &  218.194  &  8.643  &  0.038  &  108.975  &  8.603  &  0.057  &  72.567 &  8.574  &  0.076  &  54.362  &  8.564  &  0.095  &  43.439 \\
Cu-4H-HCP  &  7.312  &  0.037  &  111.196  &  7.298  &  0.074  &  55.571  &  7.295  &  0.112  &  37.029  &  7.300  &  0.149  &  27.758  &  7.294 &  0.186  &  22.196 \\

\hline \hline
\end{tabular}}
\end{center}
\caption{Drude plasmon energies ($\omega_\mathrm{p}$) in eV units, the inverse lifetimes ($\eta_\mathrm{p}$) in eV units and lifetimes ($\tau_\mathrm{p}$) in fs units of the isolated  Au, Ag and Cu polytypes at the available temperatures (in K units).}

\label{tab:DPList}
\end{sidewaystable}

\begin{sidewaysfigure}[p]
\centering

\includegraphics[width=0.3\textwidth]{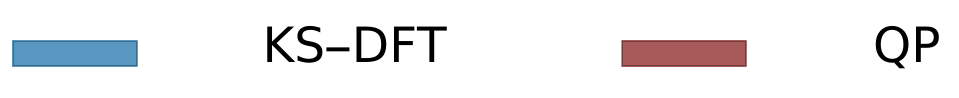}

\subfloat{ \includegraphics[width=0.33\textwidth]{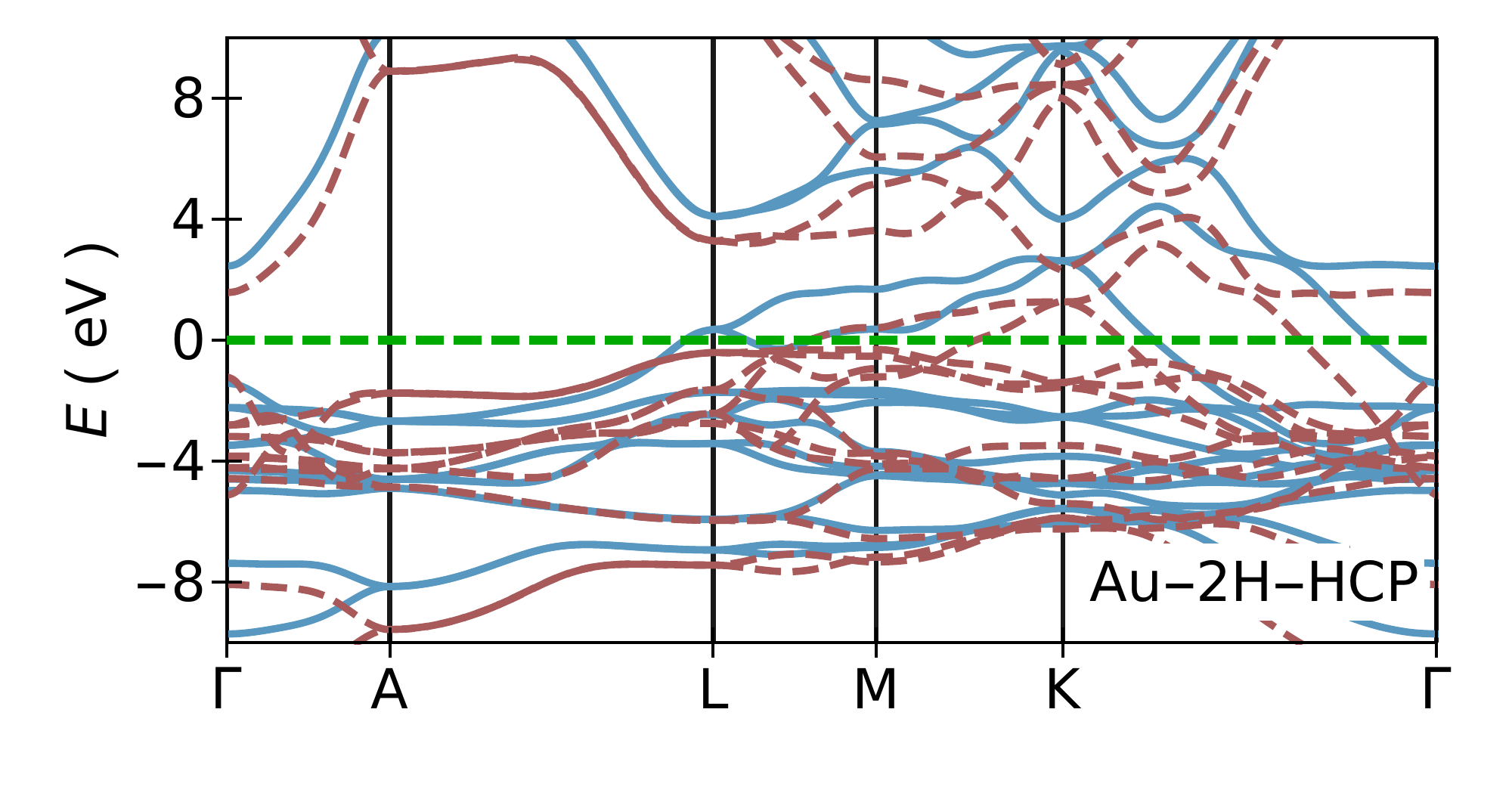}}
\subfloat{ \includegraphics[width=0.33\textwidth]{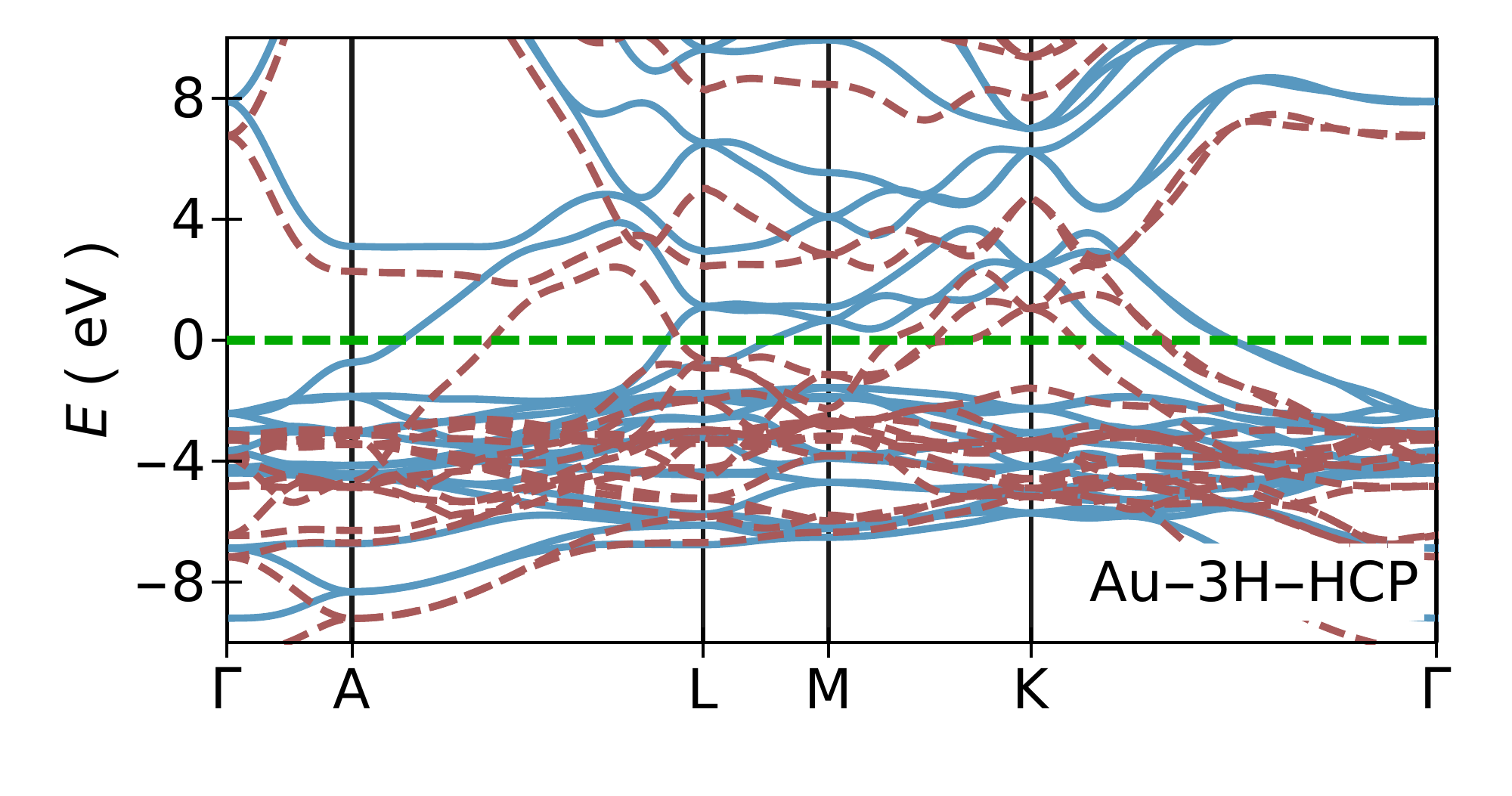}}
\subfloat{ \includegraphics[width=0.33\textwidth]{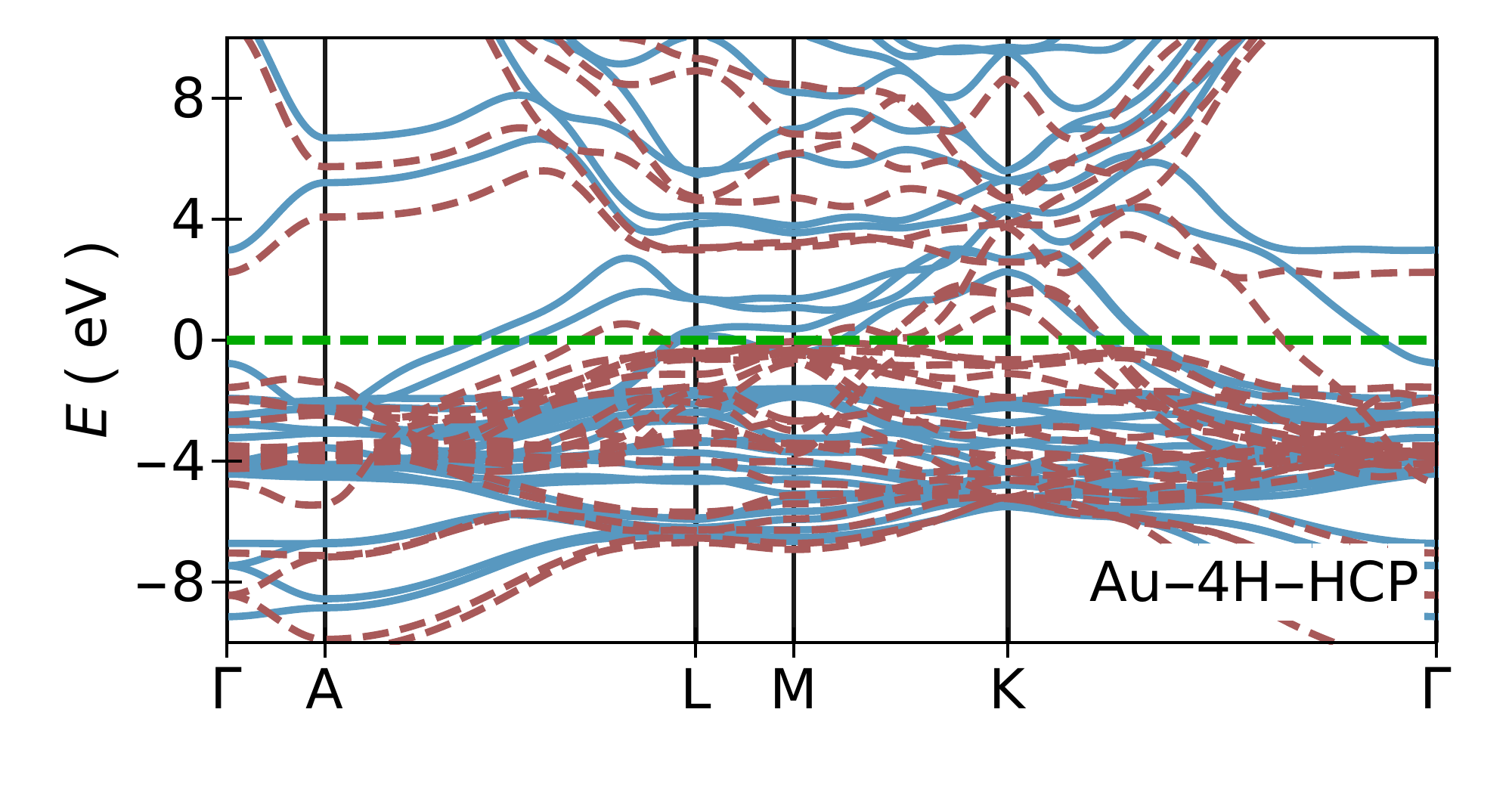}}

\subfloat{ \includegraphics[width=0.33\textwidth]{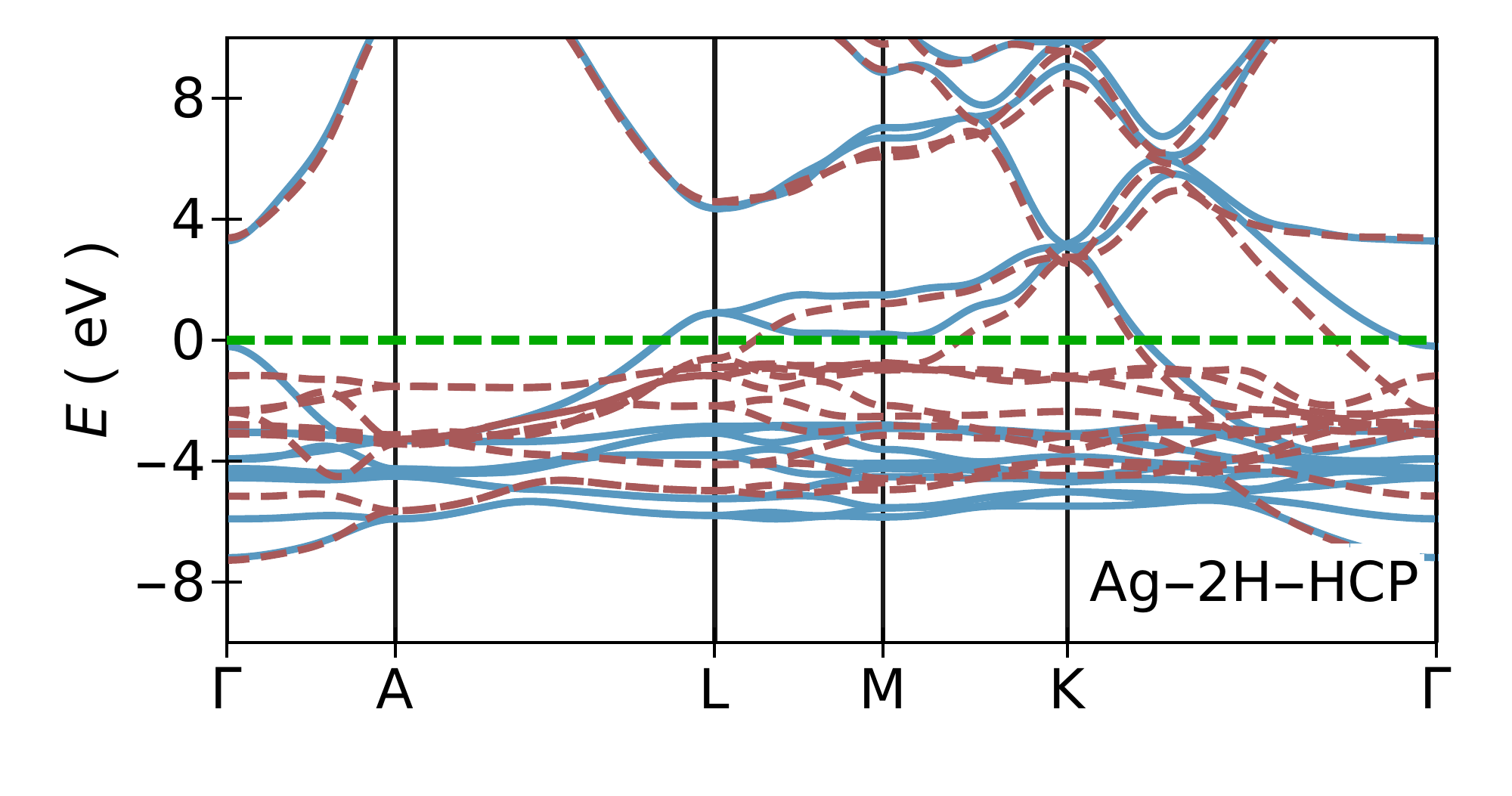}}
\subfloat{ \includegraphics[width=0.33\textwidth]{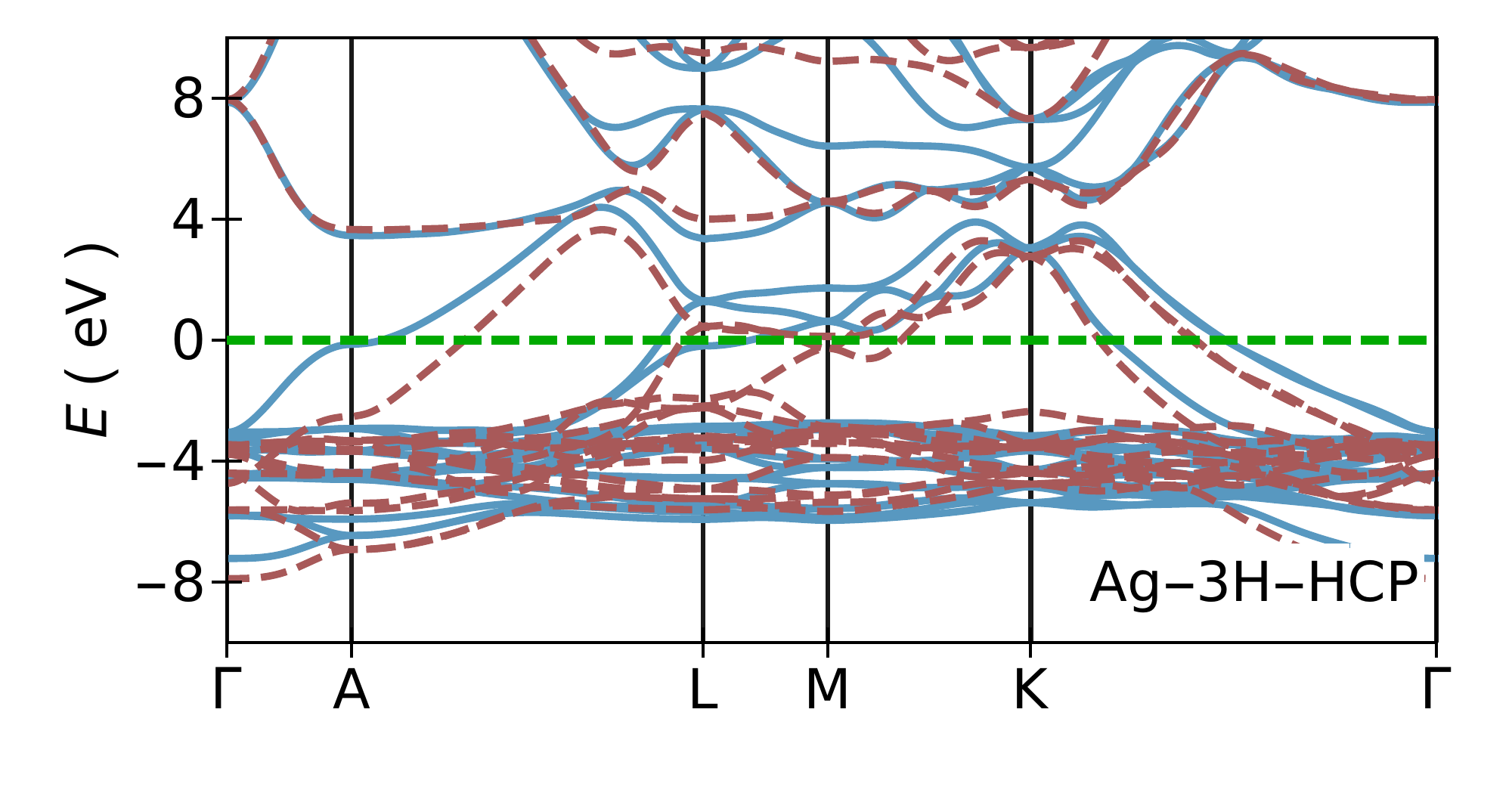}}
\subfloat{ \includegraphics[width=0.33\textwidth]{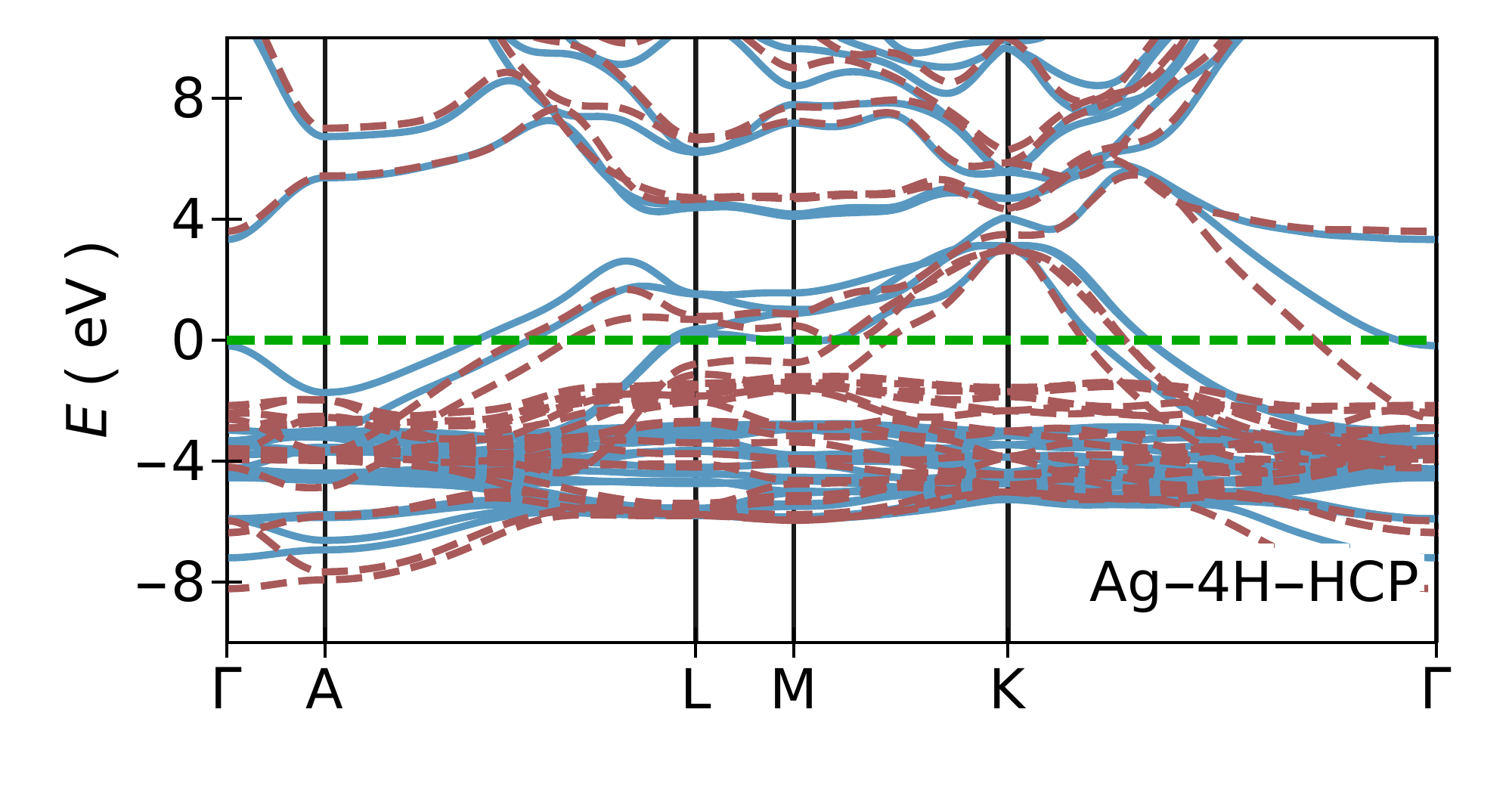}}

\subfloat{ \includegraphics[width=0.33\textwidth]{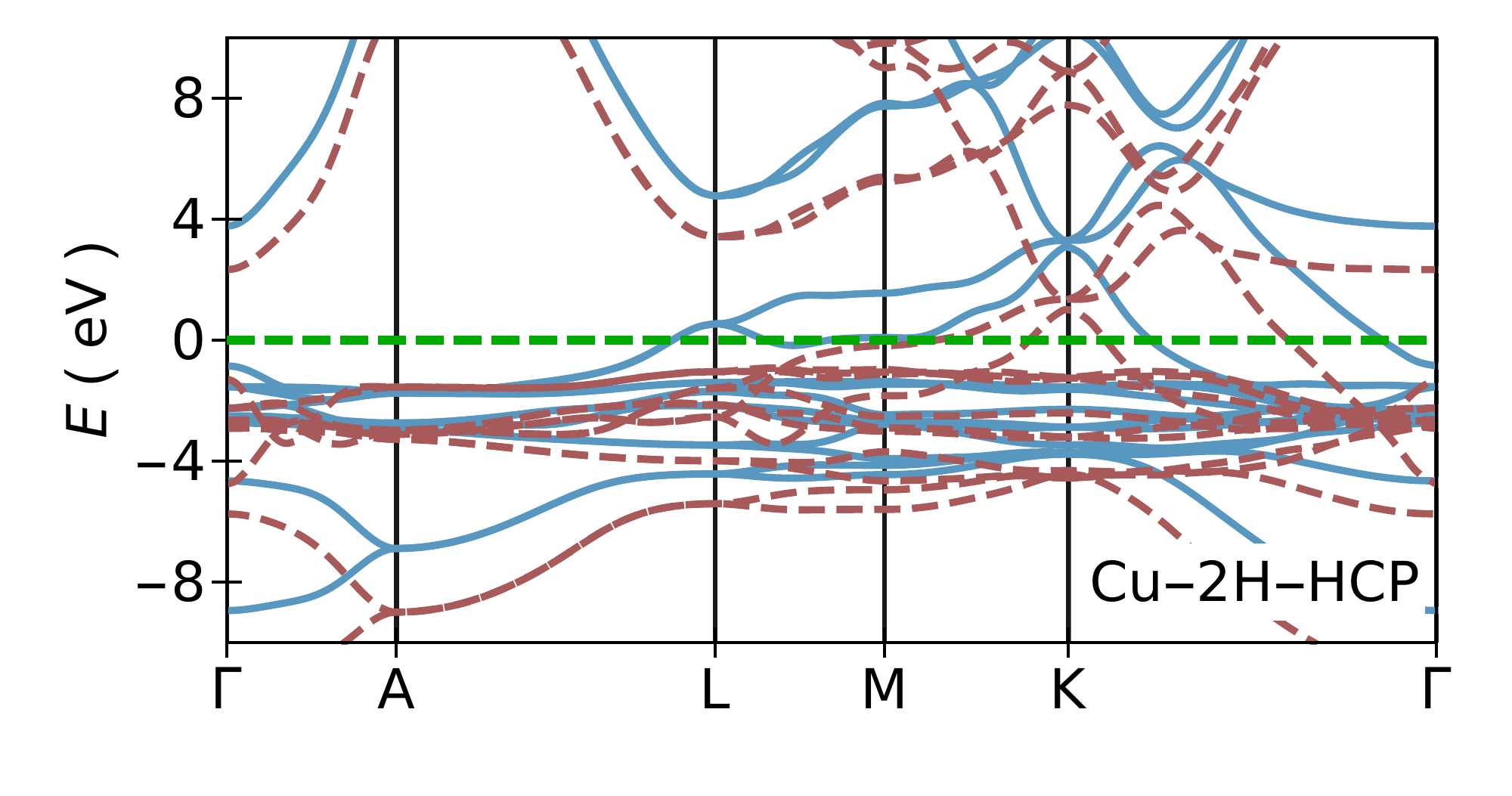}}
\subfloat{ \includegraphics[width=0.33\textwidth]{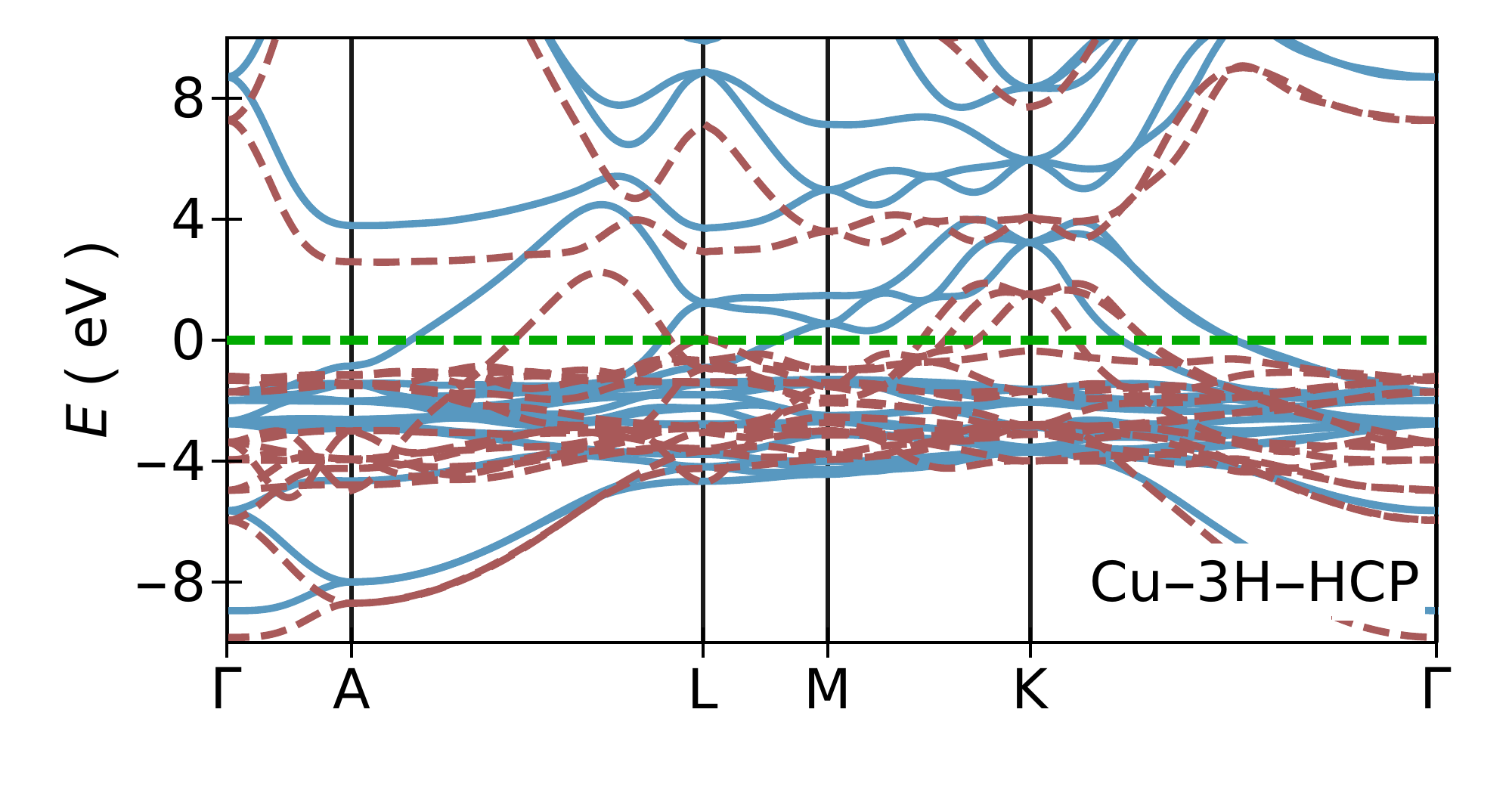}}
\subfloat{ \includegraphics[width=0.33\textwidth]{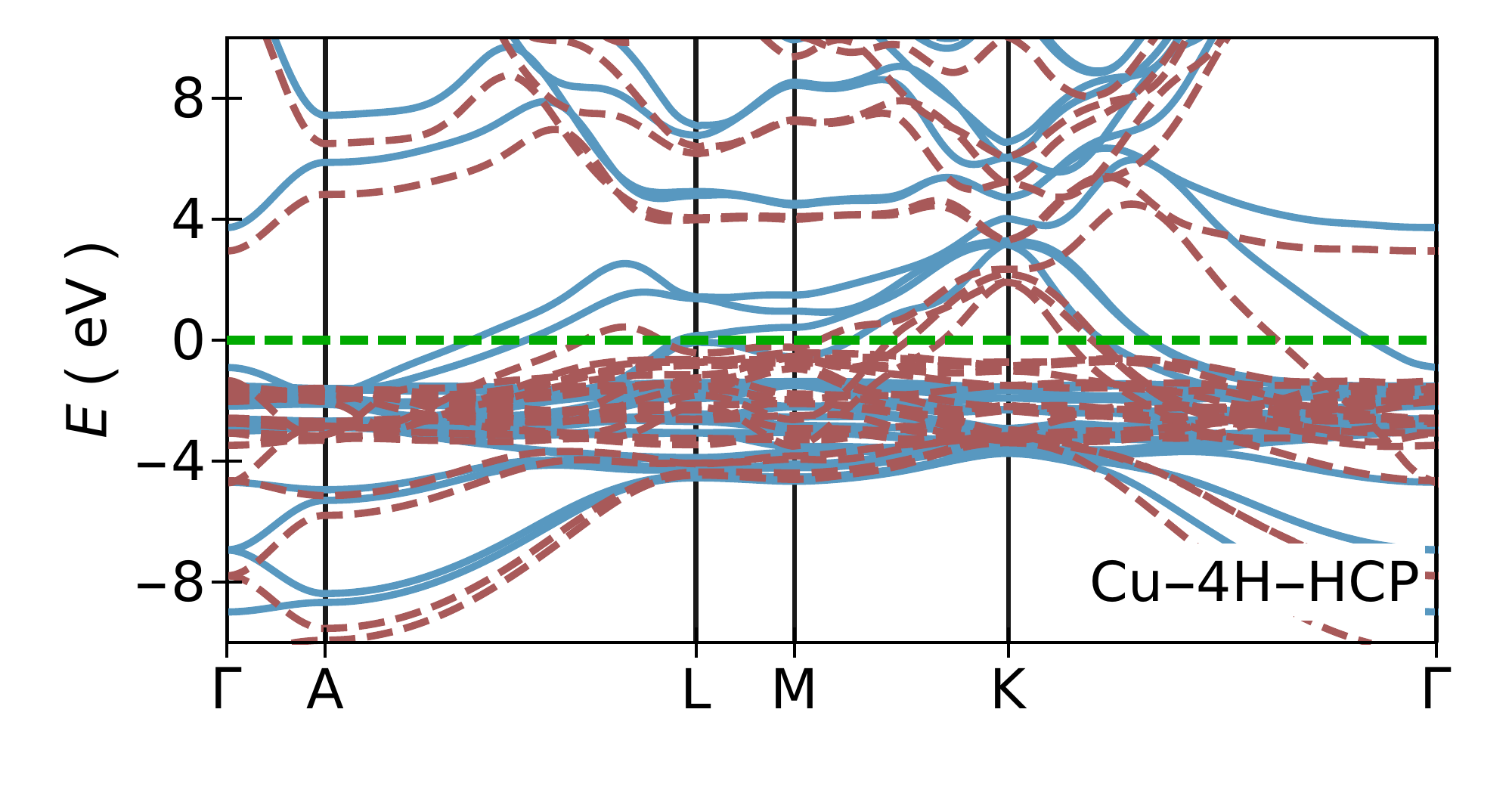}}

\caption{Kohn-Sham band-structures (blue solid lines) and the approximate quasiparticle band-structure (red dashed lines). 
The Fermi levels are set to  $0$~eV.
}
\label{fig:ElBands}
\end{sidewaysfigure}

\section{Temperature-dependence of localized surface-plasmons}

\begin{figure}[H]
\centering
\includegraphics[width=0.8\textwidth]{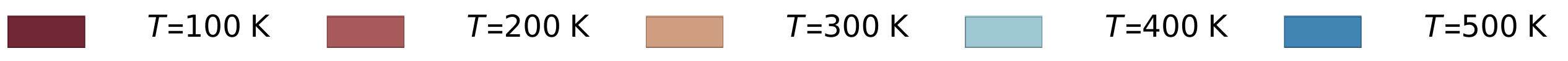}

\subfloat{ \includegraphics[width=0.3\textwidth]{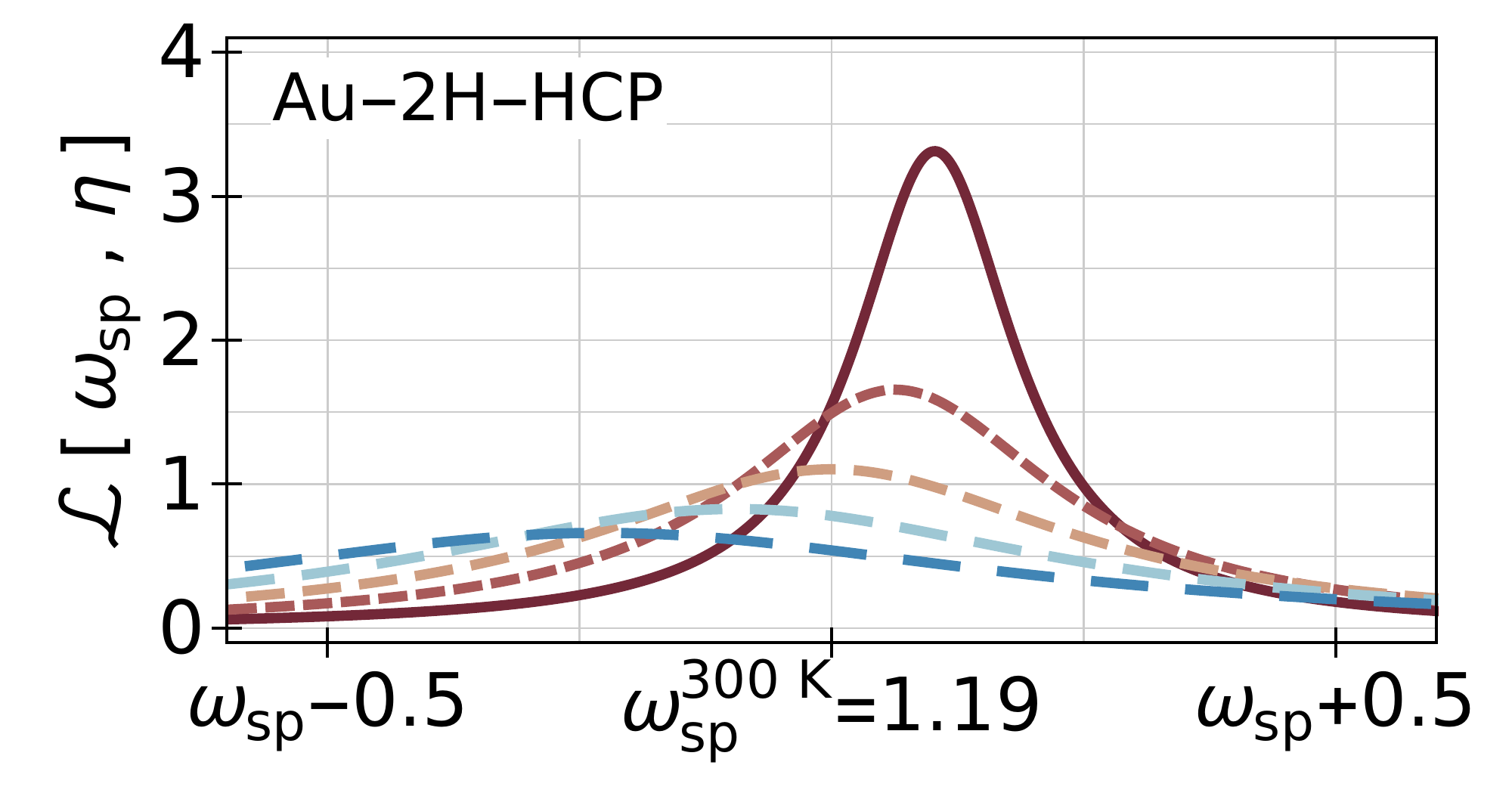}}
\subfloat{ \includegraphics[width=0.3\textwidth]{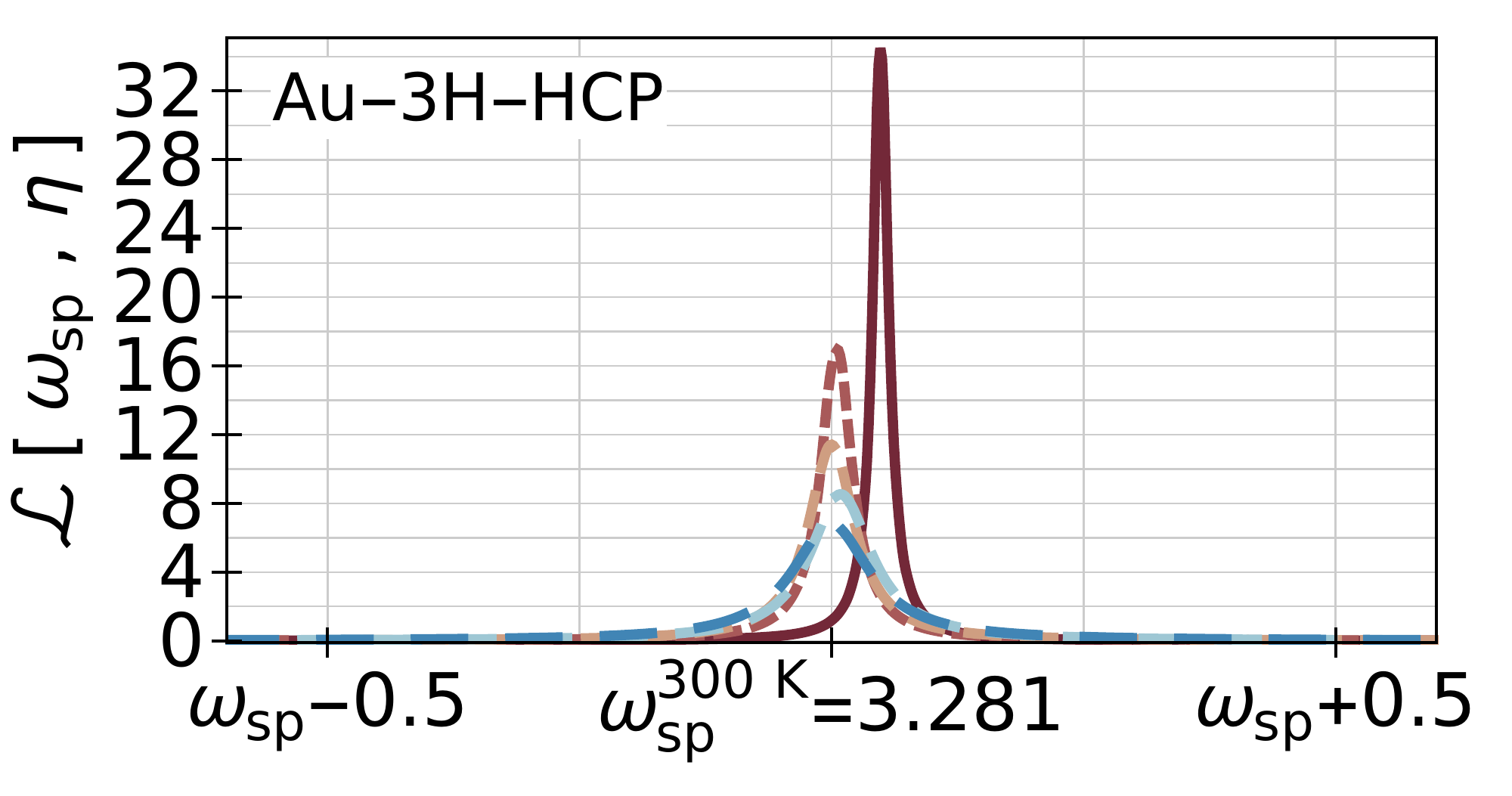}}
\subfloat{ \includegraphics[width=0.3\textwidth]{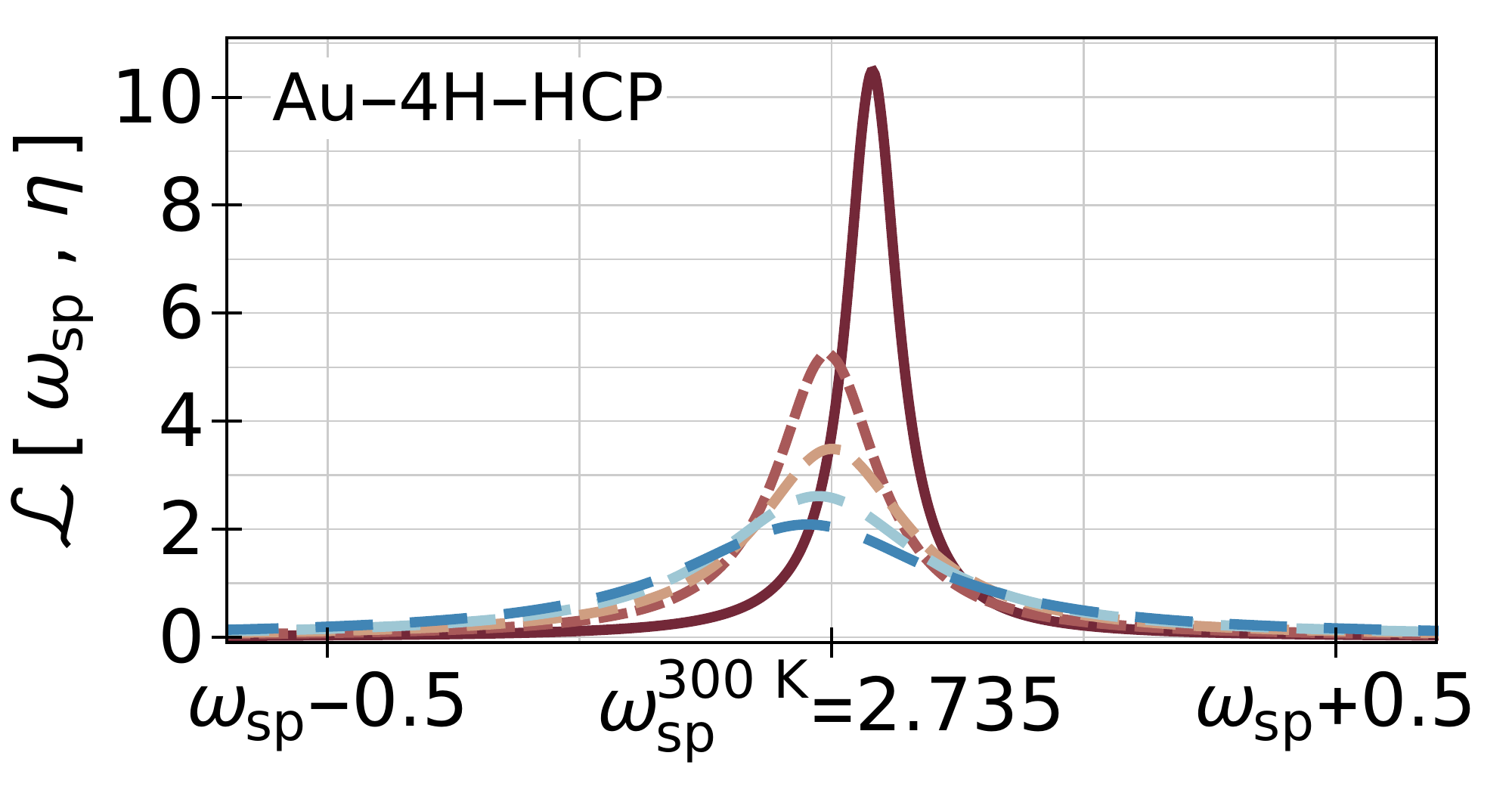}}

\subfloat{ \includegraphics[width=0.3\textwidth]{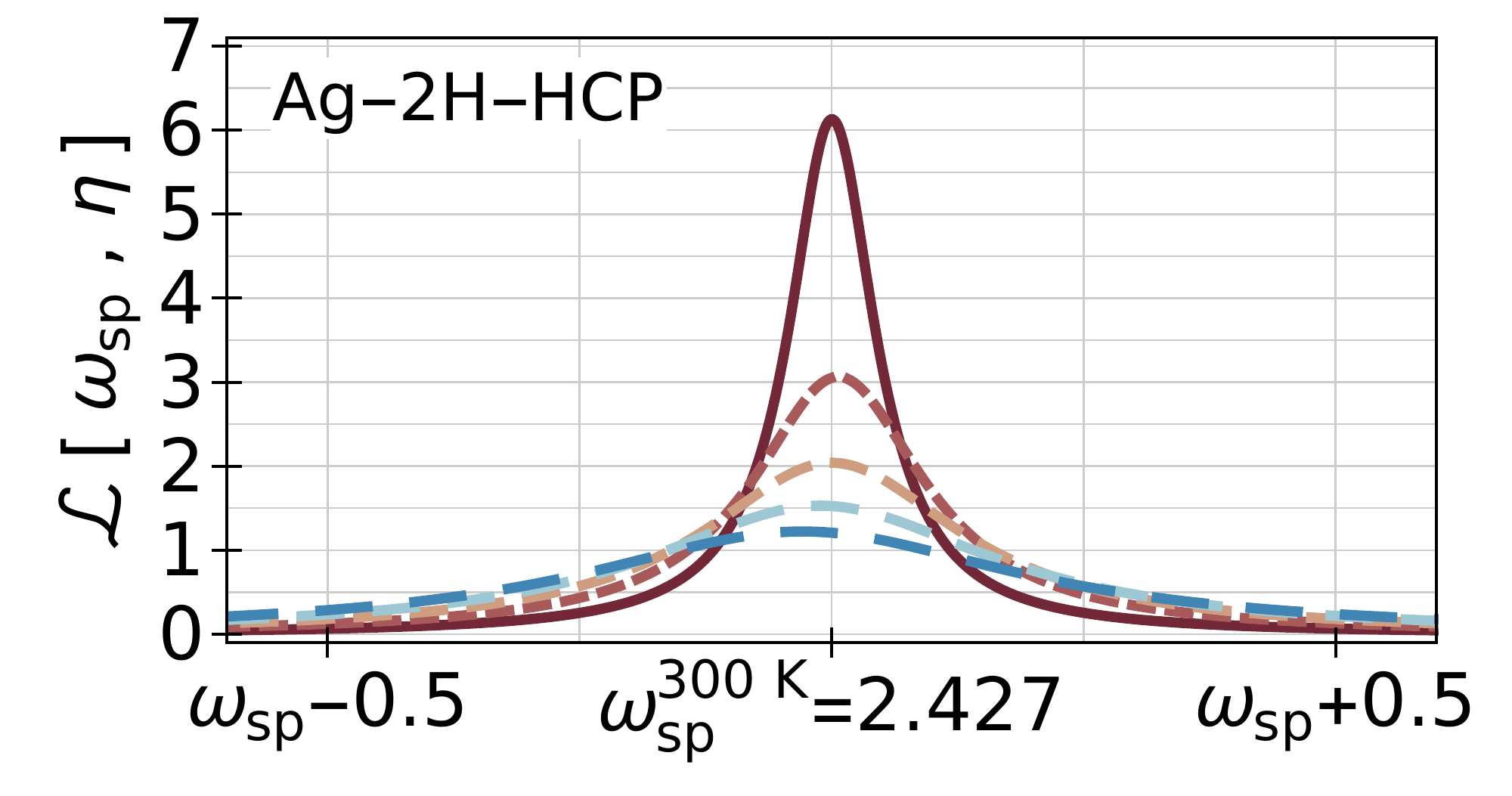}}
\subfloat{ \includegraphics[width=0.3\textwidth]{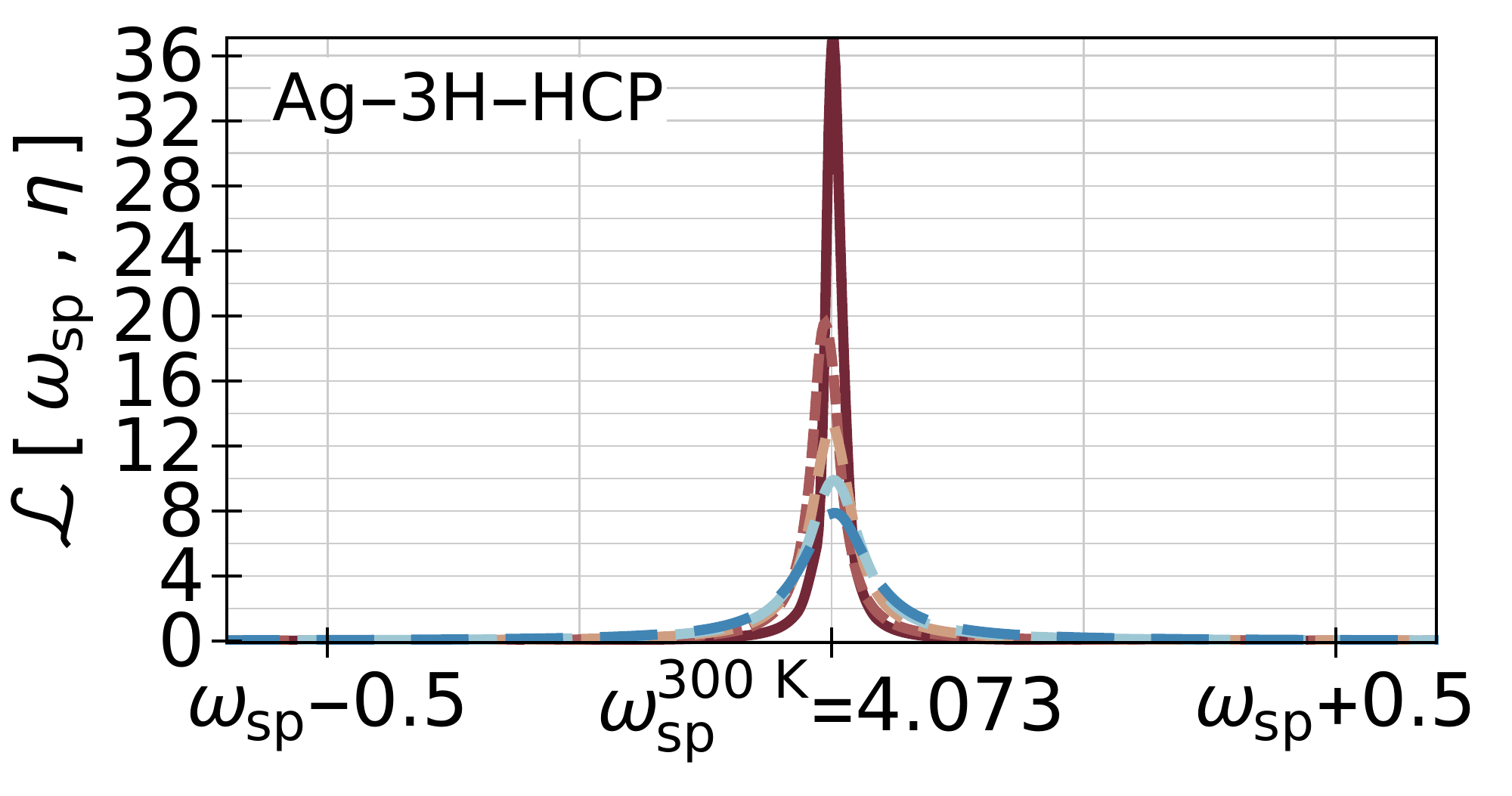}}
\subfloat{ \includegraphics[width=0.3\textwidth]{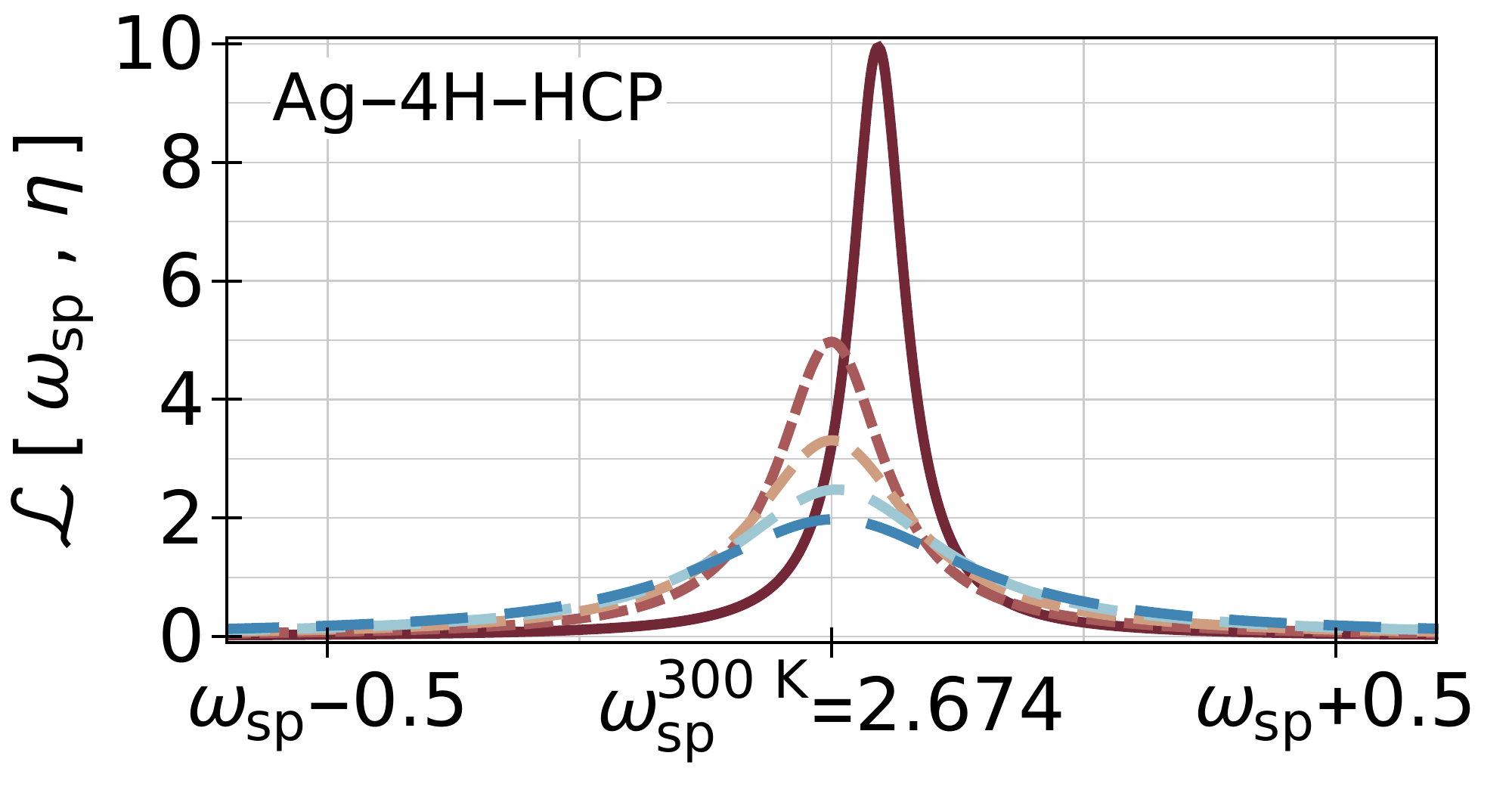}}

\subfloat{ \includegraphics[width=0.3\textwidth]{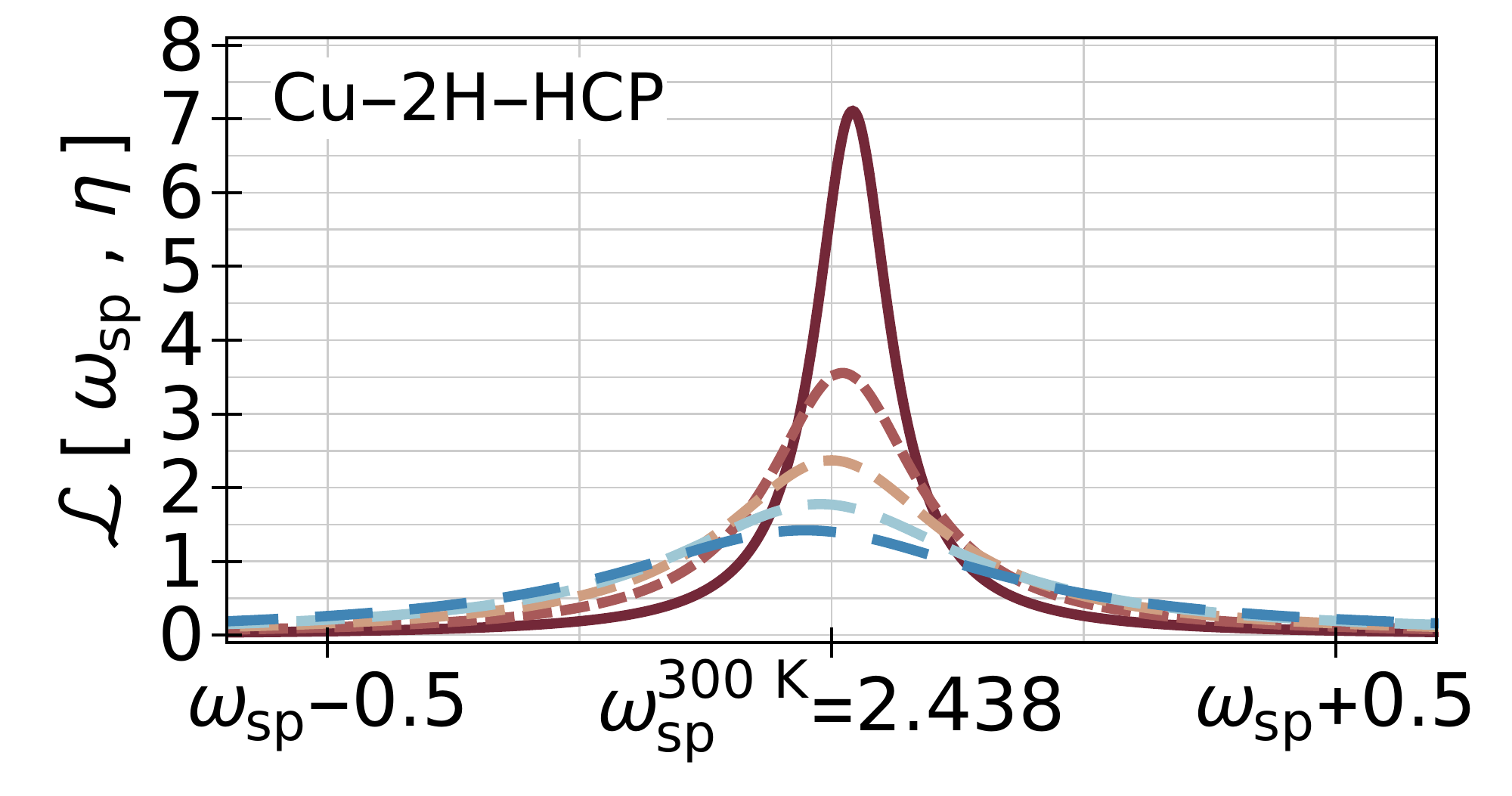}}
\subfloat{ \includegraphics[width=0.3\textwidth]{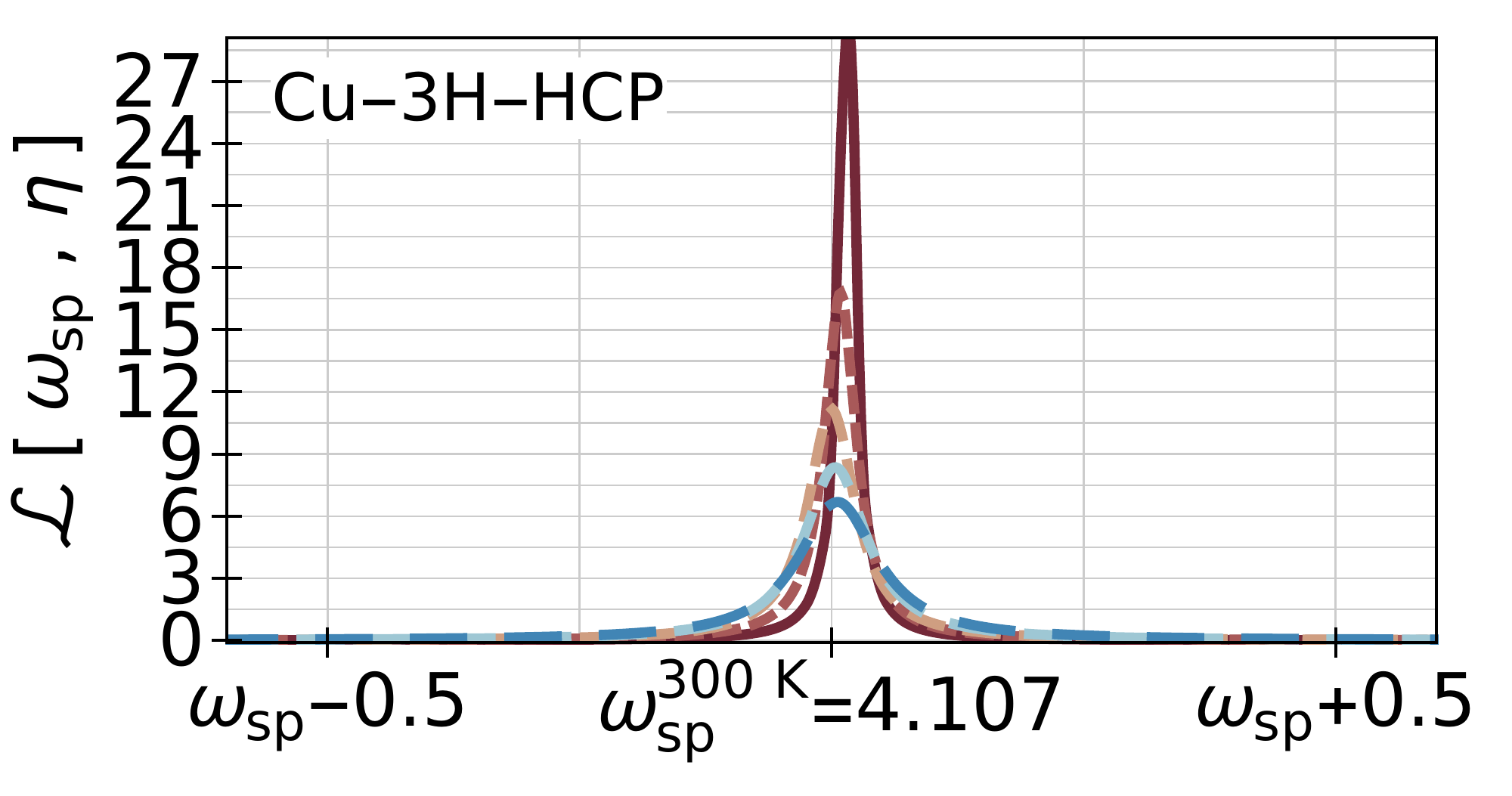}}
\subfloat{ \includegraphics[width=0.3\textwidth]{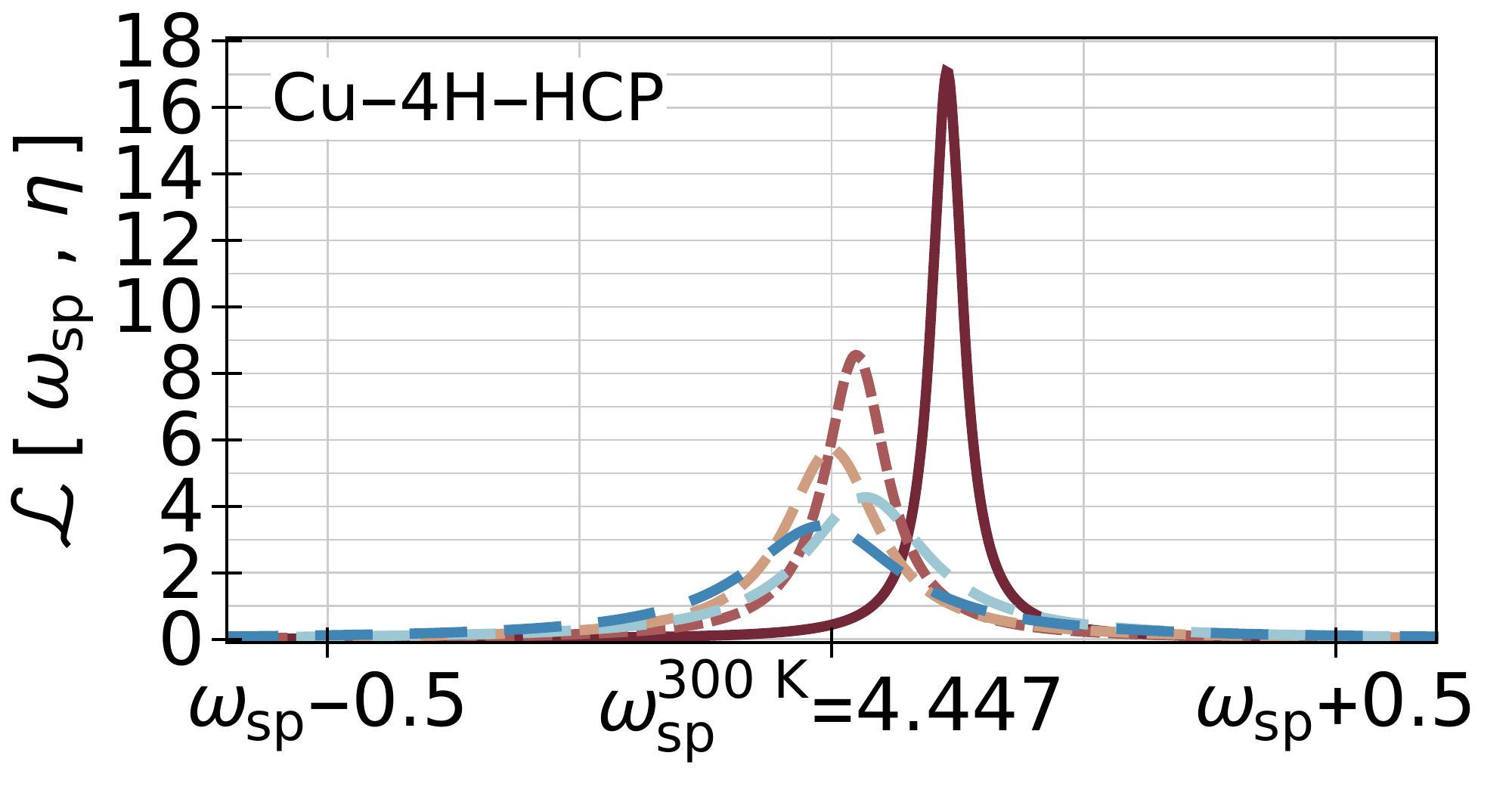}}

\caption{Temperature dependence of the localized surface-plasmon resonance of noble-metal nanoparticles on a hosting matrix with the unit dielectric permittivity of $\ep_m=1.0$ when the geometry- and size-dependence of the inverse plasmon lifetime are ignored by setting $\alpha=0.0 \rightarrow \eta=\eta_\mathrm{p}$, shown using the Lorentzian line shape ($\mathcal{L}$).}
\label{fig:SurfPlasT}
\end{figure}

\section{Hosting-matrix and size-dependence of localized surface-plasmon at 400~K}

\begin{figure}[H]
\centering
\includegraphics[width=1.0\textwidth]{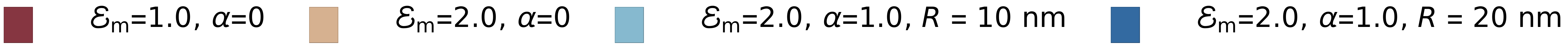}

\subfloat{ \includegraphics[width=0.3\textwidth]{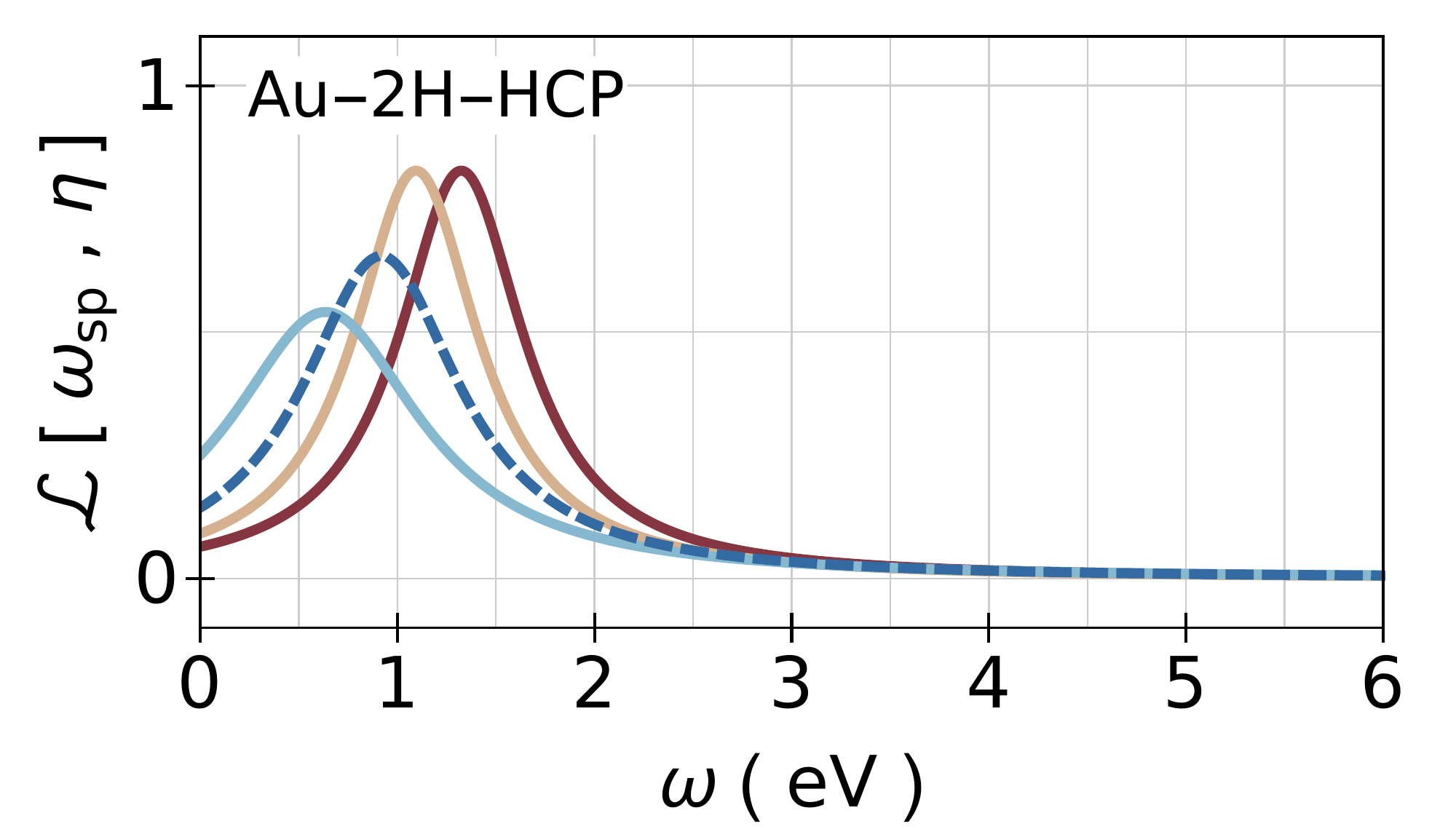}}
\subfloat{ \includegraphics[width=0.3\textwidth]{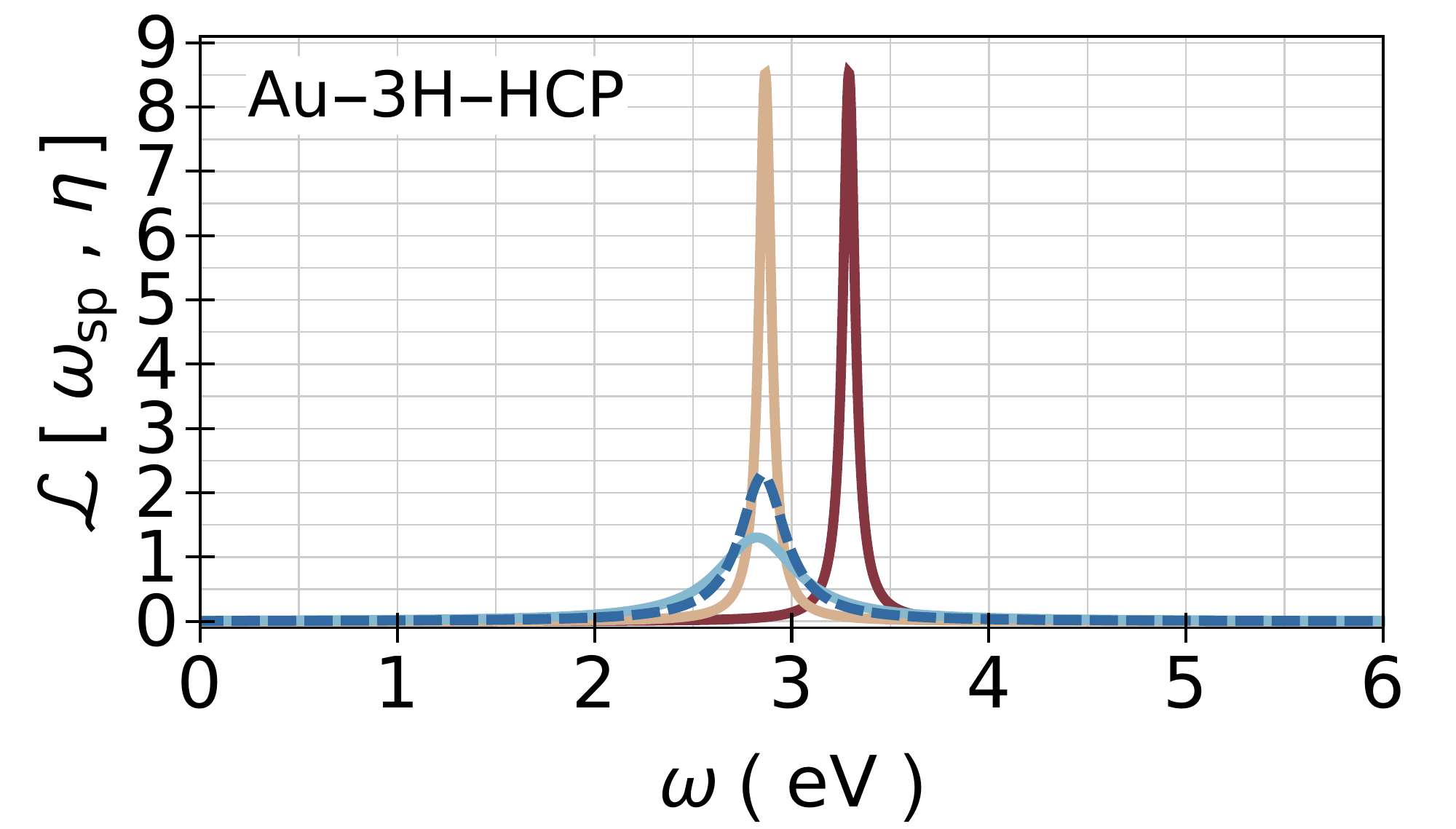}}
\subfloat{ \includegraphics[width=0.3\textwidth]{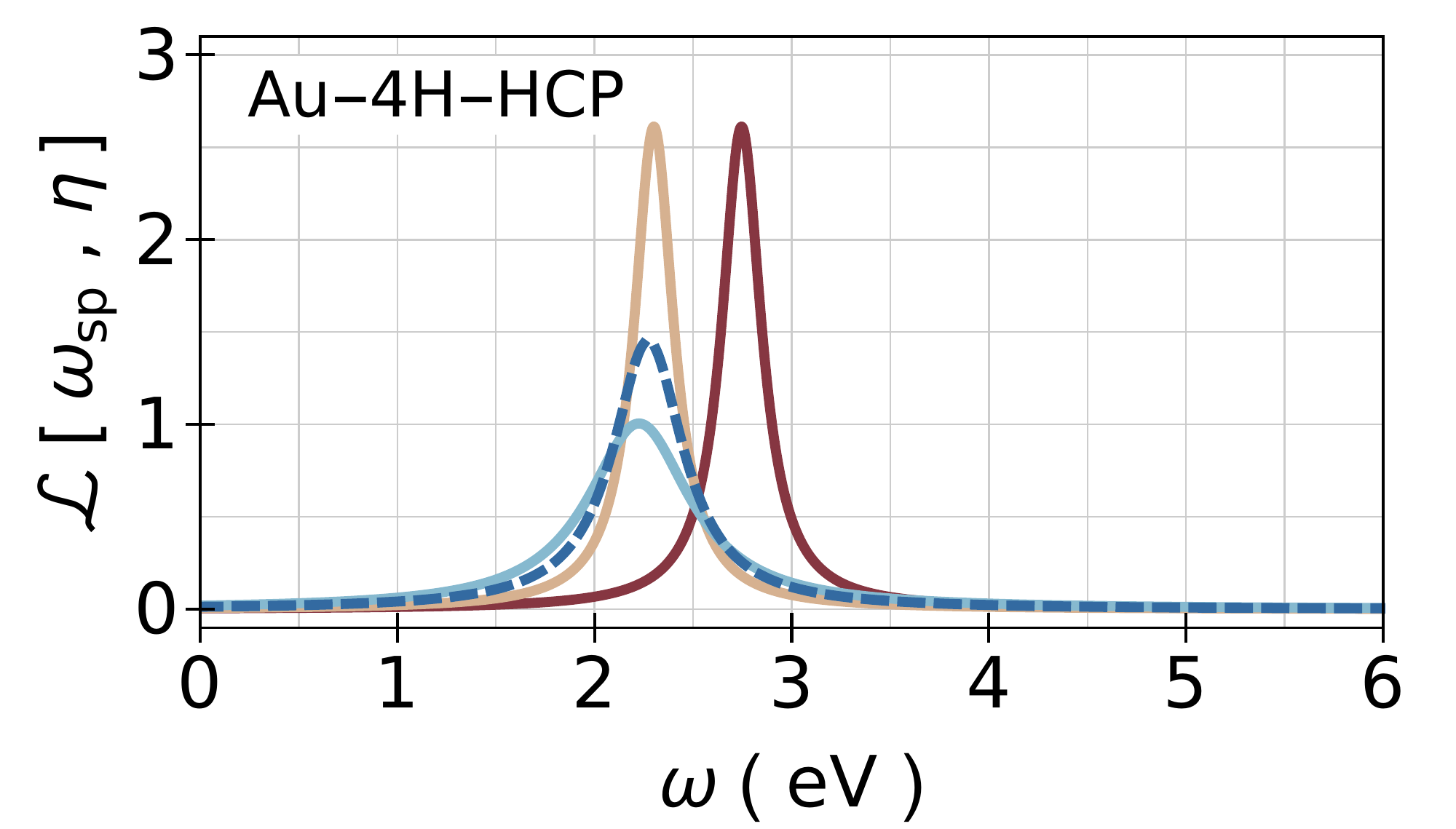}}

\subfloat{ \includegraphics[width=0.3\textwidth]{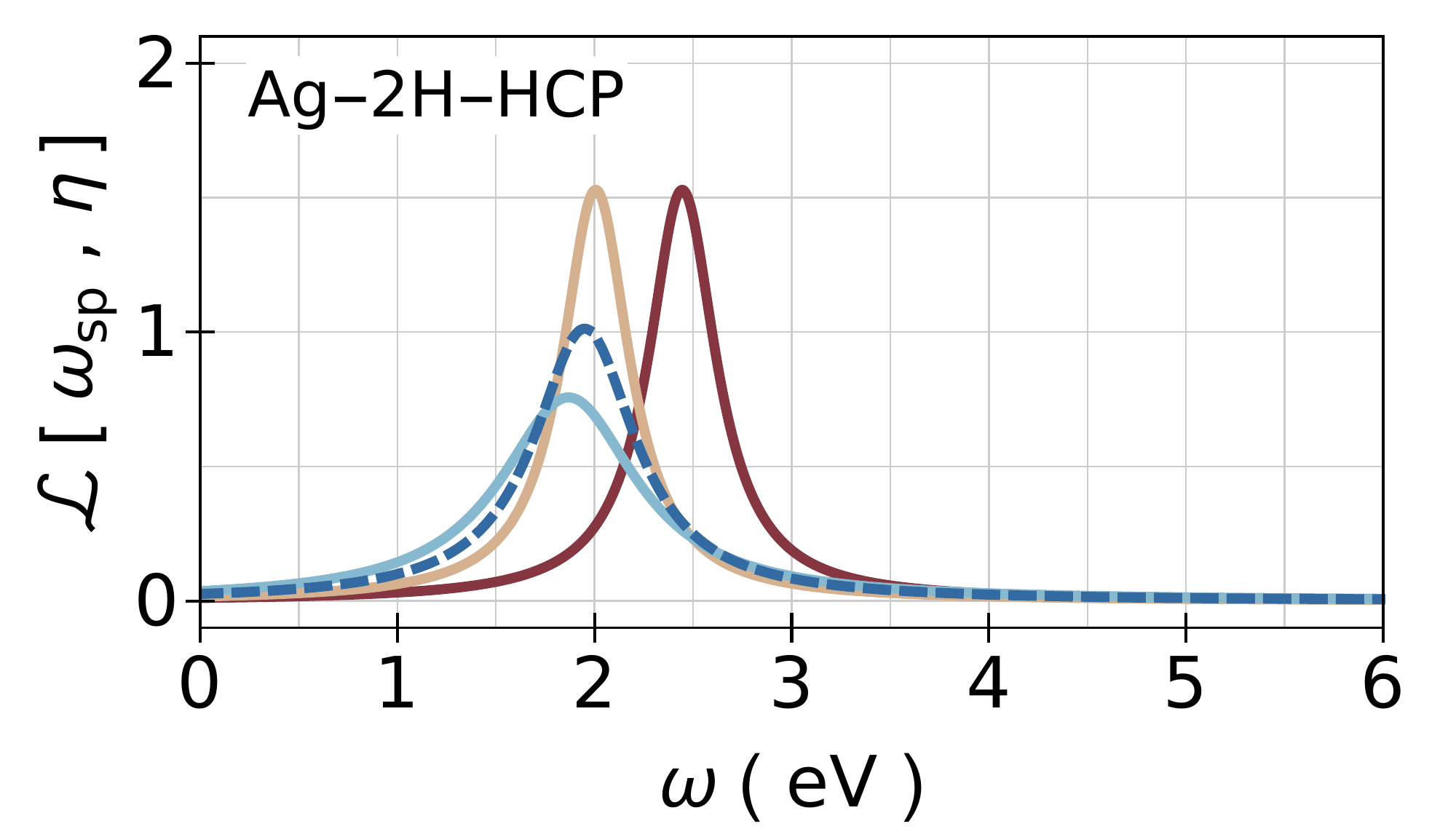}}
\subfloat{ \includegraphics[width=0.3\textwidth]{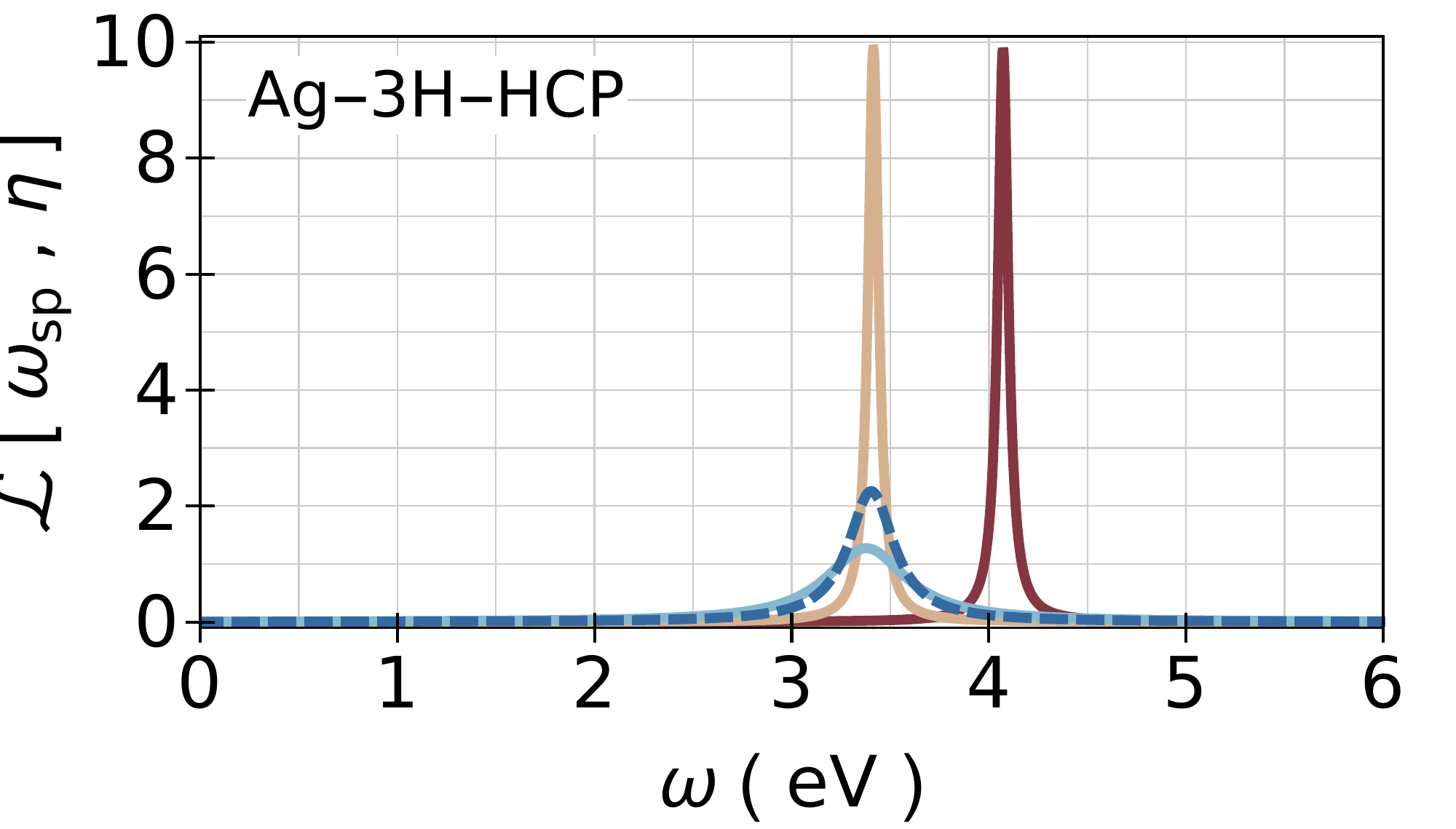}}
\subfloat{ \includegraphics[width=0.3\textwidth]{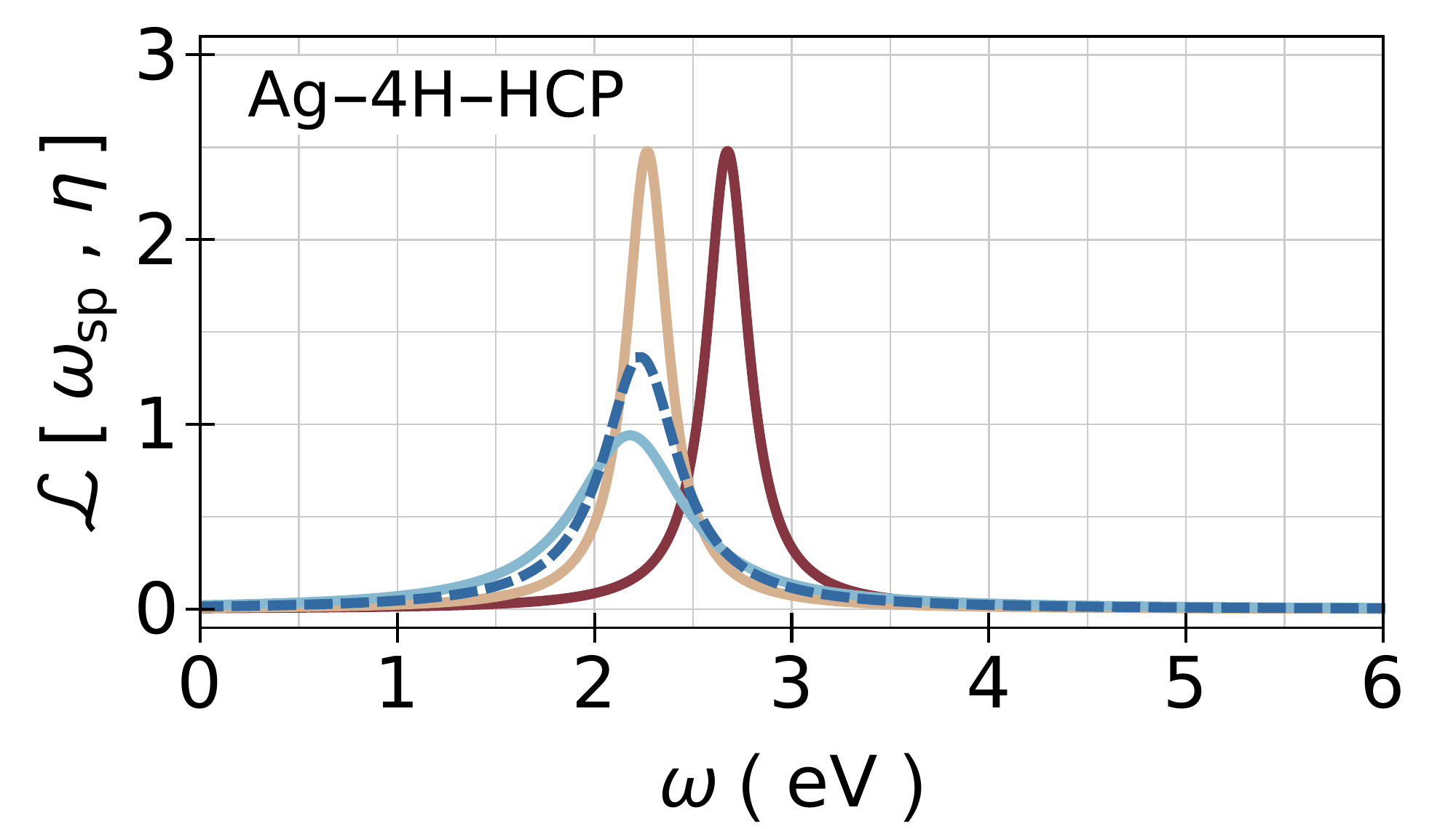}}

\subfloat{ \includegraphics[width=0.3\textwidth]{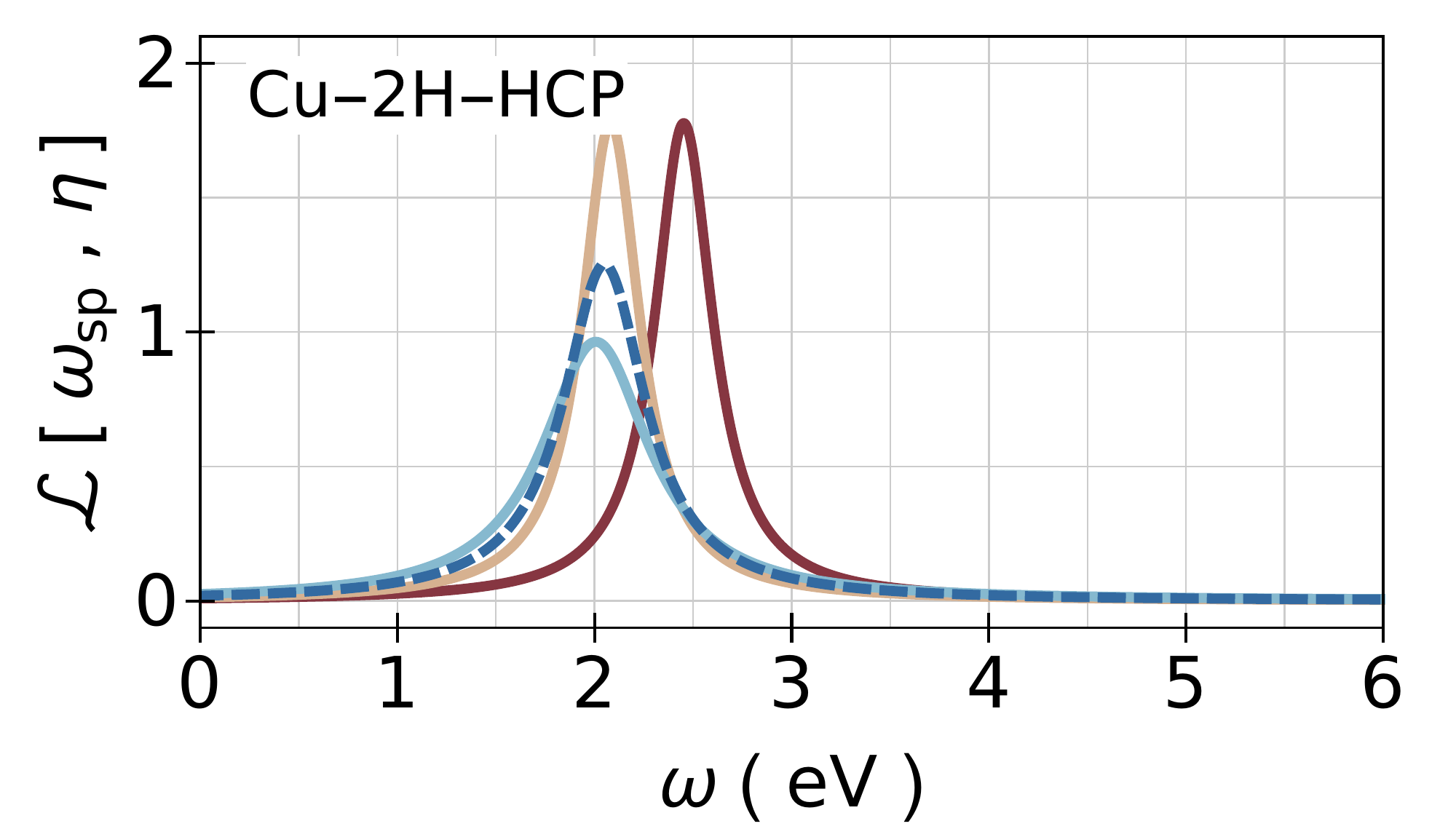}}
\subfloat{ \includegraphics[width=0.3\textwidth]{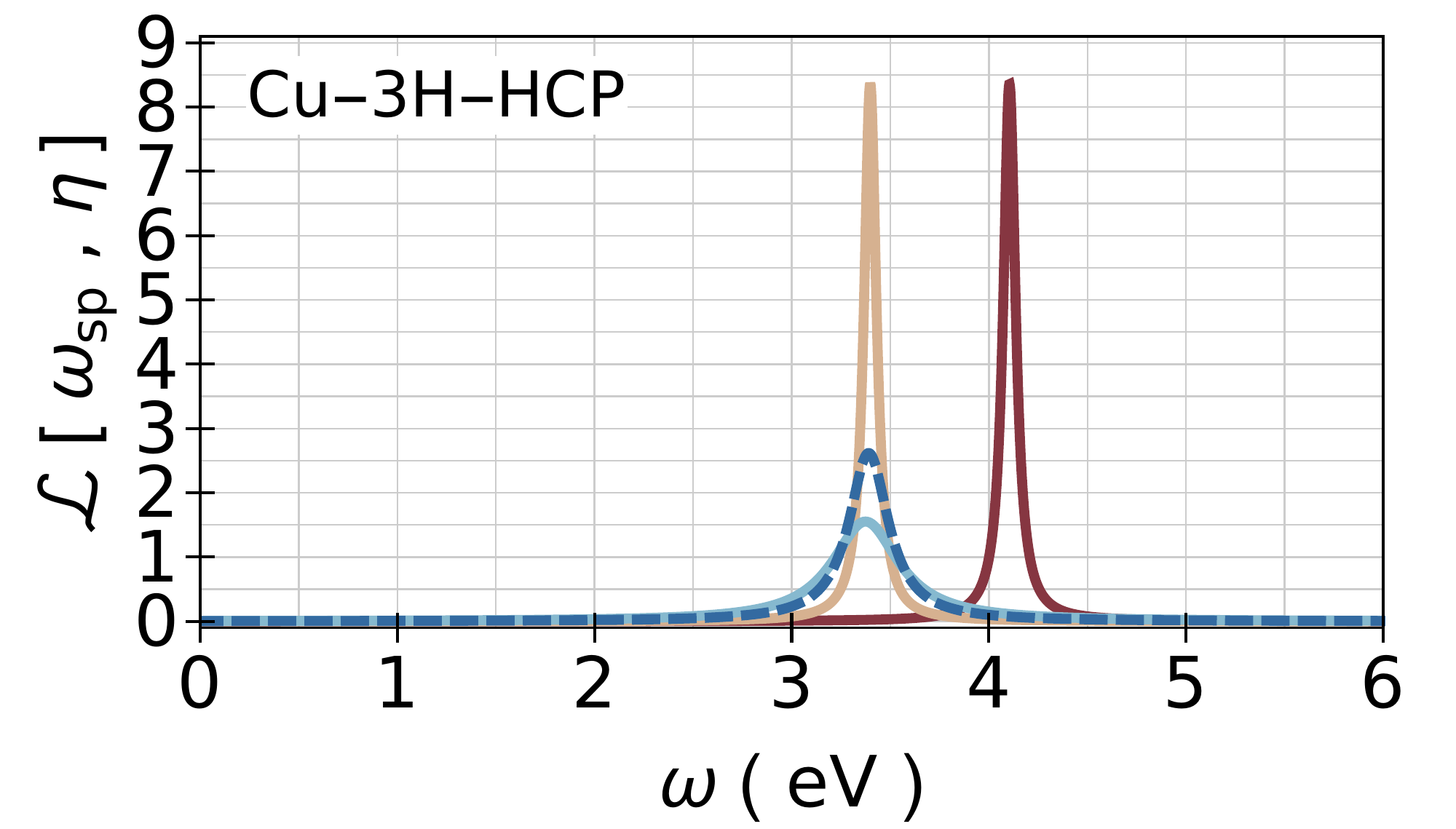}}
\subfloat{ \includegraphics[width=0.3\textwidth]{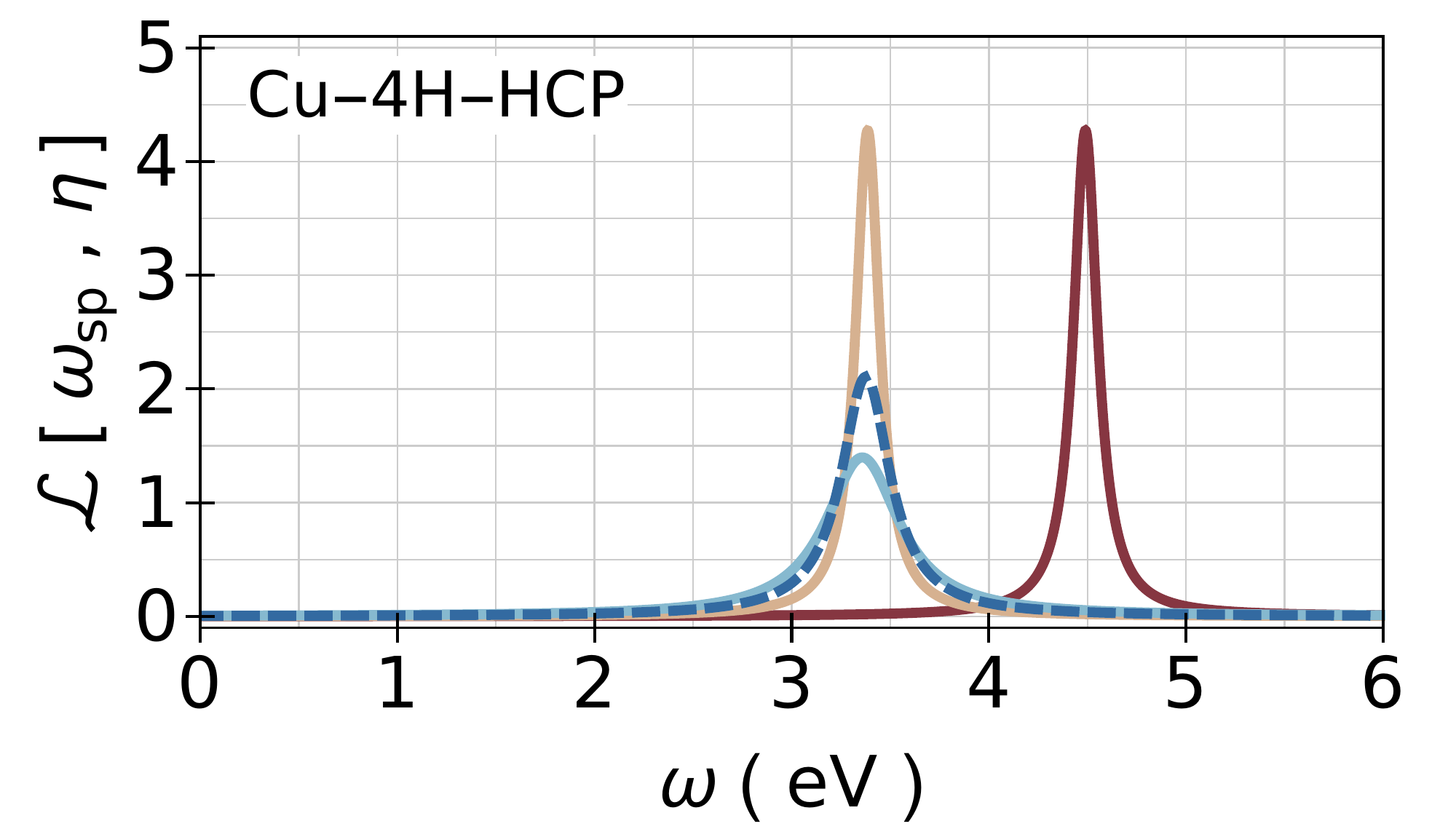}}

\caption{Hosting-matrix dielectric matrix($\ep_\mathrm{m}$), the theory-dependent  constant ($\alpha$) and the size-dependence, controlled by the nanoparticle radius ($R$), of the localized surface-plasmon resonance of noble-metal nanoparticles at  $T=400$~K.}

\label{fig:SurfPlasT400K}
\end{figure}

\begin{table}[H]
\renewcommand{\arraystretch}{1.5} \setlength{\tabcolsep}{10pt}
\begin{center}
{\footnotesize
\begin{tabular}{lccccccc} \hline \hline 

  &  $\lambda$  &  $\omega_\mathrm{p}$ (eV)  &  $\eta_\mathrm{p}$ (eV)  &  $\omega_\mathrm{sp}$ (eV)  &  $\eta$ (eV)  & $\tau$ (fs) & $\mathcal{L}[\omega_\mathrm{sp},\eta$]     \\ \cline{2-7}

Au-2H-HCP  &  1.454  &  3.887  &  0.769  &  0.895  &  0.973  &  4.249  &  0.260   \\ 

Au-3H-HCP  &  0.144  &  8.253  &  0.075  &  2.794  &  0.281  &  14.697  &  0.026  \\ 

Au-4H-HCP  &  0.473  &  6.021  &  0.244  &  2.001  &  0.438  &  9.431  &  0.229  \\[0.25cm]

Ag-2H-HCP  &  0.715  &  5.142  &  0.416  &  1.632  &  0.629  &  6.577  &  0.848  \\ 

Ag-3H-HCP  &  0.115  &  8.863  &  0.064  &  3.002  &  0.282  &  14.653  &  0.020  \\ 

Ag-4H-HCP  &  0.422  &  6.115  &  0.257  &  2.027  &  0.467  &  8.864  &  0.219  \\[0.25cm]

Cu-2H-HCP  &  0.652  &  5.711  &  0.358  &  1.875  &  0.509  &  8.119  &  0.386  \\ 

Cu-3H-HCP  &  0.16  &  8.574  &  0.076  &  2.907  &  0.243  &  17.007  &  0.019  \\ 

Cu-4H-HCP  &  0.277  &  7.300  &  0.149  &  2.465  &  0.302  &  13.682  &  0.050  \\

\hline \hline
\end{tabular}}
\end{center}
\caption{Localized surface-plasmon resonance in the spherical nanoparticles with radii of $R=20$~nm and embedded on a hosting matrices with a dielectric permittivity, $\ep_\mathrm{m}=4.0$ at the $830$~nm ($\sim1.494$~eV) wavelength and at a temperature of $T=400$~K.
}
\label{tab:SurfPlas400}
\end{table}

\newpage

\section{Mechanical properties}

\begin{table}[H]
\renewcommand{\arraystretch}{1.3} \setlength{\tabcolsep}{5pt}
\begin{center}
{\footnotesize
\begin{tabular}{lccccccccc} \hline \hline 

  &  $B$ (GPa)  &  $Y$ (GPa)  &  $S$ (GPa)  &  $\nu_\mathrm{P}$  &  $v_\mathrm{P}$ (m/s)  &  $v_\mathrm{B}$ (m/s)  &  $v_\mathrm{G}$ (m/s)  &  $v_\mathrm{D}$ (m/s)  &  $\Theta_\mathrm{D}$ (K) \\\cline{2-10} 
Au-2H-HCP  &  141  &  67  &  24  &  0.42  &  3086  &  2791  &  1141  &  1257  &  143   \\
Au-3H-HCP  &  147  &  95  &  34  &  0.39  &  3256  &  2845  &  1371  &  1537  &  175   \\
Au-4H-HCP  &  147  &  74  &  26  &  0.41  &  3166  &  2843  &  1206  &  1346  &  153  \\[0.25cm]
Ag-2H-HCP  &  87  &  86  &  32  &  0.33  &  3666  &  2999  &  1826  &  2038  &  229   \\
Ag-3H-HCP  &  89  &  77  &  29  &  0.35  &  3613  &  3022  &  1715  &  1889  &  213   \\
Ag-4H-HCP  &  88  &  94  &  36  &  0.32  &  3740  &  3012  &  1921  &  2137  &  241   \\[0.25cm]
Cu-2H-HCP  &  135  &  151  &  57  &  0.31  &  5051  &  4036  &  2630  &  2938  &  375  \\
Cu-3H-HCP  &  137  &  158  &  60  &  0.31  &  5111  &  4055  &  2695  &  2980  &  380   \\
Cu-4H-HCP  &  136  &  141  &  53  &  0.33  &  4985  &  4043  &  2525  &  2822  &  360  \\
\hline \hline
\end{tabular}}
\end{center}
\caption{Elastic modulus, the Poisson ratios, sound velocities and the Debye temperatures of Au, Ag and Cu polytypes.}
\label{tab:ElasticMod}
\end{table}

\section{Temperature dependence of the total thermal conductivity}
\begin{table}[H]
\renewcommand{\arraystretch}{1.5} \setlength{\tabcolsep}{18pt}
\begin{center}
{\footnotesize
\begin{tabular}{lccccc} \hline \hline 

  &  100  &  200  &  300  &  400  &  500 \\ \cline{2-6}
  
Au-2H-HCP  &  9  &  9  &  9  &  10  &  11   \\

Au-3H-HCP  &  263  &  323  &  342  &  351  &  349    \\

Au-4H-HCP  &  42  &  52  &  57  &  61  &  66    \\[0.25cm]

Ag-2H-HCP  &  30  &  26  &  32  &  35  &  38    \\

Ag-3H-HCP  &  280  &  409  &  452  &  465  &  469    \\

Ag-4H-HCP  &  58  &  54  &  63  &  72  &  77    \\[0.25cm]

Cu-2H-HCP  &  86  &  69  &  71  &  73  &  75    \\

Cu-3H-HCP  &  454  &  543  &  537  &  532  &  535    \\

Cu-4H-HCP  &  163  &  184  &  206  &  217  &  222    \\

\hline \hline
\end{tabular}}
\end{center}
\caption{Temperature-dependent total thermal conductivity in the (W$\cdot$m$^{-1}$ $\cdot$K$^{-1}$) units of the Au, Ag and Cu polytypes.}
\label{tab:TotThermCond}
\end{table}

\end{document}